\documentclass[12pt,preprint]{aastex}

\usepackage{times}
\newcommand{\commentpcf}[1]{}

\def\sixteen{$\mathrm{RA}=16^\mathrm{H} - 20^\mathrm{H}$}
\def\BdotR{$B \cdot R$}

\def\microG{$\mu$G}

\def\deg{$^\circ$}

\def\glon{$\ell$}
\def\glat{$b$}
\def\elon{$\lambda$}
\def\elat{$\beta$}

\def\PAcel{PA$_\mathrm{RA}$}

\def\ebv{$E$(B--V)} 

\def\OI{O$^{\rm o}$}
\def\NeI{Ne$^{\rm o}$}

\def\TiII{Ti$^{\rm +}$}
\def\HH{H$_2$}
\def\NHI{$N$(H$^{\rm o}$)}

\def\myfunct{$F_\mathrm{i} = ~\overline{sin(\theta_\mathrm{n,i})}~ $}

\def\myfuncttwo{$F^\prime_\mathrm{i} = ~\overline{sin(\theta_\mathrm{n,i})/G_\mathrm{n}}~ $}

\def\PA{$PA$}
\def\Gfact{$G_\mathrm{n}$}

\def\thetaji{$\theta_\mathrm{n,j}$}

\def\HI{H$^{\rm o}$}
\def\DI{D$^{\rm o}$}
\def\FeII{Fe$^{\rm +}$}
\def\CaII{Ca$^{\rm +}$}
\def\MgII{Mg$^{\rm +}$}

\def\HeI{He$^{\rm o}$}

\newcommand{\Pol}{$P$}

\def\P5{$P_5$}

\def\el{\hbox{$\lambda$}}
\def\eb{\hbox{$\beta$}}

\def\kms{\hbox{km s$^{-1}$}}
\def\deeg{\hbox{$^{\rm o}$}}
\def\NH{$N$(H)}
\def\Lya{\hbox{Ly$\alpha$}}

\def\cmtwo{cm$^{-2}$}

\def\cc{cm$^{-3}$}

\def\persec{s$^{-1}$}
\def\Gn{$G_\mathrm{n}$}
\def\deeg{$^\circ$}
\def\fluxnine{$F_\mathrm{3}$}
\def\PdP{$P/dP$}

\def\Bbest{${B}_\mathrm{i=best}$}
\def\Fbest{${F}_\mathrm{i=best}$}
\begin{document}


\slugcomment{}

\title{The Interstellar Magnetic Field Close to the Sun II. }
\shorttitle{}
\shortauthors{Frisch et al.}


\author{P. C. Frisch}
\affil{Dept. Astronomy and Astrophysics, University of Chicago,
Chicago, IL  60637}


\author{B-G Andersson}
\affil{SOFIA, USRA}


\author{A. Berdyugin and V. Piirola}
\affil{Finnish Centre for Astronomy with ESO, University of Turku, Finland}


\author{R. DeMajistre}
\affil{The Johns Hopkins University Applied Physics Laboratory, Laurel, MD}

\author{H. O. Funsten}
\affil{Los Alamos National Laboratory, Los Alamos, NM}

\author{A. M. Magalhaes and D. B. Seriacopi}
\affil{ Inst. de Astronomia, Geofisica e Ciencias Atmosfericas, Universidade de Sao Paulo,
  Brazil}

\author{D. J. McComas\altaffilmark{1}}
\affil{Southwest Research Institute, San Antonio, TX}
\altaffiltext{1}{also University of Texas, San Antonio, TX}



\author{N. A. Schwadron}
\affil{Space Science Center, University of New Hampshire}


\author{J. D. Slavin}
\affil{Harvard-Smithsonian Center for Astrophysics, 
  Cambridge, MA}

\author{S. J.  Wiktorowicz}
\affil{Dept. Astronomy, University of California at Santa Cruz, Santa Cruz, CA}
\author{}


\begin{abstract}
The magnetic field in the local interstellar medium (ISM) provides a
key indicator of the galactic environment of the Sun and influences
the shape of the heliosphere.  We have studied the interstellar
magnetic field (ISMF) in the solar vicinity using polarized starlight
for stars within 40 parsecs of the Sun and 90\deeg\ of the heliosphere
nose.  In Frisch et al. (2010, Paper I) we developed a method for
determining the local ISMF direction by finding the best match to a
group of interstellar polarization position angles obtained towards
nearby stars, based on the assumption that the polarization is
parallel to the ISMF.  In this paper we extend the analysis by
utilizing weighted fits to the position angles and by including new
observations acquired for this study.  We find that the local ISMF is
pointed towards the galactic coordinates \glon,\glat$=47^\circ \pm
20^\circ, 25^\circ \pm 20^\circ$.  This direction is close to the
direction of the ISMF that shapes the heliosphere, 
\glon,\glat$=33^\circ \pm 4^\circ, 55^\circ \pm 4^\circ$,
as traced by the center of the ``Ribbon'' of energetic neutral
atoms discovered by the Interstellar Boundary Explorer (IBEX) mission.
Both the magnetic field direction and the kinematics of the local ISM
are consistent with a scenario where the local ISM is a fragment of
the Loop I superbubble.  A nearby ordered component of the local ISMF has
been identified, in the region \glon$\approx 0^\circ \rightarrow 80^\circ$ and \glat$\approx 0^\circ \rightarrow 30^\circ$, where PlanetPol data show a
distance-dependent increase of polarization strength. 
The ordered component extends to within 8 parsecs of the Sun and implies a
weak curvature in the nearby ISMF of $\sim 0.25^\circ$ per parsec.
This conclusion is conditioned on the small sample of stars available
for defining this rotation.
Variations from the ordered component suggest a turbulent component of
$\sim 23^\circ$.  
The ordered component and standard relations between
polarization, color excess, and \HI\ column density predict 
a reasonable increase of \NH\ with distance in the local ISM.
The similarity of the ISMF directions traced
by the polarizations, the IBEX Ribbon, and pulsars inside the Local
Bubble in the third galactic quadrant suggest that the ISMF is
relatively uniform over spatial scales of 8--200 parsecs and is more
similar to interarm than spiral-arm magnetic fields.  The ISMF
direction from the polarization data is also consistent with
small-scale spatial asymmetries detected in GeV-TeV cosmic rays with a
galactic origin.  The peculiar geometrical relation found earlier
between the cosmic microwave background dipole moment, the heliosphere
nose, and the ISMF direction, is supported by this study.  The
interstellar radiation field at $\sim 975$ \AA\ does not appear to
play a role in grain alignment for the low density ISM studied here.

\end{abstract}

\keywords{ISM: magnetic field, clouds, 
solar system: general, 
Galaxy:  solar neighborhood,
cosmology: cosmic microwave background}

\section{Introduction \label{sec:intro}}

The characteristics of the interstellar magnetic field (ISMF) in the
warm, low density, partially ionized interstellar gas near the Sun are
difficult to study.  A new diagnostic of the magnetic field in the
Local Interstellar Cloud (LIC) surrounding the heliosphere is provided
by the giant arc, or `Ribbon', of energetic neutral atoms (ENAs)
discovered by the Interstellar Boundary Explorer (IBEX) spacecraft
\citep{McComas:2009sci,McComas:2011grlrev}. ENAs are formed by
charge-exchange between interstellar neutral hydrogen atoms and the
solar wind and other heliospheric ions.  The ENA Ribbon is observed
towards sightlines that are perpendicular to the ISMF direction as it
drapes over the heliosphere, and is superimposed on a distributed ENA
flux that traces the global heliosphere (Appendix \ref{app:ibex}
provides additional information on the Ribbon and Ribbon models).  In
\citet[][Paper I]{Frisch:2010ismf1} we showed that the measurements of
interstellar polarization towards nearby stars yields a best-fit
ISMF direction that is close to the field direction obtained from the
IBEX Ribbon.  In this paper, we further study the connection between
the magnetic field sampled by the IBEX Ribbon and the magnetic field
in the local ISM.

Starlight polarized by magnetically aligned dust grains in the ISM
shows that the magnetic field in our galactic neighborhood has two
large-scale components, a uniform field parallel to the galactic plane
that is directed towards \glon$\sim 82.8^\circ$\footnote{In this
paper, galactic coordinates are denoted by $\ell,~b$, and ecliptic
coordinates by $\lambda,~\beta$.  Plots in ecliptic coordinates are
labeled as such.  } and extends beyond a kiloparsec, and a giant
magnetic structure, Loop I, that is within 100 pc and dominates the
northern sky and probably reaches to the solar location.  The ISMF of
Loop I is traced by optical polarizations
\citep{MathewsonFord:1970,Heiles:1976araa,Santosetal:2010,Berdyuginetal:2011},
the polarized radio continuum \citep{Berkhuijsen:1973,Wolleben:2007},
and Faraday rotation measures
\citep{TaylorStilSunstrum:2010,MaoZweibel:2010ismf}.  It is a
superbubble formed by stellar winds and supernova in the Sco-Cen
Association during the past $\sim 15$ Myrs
\citep[e.g.][]{deGeus:1992,Frisch:1995rev,Frisch:1996,Heiles:1998lb,Heiles:2009}.
The interstellar magnetic field is swept up during superbubble
expansion, creating a magnetic bubble that persists through the late
stages of the bubble evolution \citep{TilleyBalsaraHowk:2006}. Optical
polarization and reddening data show that the eastern parts of Loop I,
\glon$=3^\circ - 60^\circ$, \glat$>0^\circ$, are within 60--80 parsecs
of the Sun \citep[][]{Santosetal:2010,Frisch:2011araa}.  If Loop I is
a spherical feature, the ISMF near the Sun should be associated with
the Loop I superbubble
\citep[e.g.][]{Frisch:1990,Heiles:1998whence,Heiles:1998lb}.  The
magnetic structure of Loop I provides a framework for understanding
the magnetic field in the local ISM and the relation between the galactic
magnetic field and the magnetic field that shapes the heliosphere.

In this paper we use high-sensitivity observations of the polarization
toward nearby stars as a tool for determining the direction of the
ISMF near the Sun, within $\sim 40$ parsecs. This study builds on our
earlier study (Paper I) which presents a new method for determining
the ISMF direction, based on testing for the best match between
polarization position angles and the ISMF, where we found that the best-fit
local ISMF direction in the galactic hemisphere regions agrees, to
within the uncertainties, with the ISMF direction defined by the
center of the IBEX Ribbon arc.  Our new study differs from the first
study in that the method is broadened to include weighted fits (\S
\ref{sec:method}),which is possible because of the inclusion of new
data (\S \ref{sec:data}).  The best-fit ISMF direction found from
both unweighted fits (\S \ref{sec:unweight}) and weighted fits (\S
\ref{sec:weight}) is close to the direction of the ISMF shaping the
heliosphere (\S \ref{sec:ibex}).  We also look closely at the
PlanetPol data \citep[][\S \ref{sec:pp}]{BaileyLucas:2010planetpol}
and use the upper envelope of the polarization-distance curve (\S
\ref{sec:ppdispol}, Appendix \ref{app:pol}) to show that there is an ordered component in the
local ISMF with a direction that varies weakly with distance and
extends to within 8 parsecs of the Sun (\S \ref{sec:pppa}).  The
ordered component provides information on magnetic turbulence.  An
anti-correlation of polarization with radiation fluxes suggests
ineffective radiative torques (\S \ref{sec:ppisrf}, Appendix
\ref{app:isrf}).  The best-fit ISMF direction is compared with the
field direction derived from the IBEX Ribbon, and the relation between
the LIC and G-clouds (\S \ref{sec:ibex}).  For our new ISMF direction
and the new IBEX heliosphere nose direction, the symmetry of the
cosmic microwave background (CMB) dipole moment with respect to the
heliosphere nose direction is still apparent (\S \ref{sec:cmb}).  The
similar directions of the ISMF over scales of several hundred parsecs
suggests an interarm-type field (\S \ref{sec:fr}).  The local ISMF
direction also appears to affect anisotropies observed in GeV--TeV
galactic cosmic rays (\S \ref{sec:gcr}).  Results are summarized in \S
\ref{sec:conclusion}.

\section{Method of finding the best-fit ISMF direction
  \label{sec:method}}

Optical polarization is an important probe of the direction of the
ISMF in interstellar clouds, both over large spatial scales
\citep{MathewsonFord:1970} and the very local ISM
\citep{Tinbergen:1982,Frisch:1990,Frisch:2010ismf1}.  The observed
polarization from aligned grains depends on the total column of dust
and the grain size distribution \citep{Mathis:1986}, the fraction of
asymmetrical dust grains and alignment efficiency of the grains
(Whittet et al. 2008), and the projection of the magnetic field onto
the sky.  The axis of lowest average grain opacity is parallel to the
ISMF direction
\citep{DavisGreenstein:1951,Martin:1971,Roberge:2004,Lazarian:2007rev}.
\footnote{Leverett Davis provided the first viable model for the
relation between the ISMF and starlight polarization, and also first
proposed the existence of the solar wind cavity in the ISM, now known
as the heliosphere. He proposed a solar wind cavity of radius $\sim
200$ AU partly out of the necessity of excluding the interstellar
magnetic field from the solar system \citep{Davis:1955}.}
Polarization is a pseudo-vector entity (always in the range $0^\circ -
180^\circ$), and in denser regions radiative transfer effects can
influence the observed polarization both due to turbulence in the
probed medium (Jones, Dickey \& Klebe 1992; Ostriker, Stone \& Gammie,
2001), and due to the influence of foreground depolarizing screens
\citep{AnderssonPotter:2006,Andersson:2012rev}.
\nocite{JonesKlebe:1992,OstrikerGammie:2001ismf} ISM within 10 parsecs
has low average ISM densities, \NH $\lesssim 10^{18} $ \cmtwo\ and
$<n>\sim 0.1$ \cc, suggesting there is very little nearby dust.
Therefore depolarizing foreground screens and collisional disruption
of grain alignment are less likely locally compared to distant dense
cloud regions. This low density ISM is partially ionized
\citep{Frisch:2011araa} so that gas and the ISMF will be tightly
coupled.

Two properties of polarized starlight could in principle be exploited
to obtain the ISMF direction causing the polarization: the position
angle of the polarizations, which are measured with respect to a great
circle meridian and increase positively to the east, and the
polarization strengths that reach a maximum for sightlines
perpendicular to the ISMF for a uniform medium.  The drawbacks with
using polarization strengths to determine the ISMF direction are that
polarization varies with the dust column density of the sightline, and
polarizations are weaker near the magnetic pole where the statistical
significance of detections will be lower.  Polarization directions are
insensitive to the dust column density, but sensitive to magnetic
turbulence that could distort the field over either small or large
scales.

Paper I introduces a new method for determining the ISMF direction
based on finding the best fit to an ensemble of polarization position
angles.  The analysis makes the assumption that the polarization
position angle is parallel to the direction of the interstellar
magnetic field \citep{Roberge:2004}.  This assumption is justified for
the diffuse clouds in Loop I, where the field directions determined
from optical polarization data and synchrotron emission in high-latitude regions agree
\citep{Spoelstra:1972lb}.  The ensemble of data used in
the analysis are chosen to meet a set of selection criteria.  These
data and the criteria are discussed in the following section (\S
\ref{sec:data}).  All measurements meeting the criteria are used in
the analysis, including multiple measurements of the same star by
different observers.
 
The ISMF direction is derived by assuming that a single large-scale
ISMF close to the Sun is aligning the grains, so that the polarization
direction would then be parallel to a great-circle meridian of the
``true'' magnetic field direction.  The best-fit ISMF is
calculated to be the direction that corresponds to the minimum value
of:
\begin{equation} \label{eqn:one}
 F_\mathrm{i} = F(B_\mathrm{i}) = \mathrm{N^{-1}} \sum_\mathrm{n=1}^\mathrm{N} ~|
\frac{\mathrm{sin}(\theta_\mathrm{n}(B_\mathrm{i}))}{G_\mathrm{n}}|
\end{equation}
where $\theta_\mathrm{n}(B_\mathrm{i})$ is the polarization position
angle \PA$_\mathrm{n}$ for star n, calculated with respect to the
$\mathrm{i^\mathrm{th}}$ possible interstellar magnetic field
direction $B_\mathrm{i}$, and where the sum is over $N$ stars.
\Gfact\ is the weighting factor for each star \emph{n} based on the
measurement uncertainties of the polarization position angle for that
star.  A grid resolution of 1\deeg\ in longitude and latitude is used
for the spacing of $B_\mathrm{i}$ values.  An alternate approach to
find the best-fit field direction \Bbest\ that maximizes the
cosecants, i.e.  maximize $\overline{|
G_\mathrm{n}/sin(\theta_\mathrm{n}(B_\mathrm{i})|}~$, generated
numerical glitches related to dividing by zero, and is not used.

Paper I calculated the ISMF direction with the weighting factor
\Gfact=1.  Statistical weights were ignored since the northern
hemisphere data included a large number of recent high-sensitivity
data collected at ppm accuracy, while the southern hemisphere data
were dominated by data collected in the 1970's that are less sensitive
(1$\sigma >$60 ppm, \S \ref{sec:data}).  Weighting data points by the
uncertainties yielded a magnetic field direction biased by the
difference in uncertainties in the two hemispheres.  The best-fit
ISMF direction from Paper I is listed in Table \ref{tab:ismf2}.  In
this paper, new data (\S \ref{sec:data}) are added and the unweighted
fit is recalculated (\S \ref{sec:unweight}).

New high-precision polarization data make it possible to evaluate
the best-fit ISMF direction by utilizing all of the position
angles in the sample, but weighting individual values according to
their statistical significance.  By combining data based on weighted
values, the emphasis is implicitly placed on the ISMF field in the
distant portions of the sampled ISM, where polarizations may be
stronger relative to the measurement errors (depending on the
patchiness of the ISM, \S \ref{sec:pp}).  The probability distribution
for position angles does not have a normal distribution for weak
polarizations, \PdP$<6$.  The weighting function is based on the
results of \citet{Naghizadeh-KhoueiClarke:1993} that give the
probability distribution for the position angle of linear
polarization.  For true polarization $p_\mathrm{o}$ with a measurement
uncertainty of $\sigma$, and the true position angle in equatorial
coordinates $\theta_\mathrm{o}$, the probability
$G_\mathrm{n}(\theta_\mathrm{obs};~\theta_\mathrm{o},P_\mathrm{o})$ of
observing a position angle $\theta_\mathrm{obs}$ (which will be
$0^\circ$ for the ``true'' value of a perfectly aligned \PA) is:

\begin{equation}  \label{eqn:two}
G_\mathrm{n}(\theta_{\rm{obs}} ;~\theta_{\rm{o}},P_{\rm{o}} ) ~ = ~ 
\frac{1}{\sqrt{\pi}}   ~ \{ \frac{1}{\sqrt{\pi}} + \eta_{\rm{o}}  ~
\rm{exp} (\eta^2_{\rm{o}} ) ~
[1 + \rm{erf}(\eta_{\rm{o}} )]\}
\rm{exp}({-\frac{P^2_{\rm{o}}} {2}} )  
\end{equation}

\noindent where
$\eta_\mathrm{o}~=~\frac{P_\mathrm{o}}{\sqrt{2}}~\mathrm{cos}~[2(\theta_\mathrm{obs}-\theta_\mathrm{o})]$
and $P_\mathrm{o}=\frac{P_\mathrm{o}}{\sigma}$, and the function
$erf(z)$ is the Gaussian error function $ erf (Z) ~ = ~
\frac{2}{\sqrt{\pi}} \int_0^Z {exp}({-t^2})~dt$.  The goal is then to
choose a procedure that maximizes the contributions of
high-significance (e.g. high $G_\mathrm{n}$) measurements, and
minimizes the contributions from data points with large values of
sin(\thetaji).  When \Bbest\ is calculated using weighted fits (\S
\ref{sec:weight}), the function $G_\mathrm{n}$ in eqn. \ref{eqn:two}
is adopted as the weighting factor for each star $n$ in
eqn. \ref{eqn:one}.

\section{Polarization data used in study} \label{sec:data}

\begin{figure}
\plotone{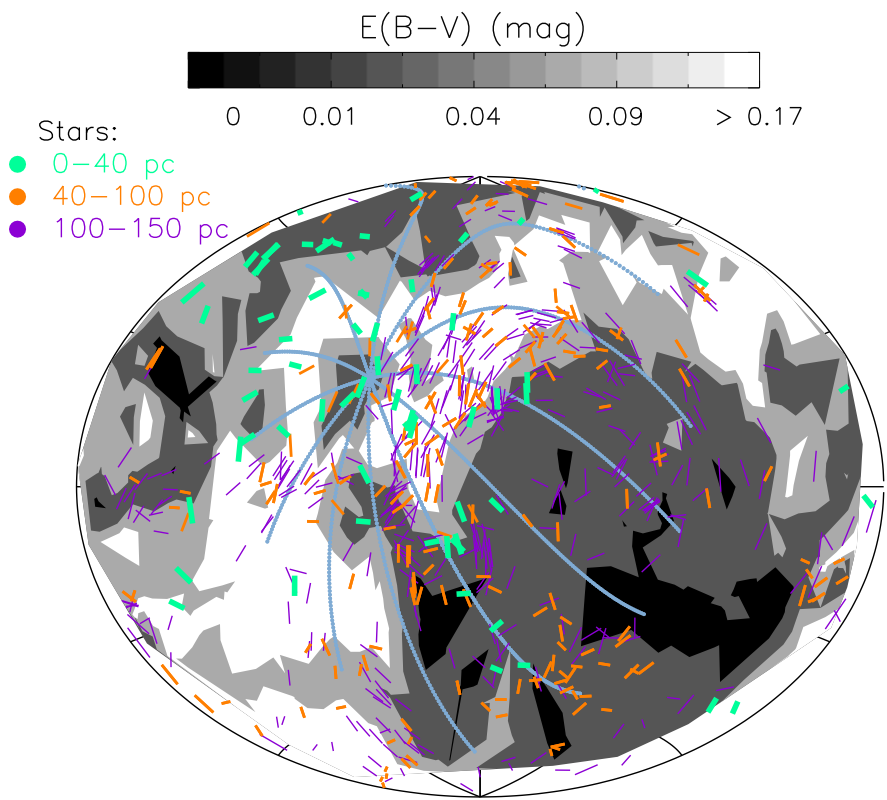}
\caption{The configuration of the ISM associated with Loop I is
displayed.  Filled contours show interstellar dust 100 within parsecs
and the short lines show interstellar polarization directions.  The
extended gray-blue curved lines show the best-fitting ISMF for the
region within 40 pc and 90\deeg\ of the heliosphere nose (\S
\ref{sec:weight}, Table \ref{tab:ismf2}).  Star distance is coded as
follows: within 40 pc (cyan blue), 40--100 pc (orange) and 100--150 pc
(purple). Line length is unrelated to polarization strength.  The Loop
I ISMF direction is clearly defined by the polarizations of distant
stars \citep[e.g.][]{Santosetal:2010}, but also appears to extend
close to the Sun (\S \ref{sec:pp}).  The lightest contours show
maximum dust color excess values, \ebv, while the darkest regions with
\ebv$\le 0.01$ mag show that the interior void of the Loop I bubble
extends to the Sun location.  The color excess \ebv\ contours are
based on the photometric and astrometric data for stars brighter than
V=9 mag in the Hipparcos catalog \citep{Perrymanetal:1997} and the
absolute stellar colors in \citet{Cox_2000}; \ebv\ values are then
smoothed spatially over $\sim \pm 13^\circ$ intervals for overlapping
distances (considering uncertainties).  Variable stars with $\delta \mathrm{V} \ge 0.06$ mag,
as indicated by
the Hipparcos variability index H6, have been omitted.  The \ebv\
contour levels of 0.01, 0.04, 0.09, 0.17 mag correspond to column
densities log($N$(\HI+$H_2$)) of 19.76, 20.37, 20.72, and 20.99
\cmtwo, for $N$(\HI+$H_2$)/\ebv$=5.8 \times 10^{\rm 21}$ atoms \cmtwo\
mag${-1}$ \citep{Bohlin_etal_1978}.  Note the third and fourth
quadrant voids (dark region between \glon$\sim$210\deeg--0\deeg)
compared to the nearby brightest interstellar radiation field (Appendix
\ref{app:isrf}).  The figure is centered on the galactic center, with
longitude increasing towards the left.  The polarization data are from
the sources listed in \S \ref{sec:data} and
\citet{Heiles:2001}. }\label{fig:loopI}
\end{figure}

\begin{figure}
\plotone{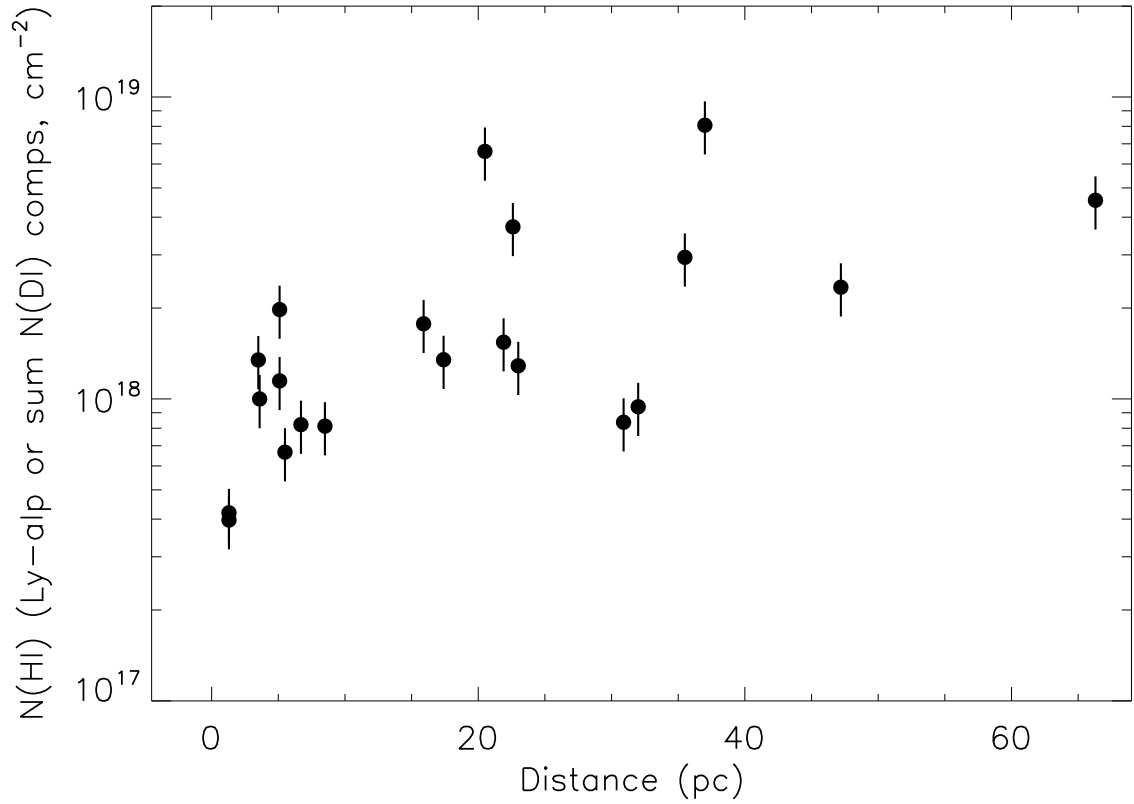}
\caption{Column densities are plotted as a function of distance toward
nearby stars within 40 pc and $90^\circ$ of the heliosphere nose.  The
\HI\ column density is found from either \Lya\ data
\citep{Woodetal:2005lya}, or the the sum of \DI\ components using
\DI/\HI$=1.5 \times 10^{-5}$ \citep{RLII}.  The uncertainties are
plotted as 20\% of the column density. }\label{fig:hIdI}
\end{figure}

The Local Bubble void appears as an absence of both interstellar gas
and dust within $\sim 70$ parsecs, particularly in the third and
fourth galactic quadrants.  Figure \ref{fig:loopI} shows the
distribution of interstellar dust within 100 pc, as traced by color
excess \ebv.\footnote{Stars with $\delta \mathrm{V} \ge 0.06$ mag, according to
the Hipparcos variability index H6, are omitted.}  The interior of the Loop I superbubble appears as a void
in the distribution of interstellar dust within 100 parsecs in the
fourth galactic quadrant (\glon$>270^\circ$).  Models of Loop I as a
spherical object place the Sun in or close to the rim of the shell
(\S \ref{sec:loopI}). This combination of
properties is the basis for selecting polarized stars for this
analysis within 90\deeg\ of the heliosphere nose and 40 parsecs of the
Sun.  The heliosphere nose is remarkably close to the direction of the
galactic center.  IBEX-LO measurements of the $23.2 \pm 0.3$ \kms\
flow of interstellar \HeI\ through the heliosphere provide the nose
direction of \glon$=5.25^\circ \pm 0.24^\circ$, \glat$=12.03^\circ \pm
0.51^\circ$ \citep{McComas:2012bow}.  Most interstellar gas within 40
parsecs is also clumped within 20 parsecs of the Sun, as shown by
interstellar \HI\ and \DI\ (Figure  \ref{fig:hIdI}) and the fact that
the average number of interstellar \FeII, \CaII, and \MgII\ velocity
components is distant-independent for stars within 30 parsecs, and
rises only slightly ($\sim 20$\%) for stars that are 30--40 parsecs
\citep{Frisch:2011araa}.  This spatial interval provides for a
well-defined region for study, accepts most stars in the region of
Tinbergen's original ``patch'' of nearby polarization \citep[][Paper
I]{Tinbergen:1982}, includes the region where PlanetPol
\citep{BaileyLucas:2010planetpol} obtained very high sensitivity
polarization data (\S \ref{sec:data}), and includes nearby stars
observed towards Loop I by \citet{Santosetal:2010}.

These data show that the polarizations in the southern galactic
hemisphere tend to be larger than polarizations in the northern
galactic hemisphere.  For the special case where the star sample is
restricted to stars meeting the spatial and spectral criteria above,
and to recent higher precision data, then the polarizations of
southern stars with \PdP$>2.5$, are $0.024\% \pm 0.017\%$, while for
northern stars they are $0.003\% \pm 0.005\%$.  Whether this
difference, originally noted by Tinbergen (1982), is accurate will
require additional high-sensitivity data (where \PdP\ gives the
polarization strength divided by the $1\sigma$ uncertainty of the
polarization).

New polarization measurements were also collected for this study using
telescopes of the Laboratorio Nacional de Astropfisica (LNA) in Brazil
and the Nordic Optical telescope (NOT) on the island of La Palma in
the Canary Islands.  The LNA data were acquired with the CCD
polarimeter \citep{Magalhaes:1996} on the 0.6m telescope at Pico de
Dios, during the months of June 2008, August 2010, September 2010, and
May 2011.  The NOT data were acquired using the TurPol polarimeter
\citep{Piirola:1973,ButtersPiirola:2009} on the 2.52m telescope on the
island of La Palma in the Canary Islands during June 2010.  Optical
polarimetry is a differential measurement and good observing
conditions are not as critical as for high accuracy photometry.  ISM
polarization data also sample a relatively constant polarization
source, allowing repeated observations over different days or from
different sites to maximize sensitivity and minimize systematic
uncertainties.  These new data are listed in Table
\ref{tab:lnanot}. The columns in this table are HD number, galactic
longitude and latitude, star distance, the polarization in units of
$10^{-5}$, and the polarization position in the equatorial coordinate
system.  The last column gives the observatory where the data were
acquired.

The CCD imaging polarimeter at Pico dos Dias provided polarizations in
the B-band.  The instrument acquires CCD images of the orthogonal
polarizations, in sets of 8 waveplate positions, with several images
at each plate position.  In the case of constant polarizations such as
interstellar polarizations, individual observations can be combined to
reduce uncertainties.  Stars brighter than $\sim 7$ mag were observed
through a neutral density filter.

The TurPol data were obtained using the UBV mode of the polarimeter
with high voltage for the R and I photomultipliers switched off.  The
effective wavelengths of the UBV bands are 3600, 4400, and 5300 \AA,
respectively. With this mode, typical count rates of $10^6$ counts
s$^{-1}$ were achieved giving a precision of 0.01\% for a $3 \sigma$
detection.  Observations of two zero-polarization standards were
acquired each night at the same precision in order to determine the
instrumental polarization.  The position angle calibration was
obtained with observations of two high polarization standard stars.
For interstellar polarizations, position angles do not depend
significantly on the wavelength, allowing the use of weighted-averages
from the UBV bands to achieve the highest possible precision; these
'broad-band' position angle measurements are listed in Table
\ref{tab:lnanot}.


The early \citet{Tinbergen:1982} survey, made during 1973-1974 from La
Silla (Chile) and Hartebeespoortdam (S. Africa) in the southern
hemisphere, and Leiden (Netherlands) in the northern hemisphere,
remains the most extensive all-sky survey of weak polarizations, with
$1 \sigma$ sensitivity on the $Q$ and $U$ Stokes parameters of
0.006--0.009 \%, depending on the instrument and data set.  These weak
polarizations are biased towards larger polarization strengths, since
$P^2 = Q^2 + U^2$ is always positive.
\citet{NaghizadehClarke:1993stats} evaluated the statistics of stars
observed from each observing station in the Tinbergen data set, and
concluded that residual instrumental polarizations may remain in the
K-star data collected from Hartebeespoortdam.  We have examined the
data from Hartebeespoortdam for non-active K-stars that also have
\PdP$ >2.5$.  The seven stars that match the data criteria are located
at angles of $86^\circ \pm 32^\circ$ from the ISMF direction defined
by the center of the IBEX Ribbon arc.  If the direction of the ISMF at
the heliosphere is the same as in the local ISMF direction, then for a
uniform ISM the maximum polarizations are expected at angles of
90\deeg\ from the ISMF pole.  We therefore assume that any
contribution of instrumental polarization to the Hartebeespoortdam
data is insignificant, while acknowledging that high-sensitivity
observations and temporal monitoring of the polarizations of these
stars are needed.

This study also includes the \citet{BaileyLucas:2010planetpol}
polarizations of nearby stars obtained with the PlanetPol instrument
at the William Herschel telescope.  These linear polarization data
were collected over a broad red band (maximum sensitivities at
7000--8000\AA), and achieved typical accuracies of parts-per-million.
PlanetPol data were also included in the Paper I analysis, but without
screening for stellar activity.  Twenty-six stars in the PlanetPol
catalog that passed the selection criteria above were incorporated
into our analysis of the local ISMF direction and are listed in
Appendix \ref{app:A}.  \citet{BaileyLucas:2010planetpol} measured 18
northern hemisphere stars that were also observed by Tinbergen, and
detected weak polarizations towards roughly half of the stars that
were unpolarized at the sensitivities of the Tinbergen
data.\footnote{Note that the data listed for several of the Tinbergen
measurements in Table 2 of \citet{BaileyLucas:2010planetpol} contains
typographical errors (Philip Lucas, private communication).}
PlanetPol data showed that polarizations in the $\mathrm{RA}>17^\mathrm{H}$ 
region increased with distance, and measured
a weak interstellar polarization of $17.2 \pm 1.0$ ppm towards
$\alpha$ Lyr (Vega, HD 172167), 8 parsecs away.  Bailey et al.  argue
that the face-on debris disk around Vega is unlikely to contribute to
the observed polarization signal since the instrument aperture is 5"
and the radius of the central hole in the disk is $11" \pm 2"$.  We
show that the $\alpha$ Lyr polarization data have follow polarization
trends of other nearby stars (\S \ref{sec:pp}) so those data are
retained in this sample.

\citet{Santosetal:2010} surveyed interstellar polarizations towards
Loop I for stars out to distances of 500 parsecs, using the imaging
polarimeter at LNA.  Typical mean accuracies 
$1\sigma=0.05$\% were achieved in the V-band.  Forty stars in the
Santos et al. catalog passed the regional and activity selection
criteria above and are listed in the Appendix \ref{app:A}.

\begin{figure}
\plotone{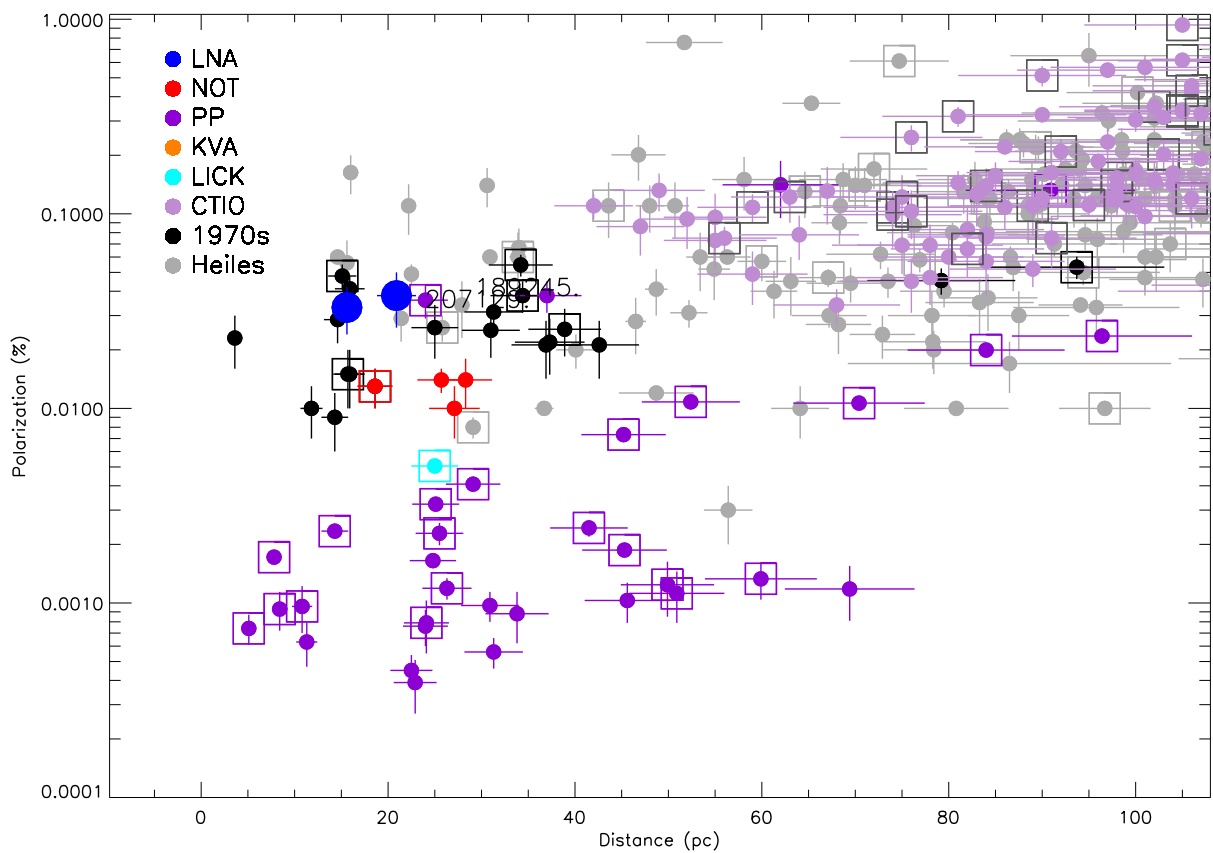}
\caption{Interstellar polarizations with \PdP$> 3.0$, and for stars
within $90^\circ$ of the heliosphere nose, are plotted as a function
of distance.  Data sources are color-coded, according to the scheme in
the figure.  Dark purple points are the PlanetPol data.  The boxes
indicate stars in region \sixteen\ where
\citet{BaileyLucas:2010planetpol} found that polarizations increase
with distance.  This region overlaps the North Polar Spur region,
which forms the tangential sections of Loop I.  PlanetPol requires
bright stars to achieve high accuracy, which suggests that
systematically lower PlanetPol polarizations beyond 40 pc (and perhaps
closer) may be due to selection effects related to the patchiness of
foreground dust. 
Data are from Table \ref{tab:lnanot}, the references listed in \S
\ref{sec:data}, and \citet{Heiles:2000pol}.
These data were collected over a variety of
wavelength ranges (\S \ref{sec:data}), and differences in polarization strengths will occur because
of the wavelength dependence of polarization \citep{Serkowskietal:1975}.
}\label{fig:dvrspolcent}
\end{figure}

Polarizations of stars used in the determination of the local ISMF
direction (within 40 parsecs) are plotted against the distance of the
star in Figure \ref{fig:dvrspolcent}.  For comparison, the polarizations
of more distant stars (40-110 parsecs) within 90\deeg\ of the
heliosphere nose are also shown \citep[from the catalogs
of][]{Santosetal:2010,Heiles:2000pol,Leroy:1993lism,Leroy:1999}.  A
general increase in polarization strength with distance is seen (this
is discussed in more detail in Appendix \ref{app:pol}).  The
polarizations shown in Figure \ref{fig:dvrspolcent} were measured in
different bandpasses. For the PlanetPol bandpass covering red
wavelengths \citep{BaileyLucas:2010planetpol}, the Serkowski curve of
polarization versus $\lambda/\lambda_\mathrm{max}$ predicts
polarizations that are up to 30\% below the peak polarizations at
$\lambda_\mathrm{max}$ \citep[$\lambda$ is the wavelength of
observation and $\lambda_\mathrm{max}$, which has a median value $\sim
5500$ \AA, is the wavelength of maximum
polarization,][]{Serkowskietal:1975}.  The PlanetPol stars, located in
the northern galactic hemisphere, includes the largest set of
detection of very weak polarizations, while the early
\citet{Tinbergen:1982} dataset provide the most detections in the
southern hemisphere.  For this reason, polarization strengths are not
used in Paper I to trace the ISMF direction.  With the addition of new
high-sensitivity data in the southern hemisphere, all polarization
position angle data can be included in the study using weighted fits
(\S \ref{sec:weight}).

Cool stars are included as target stars to provide adequate spatial
coverage, but not all cool stars are suitable for the high-sensitivity
interstellar polarizations of $\sim 0.01$\% needed for this study.
Magnetic activity is observed in cool stars, especially late K and M
stars.  With the sensitivity levels for most observations used in this
study there is no \emph{a priori} reason to omit stars with active
chromospheres from the sample, however some cool stars are known to
show polarizations in chromospheric lines
\citep{ClarkeFullerton:1996sun}.  Stars of spectral types G and
cooler, and with known active chromospheres, are usually omitted from
the data sample used to evaluate the local ISMF direction.  Screening
for active stars was performed using the lists in
\citet{Gray:2006}. For the same reason, variable stars with variations
$\delta$V $\ge 0.06$ mag are omitted to avoid other intrinsic
magnetic phenomena, by requiring the coarse variability flag, VarFlag word H6,
in the Hipparcos catalog \citep{Perrymanetal:1997}\footnote{See the
VizieR catalog I/239/.} to be H6=1.  The analysis in Paper I did
not screen for stars with known chromospheric activity.

\section{Results} \label{sec:results}

\subsection{ISMF direction from unweighted fit}\label{sec:unweight}

The first evaluation of the best-fit ISMF direction uses the same
basic method as in Paper I, but applied to the larger set of data in this
paper (\S \ref{sec:data}).  Equation \ref{eqn:one} was evaluated for
$F_\mathrm{i}$ with \Gn=1, using data points where \PdP$>2.5 $.  All
observations meeting the selection criteria (\S \ref{sec:data}) are
used in the fit, including measurements of the same star by different
observers.  The application of eqn. \ref{eqn:one} to the polarization
data set (\S \ref{sec:data}) yields an ISMF direction toward
\glon,\glat=$37^\circ \pm 15^\circ,22^\circ \pm 15^\circ$ (see Table
\ref{tab:ismf2} for ecliptic coordinates).  There is no significant
change in the direction obtained from the unweighted fit when compared
to the results of Paper I. The values of $F_\mathrm{i}$ are displayed
in Figure \ref{fig:unweight}, in both both ecliptic and galactic
coordinates, and for each point in the possible ISMF grid using the
1\deeg\ grid spacing discussed in \S \ref{sec:method}.

Because of the underlying data sample, the unweighted fit emphasizes
southern hemisphere data acquired in the 1970's.  For the best-fit
ISMF direction, \Bbest, 48 stars contributed to give \Fbest$ = 0.51$
for the function being minimized.  Of these 48 stars, 20\% have new
NOT and LNA data and they contributed 25\% of \Fbest, 35\% have
pre-1980 data from Tinbergen (1982) or Piirola (1977) and they
contributed 29\% of \Fbest, 31\% have PlanetPol data and they
contributed 32\% of \Fbest, and 8\% were observed by Santos et
al. (2010) and they contributed 12\% of \Fbest.  Note that no single
set of observations appears to dominate this result.

\subsection{ISMF direction from weighted fit }\label{sec:weight}

The new high-precision polarization data in the southern hemisphere,
as well as additional high-precision data in the northern hemisphere,
(\S \ref{sec:data}) makes it possible to evaluate the best-fit
ISMF direction by utilizing all of the position angles in the sample,
by weighting each value according by the position angle probability
distribution in eqn. \ref{eqn:two}.

If the polarizing medium were to be homogeneous over the nearest 40
parsecs in the galactic hemisphere (which is not the case), combining
polarization data using weighted values implicitly weights the
resulting field direction towards more distant points where the
signal-to-noise (S/N) is larger.  If in contrast there is no
statistically significant field in the same spatial region, the
resulting field direction based on weighted fits is close to the
center of the spatial interval for randomly distributed polarization
directions.

For the weighted fits, the polarization strengths in the combined set
of all data were capped at $50 \times 10^{-5}$ in order to minimize
the possible biases due to unrecognized intrinsic stellar
polarization.  Fewer than 5\% of the significant (\PdP$>2.5$) data
points in the fit were affected by this cap.  With the use of weighted
fits, the best-fit direction is biased towards data sets that have
low quoted uncertainties.  To avoid over-reliance on the PlanetPol
data, which are quoted to a typical accuracy of ppm and are spatially
clustered in the northern hemisphere, the \PdP\ value input to \Gn\
was capped at 3.5, and a lower limit of $10^{-5}$ was set on \Gn.
With these constraints, the best-fit ISMF direction is oriented
towards \elon,\elat$~=~260^{+15}_{-20}$,~ $49 \pm 15$ or
\glon,\glat=$47^\circ\pm 20^\circ, 25^\circ\pm 20^\circ $ (Table
\ref{tab:ismf2}).  The values of $F_\mathrm{i}$ (eqn. \ref{eqn:one})
are plotted in Figure  \ref{fig:weight}, using \Gfact\ given by
eqn. \ref{eqn:two}. It is seen that the minimum value of
$F_\mathrm{i}$ is more clearly defined for \Gfact=1 than for \Gfact\
given in eqn. \ref{eqn:two}.

One hundred and seventy-seven stars have contributed to this
best-fit ISMF direction, \Bbest.  For this fit the mean value of
eqn. \ref{eqn:one} gives \Fbest =485, using \Gn\ from
eqn. \ref{eqn:two}. Of these 177 stars, 26\% have new NOT and LNA data
and they contributed 11\% of \Fbest; 39\% have pre-1980 data from
Tinbergen (1982) or Piirola (1977) and they contributed 24\% of
\Fbest; 12\% have PlanetPol data and they contributed 60\% of \Fbest;
and 20\% were observed by Santos et al. (2010) and contribute 0.3\% of
\Fbest.  The PlanetPol contribution to the function being minimized,
$F_\mathrm{i}$, is dominated by four stars that have position angles
nearly orthogonal ($90 _{-15}^{+25}$ degrees) to the meridians of
\Bbest\ and that make 86\% of the PlanetPol contribution.  Those four
stars are HD 150680 at 10 parsecs, and three high-latitude stars at
24--31 parsecs (HD 113226, HD 116656, and HD 120315).

This method of using weighted fits produces an ISMF direction that is
biased by the distribution of measurement errors for diverse
instruments.  We have tested the effect of the weighting factor by
raising the lower limit of \Gn\ to $10^{-4}$ for \PdP$\le 3.5$.  This
test increases the homogeneity of low significance data points.
Although the resulting ISMF direction is within 25\deeg\ of \Bbest, it
is closer to the center of the spatial interval indicating increased
reliance on statistically insignificant measurements.  The cap
\PdP$\le 3.5$ in \Gn\ (eqn. \ref{eqn:two}) has the effect of
deemphasizing the PlanetPol data where the ppm-level uncertainties
otherwise dominate the result.  If instead, we take $G>10^{-5}$ and a
\PdP\ cap of 4--6, the magnetic pole is then directed towards
\glon$=337^\circ \pm 10^\circ$ and \glat$=28^\circ \pm 7^\circ$ and
the $F$ minimum (eqn. \ref{eqn:one}) becomes very poorly defined.  We
are quoting the best-fit ISMF direction for the weighted fit
(Table \ref{tab:ismf2}) for values \PdP$\le 3.5$ and \Gn$ \ge
10^{-5}$, since these limits appear to present the best overall
representation of the data given the differences in measurement
sensitivities.  With future high-sensitivity data, it should be
possible to treat the analysis utilizing a broader range of the
weighting factor.

\section{Discussion} \label{sec:discussion}

\subsection{Comparing the ISMF directions traced by the IBEX Ribbon and polarizations } \label{sec:ibex}

The very local ISMF direction provided by the center of the IBEX
Ribbon arc is separated by $32^{+27}_{-30} $ degrees from the
direction of the best-fit ISMF (weighted fit in Table
\ref{tab:ismf2}).  MHD heliosphere models that sucessfully predict the
location of the IBEX Ribbon yield offsets between the center of the
Ribbon arc and the interstellar magnetic field that shapes the
heliosphere of between $0^\circ$ and $15^\circ$
\citep{HeerikhuisenPogorelov:2011}, so the direction of the very local
ISMF found here is marginally consistent with the direction of the
ISMF that shapes the heliosphere.  Polarization of stars within 40 pc
and 90\deeg\ of the heliosphere nose have contributed to this result.
Up to fifteen interstellar clouds have been identified towards stars
in this region of space
\citep{Lallementetal:1986,FGW:2002,Frisch:2003apex,RLIV:2008vel,Frisch:2011araa}.
The magnetically aligned dust grains causing the polarizations
presumably are located in these clouds.  Although local color excess
values are too small to be measured, the variable gas-phase abundances
of the refractory elements that make up the grains show that several
types of interstellar grains must be present in these clouds
\citep{FrischSlavin:2012aogs}.  The clouds are distinguished by the
gas kinematics using the assumption of solid-body motion, and in some
cases by temperature.  Since most of the neutral ISM that is within 40
parsecs is also within 20 parsecs (\S \ref{sec:data}), the best-fit
ISMF direction then may represent primarily the ISMF direction within
$\sim 20$ parsecs of the Sun.  This would help explain the remarkably
similar directions of the ISMF traced by the IBEX Ribbon and the
polarization data.

If the IBEX ENA Ribbon and interstellar polarizations trace the same
ISMF direction, then polarization strengths should increase with ENA
fluxes because interstellar polarizations and the IBEX Ribbon both
reach maximum strengths in sightlines that are perpendicular to the
magnetic field, e.g. where \BdotR=0 for magnetic field $B$ and radial
sightline $R$.  Figure \ref{fig:pvrsena} shows a plot of the mean
value of 1.1 keV ENA fluxes that are in a $12^\circ$ diameter region
centered on stars that are within 40 parsecs, 90\deeg\ of the heliosphere
nose, and have \PdP$>2.5$.  ENA fluxes with signal-to-noise S/N$>3.5$
are included.  The data points are color-coded according to the data
source in the left figure.  Data in Figure \ref{fig:pvrsena}, left,
tend to be divided into four groups with different relations between
the polarization and ENA fluxes.  Group A consists of the highest
quality data, such as data collected for this study and the PlanetPol
data (\S \ref{sec:data}), and show a general increase of polarization
with ENA fluxes.  The PlanetPol stars in group A are located in the
\sixteen\ interval.  Region D primarily represents 1970's polarization
data, where uncertainties are substantially larger.  The trend in
region B is ambiguous.  The four groups of data in Figure
\ref{fig:pvrsena}, left, are plotted in galactic coordinates in Figure
\ref{fig:pvrsena}, right using a color-coding that shows the data
group.  The increase of ENA fluxes with polarization for stars in
group A supports the hypothesis that the ISMF sampled by the
polarization data and Ribbon ENA fluxes are tracing the same, or
similar, ISMF directions.  Results for the remaining stars are
ambiguous.

The difference between the ISMF directions from the polarization and
ENA Ribbon data may reveal structural components of the local ISM.
The polarization data set we use to determine the field direction (\S
\ref{sec:data}) may not sample the interstellar cloud surrounding the
Sun, the LIC.  The Sun is within 20,000 AU of the edge of the LIC
\citep{Frisch:2011araa}, and the LIC includes a minimal amount of ISM
within 90\deeg\ of the heliosphere nose \citep[e.g.][]{RLIV:2008vel}.
It is possible that the ISMF direction traced by the center of the
IBEX Ribbon arc represents the ISMF in the LIC, while the ISMF traced
by the polarization data represents the ISMF traced by the next cloud
in the upwind direction, the G-cloud that is observed towards the
nearest star $\alpha$ Cen.  The LIC is $\sim 22$\% ionized
\citep[][Appendix \ref{app:ibex}]{SlavinFrisch:2008}, and the G-cloud
is likely to be partially ionized, so that the ISMF will couple
tightly to the gas in these clouds.  The possible location of the ISMF
traced by polarization data is discussed further below where a small
rotation with distance of the local ISMF is found (\S \ref{sec:pppa}).
This rotation suggests that some of the small difference between the
ISMF directions found from the IBEX Ribbon and polarization data may
result from the volume-averaged ISMF direction implicit in our method
for fitting the polarization data.

\subsection{Local ISMF and Loop I} \label{sec:loopI}

The relation between the Loop I superbubble and the ISM
within 20 pc has been known for some time
\citep[see review of ][]{Frisch:2011araa}.  Recently,
\citet{Wolleben:2007} modeled the polarized low-frequency 1.4 GHz and
23 GHz radio continuum of Loop I, including the North Polar Spur
emission, as arising from two spherical shells, denoted ``S1'' and
``S2'', each with a swept up magnetic field traced by polarized
synchrotron emission.\footnote{Loop I is sometimes referred to as the
`North Polar Spur', which historically is the brightest part of the
radio continuum loop and juts vertically from the galactic plane near
longitudes of $\sim 30^\circ$ \citep[e.g.,][]{HanburyBrown:1960nps}.}
An ordered magnetic field structure is implicit to the Wolleben model.
His shell parameters suggest that the Sun is in the rim of the S1
shell, which is centered $78 \pm 10$ parsecs away towards
$\ell=346^\circ \pm 5^\circ, b=3^\circ\pm 5^\circ$.  Previous studies
reached similar conclusions, although with a single superbubble so
that the center of the \HI\ shell is $\sim 12^\circ$ south of the
center of the radio continuum shell \citep[e.g.][]{Heiles:1998lb}.
The direction of the swept-up ISMF in the S1 shell (which has generous
uncertainties) is consistent with the magnetic field direction in the
local polarization patch found by \citet{Tinbergen:1982}; any local
fragments of the S1 and S2 shells are not spherical, however, since
the \FeII\ column densities in local ISM do not trace these features
defined by Wolleben \citep{Frisch:2010s1}.  The Wolleben model
provides a convenient quantitative relation between the large-scale
ISMF defined by Loop I and the ISMF close to the Sun.

Local cloud kinematics, the local ISMF, and local gas-phase abundances
all suggest an origin related to Loop I.  In the local standard of
rest (LSR), the local cluster of clouds flows past the Sun at a bulk
velocity of $V= -17$ \kms, and towards us from the direction
\glon,\glat$\sim 335^\circ, -5^\circ$ \citep{Frisch:2011araa}.
Solid-body motion is assumed in determining this vector.  The flow
direction is $\sim 14^\circ$ from the center of the S1 shell, which is
located at \glon,\glat$\sim 335^\circ, -5^\circ$.  Although the S1 shell
is an evolved superbubble remnant, the local fragment of the shell
appears to still be radially expanding into the low density region of
the Local Bubble.  The best ISMF direction from the weighted fit
(Table 1) makes an angle of 76\deeg\ with the local ISM bulk flow
vector.  These numbers are consistent with a scenario where the normal
to the shell rim is parallel to the ISM velocity, and the magnetic
field is compressed in the shell rim so that the field direction is
nearly perpendicular to the normal of the shell surface.  An
association of local ISM with a supernova shell is consistent with the
high and variable abundances of the refractory elements Fe, Mg, and Si
in local ISM, which suggest inhomogeneous grain destruction by
interstellar shocks \citep{Frisch:2011araa,FrischSlavin:2012aogs}.

\subsection{Ordered ISMF near Sun } \label{sec:pp}

A detailed study of the behavior of interstellar polarizations and
ISMF behavior with distance is possible with the sensitive PlanetPol
data for stars located in the interval \sixteen, which
contains the RA$>17^\mathrm{H}$ region where \citet{BaileyLucas:2010planetpol}
found an increase of polarization with distance. This region overlaps
the low latitude part of the North Polar Spur, where aligned optical
polarization vectors define the ISMF for distances 60--120 parsecs
\citep{MathewsonFord:1970,Santosetal:2010}.  The behavior of the ISMF
for distances $<60$ parsecs in the North Polar Spur direction is
provided by PlanetPol data.

\subsubsection{Polarization versus distance and column density in 
$16^\mathrm{H} - 20^\mathrm{H}$ region} \label{sec:ppdispol}

\citet{BaileyLucas:2010planetpol} found that the increase
of polarization with distance was greater for stars with
RA$>17^\mathrm{H}$ than for stars in other spatial intervals.
The increase of polarization with star distance for PlanetPol stars
in the \sixteen\ region and with \PdP$>3.0$ is shown in Figure \ref{fig:pp1}, right.
For this set of uniform quality data a maximum value for
the polarization is found for each distance interval within $\sim 60$ pc,
and there are no observed points that lie above the upper envelope marked
by the filled points.
Polarizations of stars beyond that follow this trend.  
Stars which define this maximum are often refereed
to as the polarization 'upper envelope' 
\citep[e.g.][also see Appendix \ref{app:pol}]{Andersson:2012rev},
and we use that term here.  In general, polarization data
tend to have an upper envelope when plotted against
either distance or color excess due to either depolarization 
or patchy ISM.  A rough estimate of the dependence
of polarization on distance can be found by first
binning stars into 10 parsec intervals, and then selecting the stars
with the maximum polarization in each interval.  Omitting bins with no
stars, seven stars (filled circles in Figure \ref{fig:pp1}) can be used
to define the upper limit of polarization as a function of
distance. A linear fit though the envelope stars gives the
approximate distance dependence of the maximum polarization in the
\sixteen\ region of

\begin{equation}\label{eqn:pd}
\mathrm{log} P \sim 1.200 + 0.013 *D
\end{equation}
where \emph{P} is polarization in ppm and $D$ is the distance in
parsecs. The closest envelope star is $\alpha$ Lyr at 8 parsecs
(Table \ref{tab:seven} lists the envelope stars).  The locations of the
\sixteen\ stars are plotted in galactic coordinates in Figure
\ref{fig:pp1}, left.  Two unpolarized PlanetPol stars in this region
are not plotted, HD 163588 and HD 188119, both of which are located at
larger longitudes than the other stars.  Stars that form the upper
envelope of the polarization-distance relation are loosely confined
between galactic longitudes of 30\deeg\ and 80\deeg, and latitudes of
--5\deeg\ to 30\deeg.

The upper envelope of the polarization-distance relation
(Figure \ref{fig:pp1}) can be used to estimate the gas density gradient
along the ISM holding the ISMF, since dust and gas have
similar spatial distributions in diffuse clouds.  Few of the PlanetPol
stars have measured \HI\ or \DI\ column densities
\citep[e.g.][]{Woodetal:2005lya}, so column densities are estimated by
converting polarization to color excess, and then converting color
excess to \NH, using standard relations for diffuse clouds.  The
polarization for dust grains aligned by an ordered ISMF, with no
random component, provides an upper envelope of the \Pol\ vs.
\ebv\ relation of $ P(\%) = 9 E(B-V)$ \citep{FosalbaLazarian:2002}.
Interstellar gas and dust have similar spatial distributions in the
low opacity ISM.  \emph{Copernicus} observations of the saturated \HI\
\Lya\ absorption line and molecular hydrogen lines showed that
$N$(H)/\ebv=$N$(\HI + \HH)/\ebv$=5.8 \times 10^{21} \cdot $ \cmtwo\
mag$^{-1}$ \citep[][]{Bohlin_etal_1978}.\footnote{We use the ratio for
the ISM towards all stars, rather than the ratio for the intercloud
medium, $N$(\HI + \HH)$=5.0 \times 10^{21} \cdot $\ebv\ \cmtwo, that
might be relevant to the nearest ISM \citep{Bohlin_etal_1978}.}  The
\emph{Copernicus} value underestimates hydrogen column densities for
partially ionized gas since independent measurements of ionized gas
are not available for most sightlines.  The relation between
polarization and column density is estimated by combining the
distance-polarization fit with the empirical \Pol\ vs. \ebv\ and \ebv\
vs. \NH\ relations, giving
\begin{equation}\label{eqn:hId}
\mathrm{log} N(\mathrm{H})=18.01+0.013*D.
\end{equation}
for stars tracing the upper envelope of the  $P$-\ebv\ relation and
$P$-distance relation.  For other stars, foreground depolarization
can lead to lower values for \NHI.
This gas-distance relation can be checked against three nearby stars
in the \sixteen\ region with \NHI\ values determined by
\citet{Woodetal:2005lya}, HD 155886 (36 Oph), 165341 (70 Oph), and HD
165185.  The ratio of the predicted \NH\ (eqn. \ref{eqn:hId}) to the observed \NHI\ for
these three stars is $1.35 \pm 0.35$, indicating depolarization
and/or inhomogeneous ISM is present nearby.

\subsubsection{Polarization versus position angle in $16^\mathrm{H} - 20^\mathrm{H}$ region}  \label{sec:pppa}

\begin{figure}
\plottwo{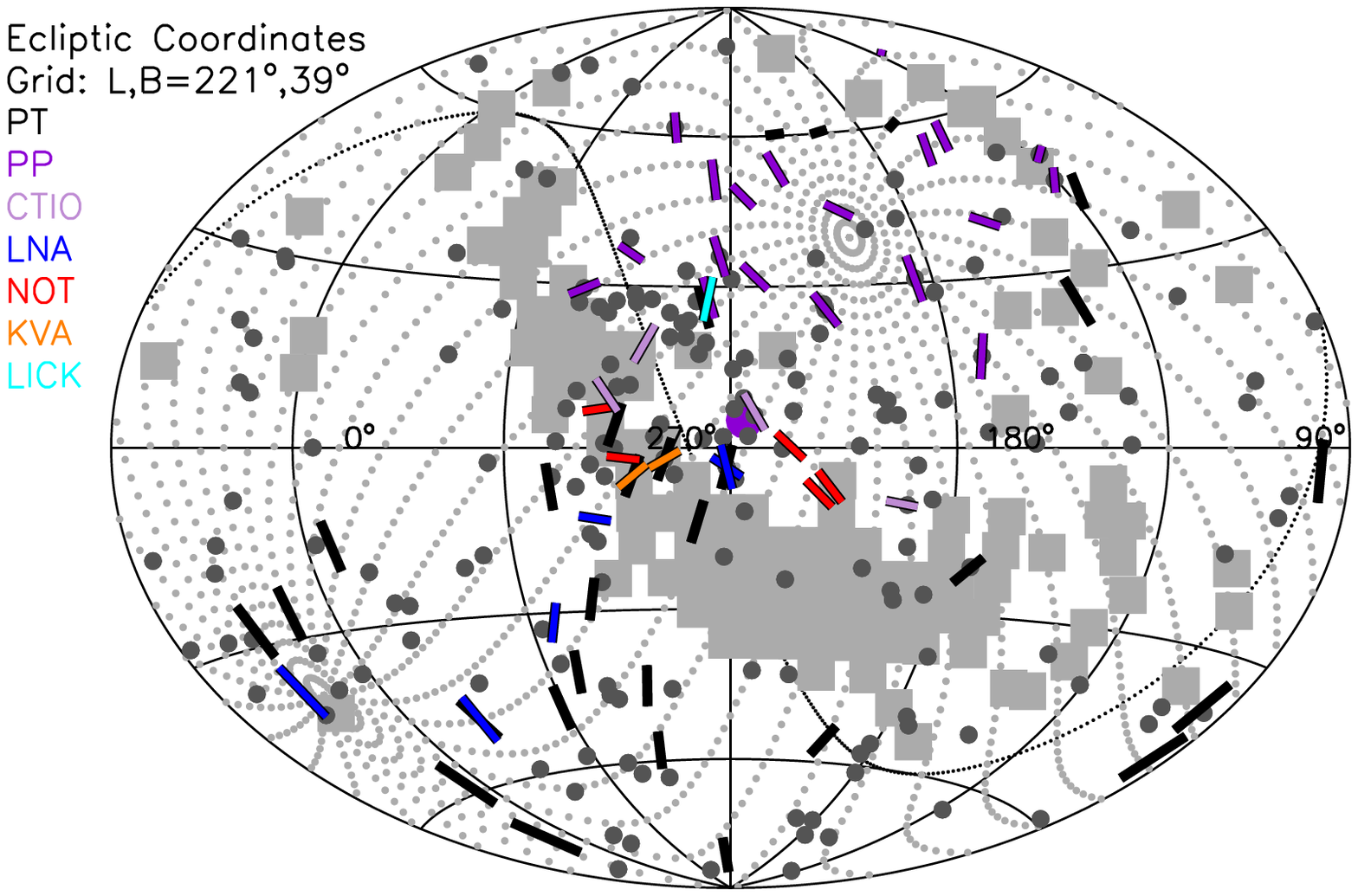}{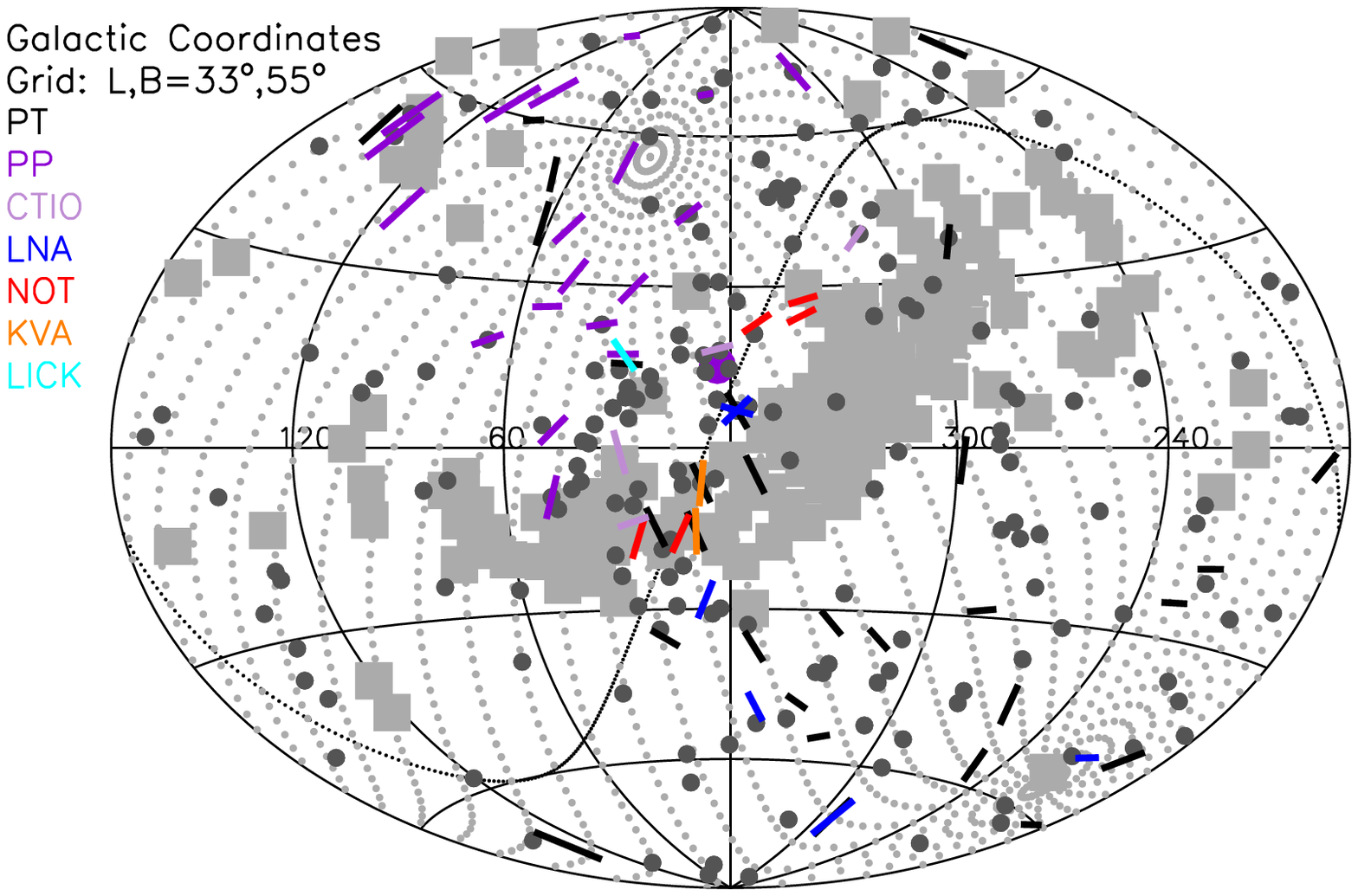}
\caption{The polarization vectors of stars within $\sim 40$ pc are
shown in both ecliptic and galactic coordinate systems, and color-coded for the data
source, as indicated in the figure legend.  The polarization data are
from this paper (LNA and NOT, \S \ref{sec:data}), PlanetPol
\citep[][PP]{BaileyLucas:2010planetpol}, Paper I (LICK, KVA),
\citep[][CTIO]{Santosetal:2010}, and 20th century data
\citep[][PT]{Piirola:1977,Tinbergen:1982}.  Both plots are centered
close to the heliosphere nose located at ecliptic coordinates (purple
circle) of \el=259\deg, \eb=5\deg; longitude increases towards the
left.  Data with \PdP$<2.5$ are plotted as dots, and vector lengths do
not indicate polarization strength.  The Compton-Getting corrected ENA
fluxes at 1 keV are plotted for directions where the ENA count rates
are larger than 113 counts cm$^{-2}$ s$^{-1}$ sr$^{-1}$ keV$^{-1}$,
which is $\sim 1.5$ times the mean ENA flux at 1 keV as measured by the
IBEX-HI instrument \citep{McComas:2009sci}.  The grid of dotted lines
show the ISMF determined from the center of the Ribbon arc
\citep[][Table \ref{tab:lnanot}]{Funsten:2009sci}.  }\label{fig:aitoff}
\end{figure}

\begin{figure}
\plottwo{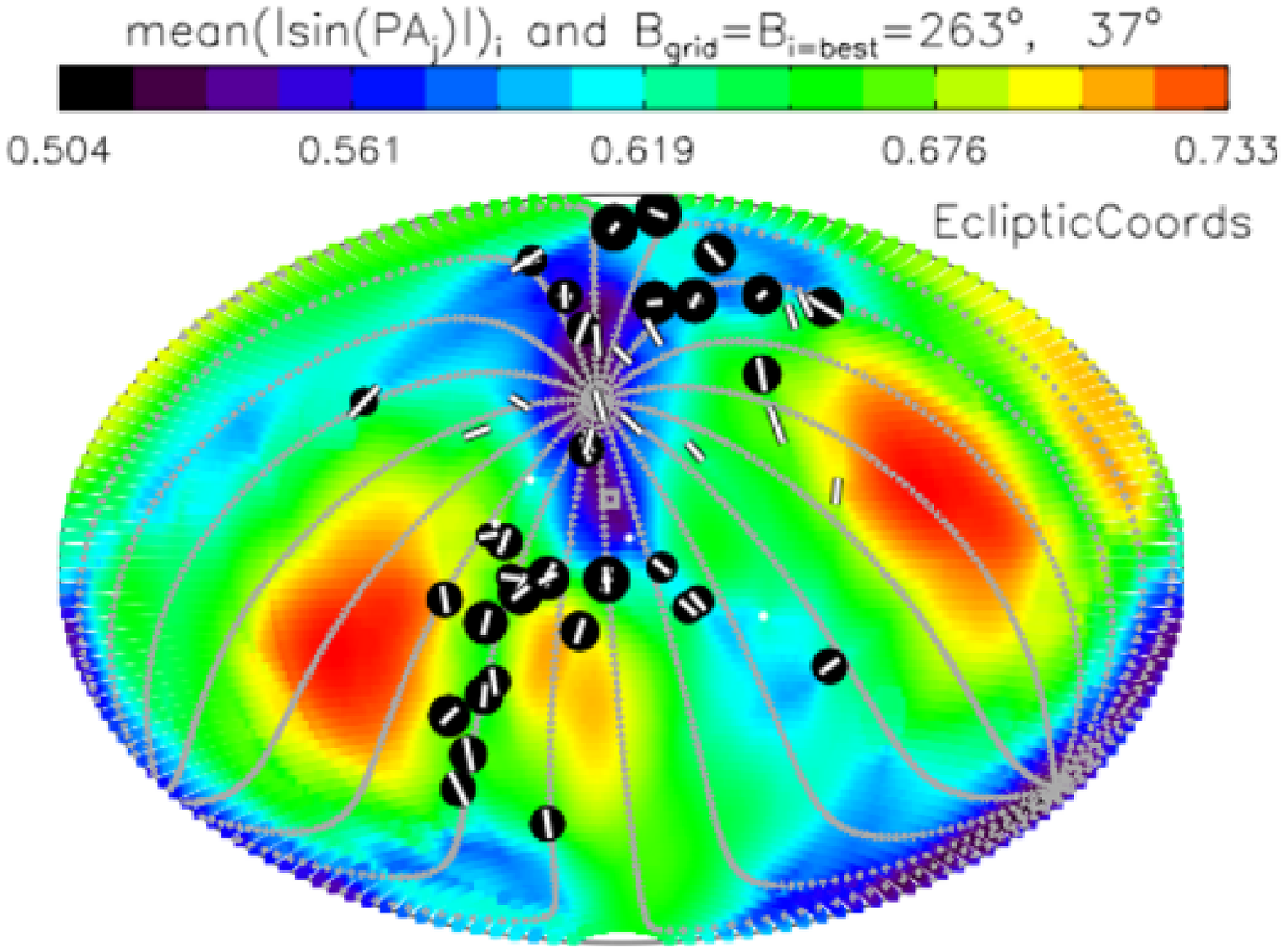}{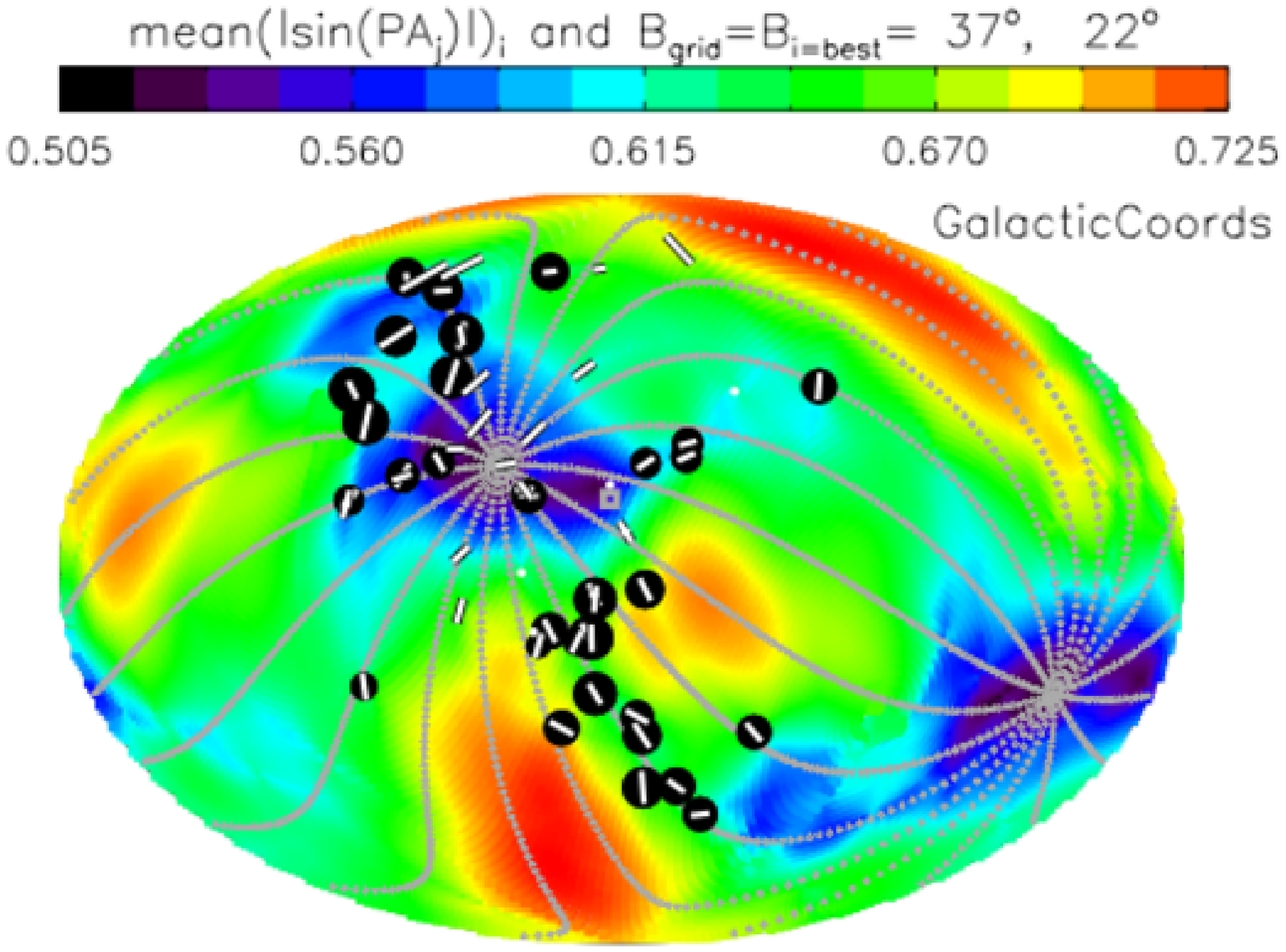}
\caption{The value of the function \myfunct\ for the unweighted
fit (\Gn=1, eqn. \ref{eqn:one})
is plotted over a regular
grid of $i$ possible interstellar field directions, based on
unweighted position angles (\S \ref{sec:unweight}).  The function is
color-coded and plotted in the ecliptic coordinates (left) and
galactic coordinates (right), with the left plot centered on the
ecliptic nose at $\lambda \sim 259^\circ$, and the right plot centered
on the galactic center ($16^\circ$ from the ecliptic nose).  The gray
dotted grid shows the best-fitting ISMF direction for the
unweighted fit listed at the top of the figures (Table \ref{tab:ismf2}).  Stellar polarization vectors are shown
superimposed on black dots for visual clarity. }\label{fig:unweight}
\end{figure}

\begin{figure}
\plottwo{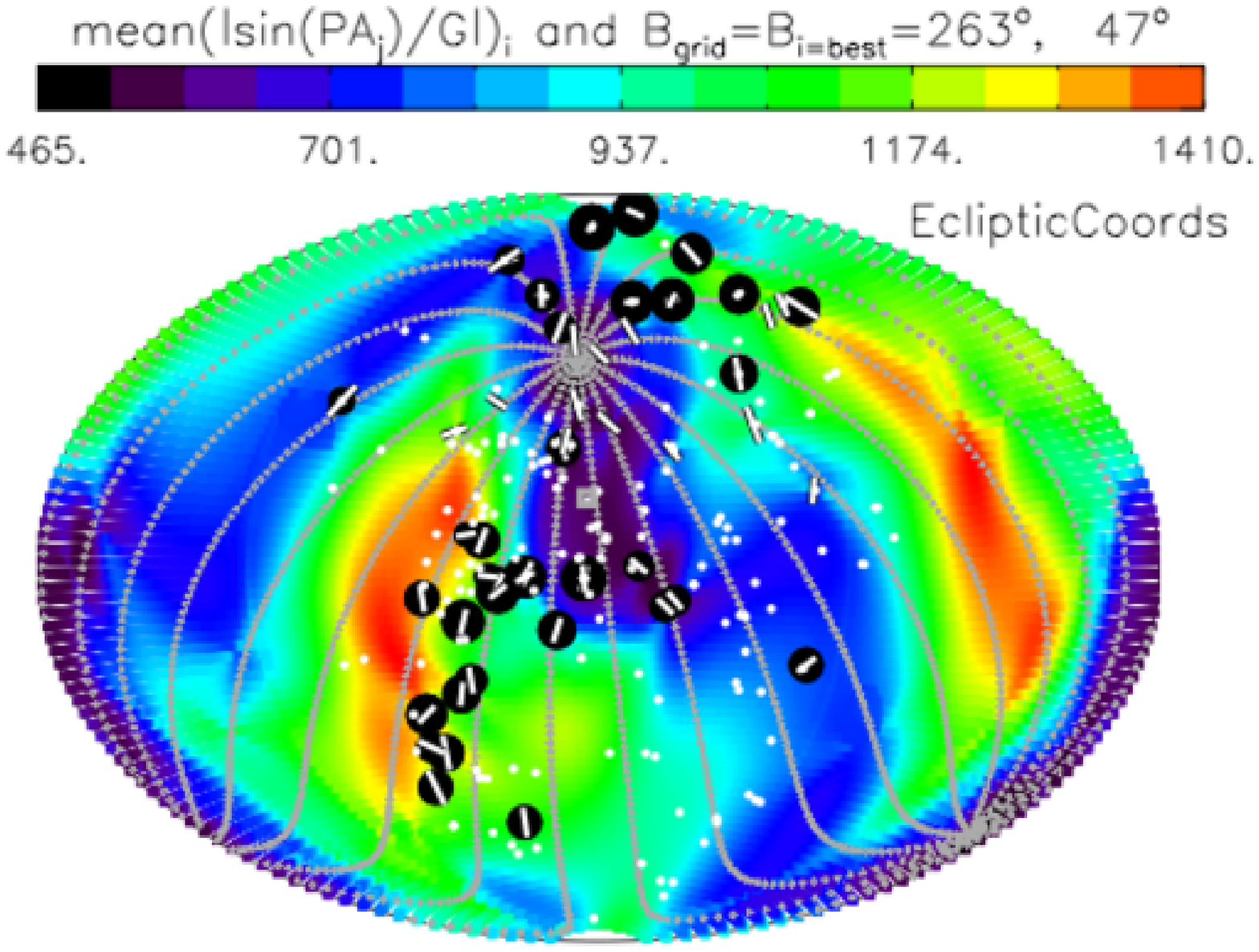}{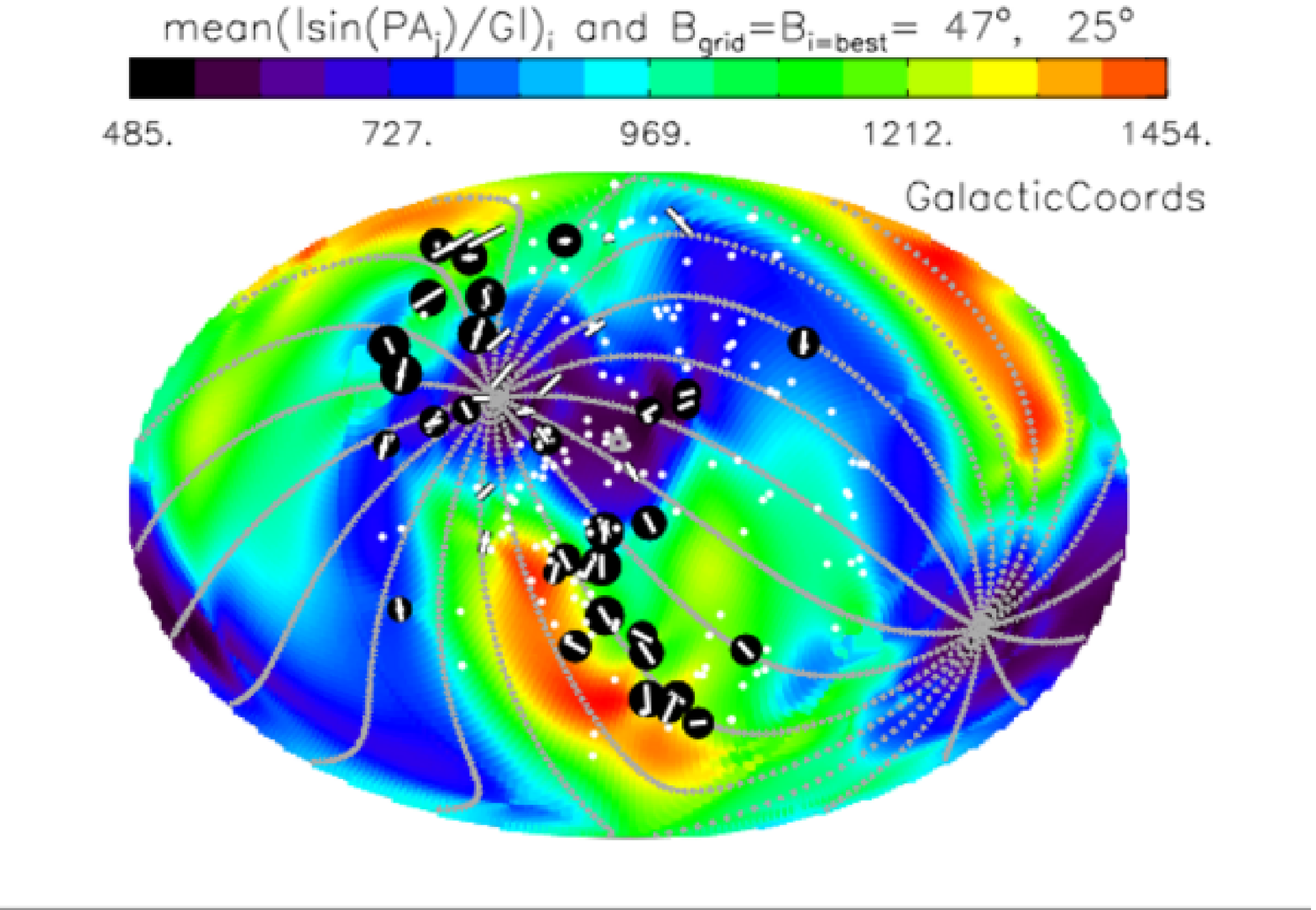}
\caption{The value of the function \myfuncttwo\ (eqn. \ref{eqn:one})
evaluated over a regular
grid of $i$ possible interstellar field directions, based on weighted
position angles with \Gn\ given by eqn. \ref{eqn:two}.  The symbol size is coded to
increase with polarization strength, and the dots represent stars
where the observed polarization is not statistically significant.  The
function is color-coded and plotted in ecliptic (left) and galactic
(right) coordinates.  The gray dotted grid shows the best-fit ISMF
direction given at the top of the figures (Table \ref{tab:ismf2}).  
The secondary weaker minimum centered near the heliosphere nose, at plot
centers, is dominated by the randomly distributed
position angles contributed by low significance data points.
See Figure \ref{fig:unweight} for additional information. }\label{fig:weight}
\end{figure}

\begin{figure}
\plotone{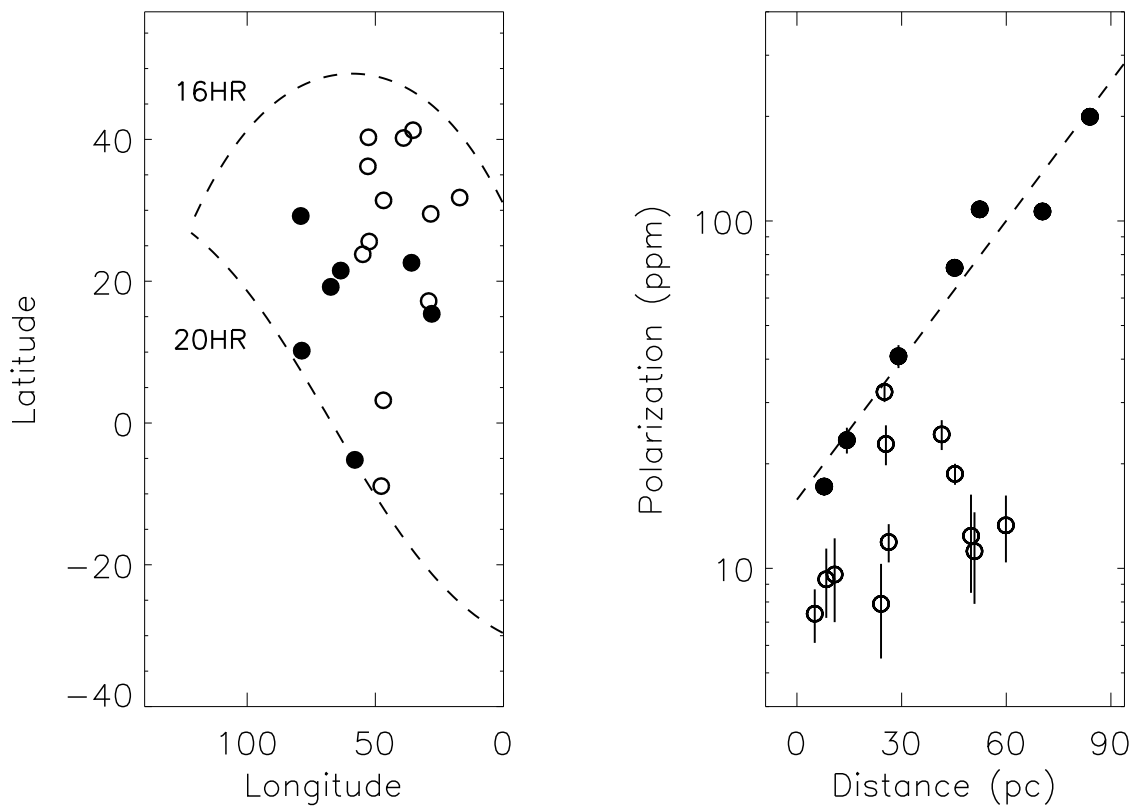}
\caption{Right: The polarizations of the PlanetPol stars with \PdP $>
3.0$, distance $D< 94$ pc, and inside the \sixteen\ region are
plotted as a function of distance (see text).  The upper envelope of
the increase of polarization with distance is shown by the filled circles.
This upper envelope is analogous to the
increase of polarization with color excess \citep{Serkowskietal:1975}.
A linear fit to the
seven stars in the upper envelope (listed in Appendix \ref{app:A}) 
yields 
$\mathrm{log}P= 1.200 + 0.013*D$, where \emph{P} is in ppm and \emph{D} is in parsecs.
Left: The positions of these stars are plotted in galactic coordinates.}\label{fig:pp1}
\end{figure}

\begin{figure}
\plottwo{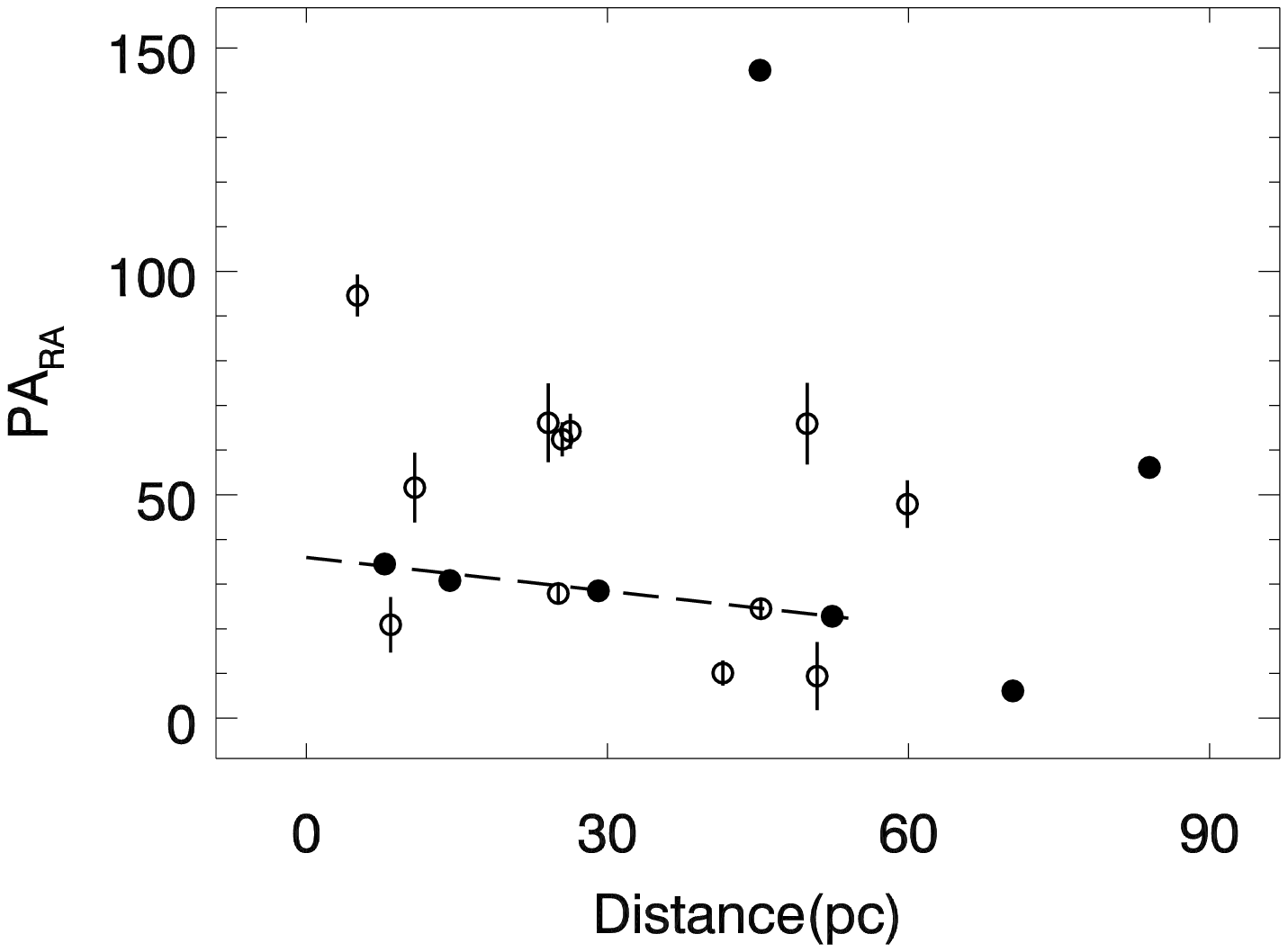}{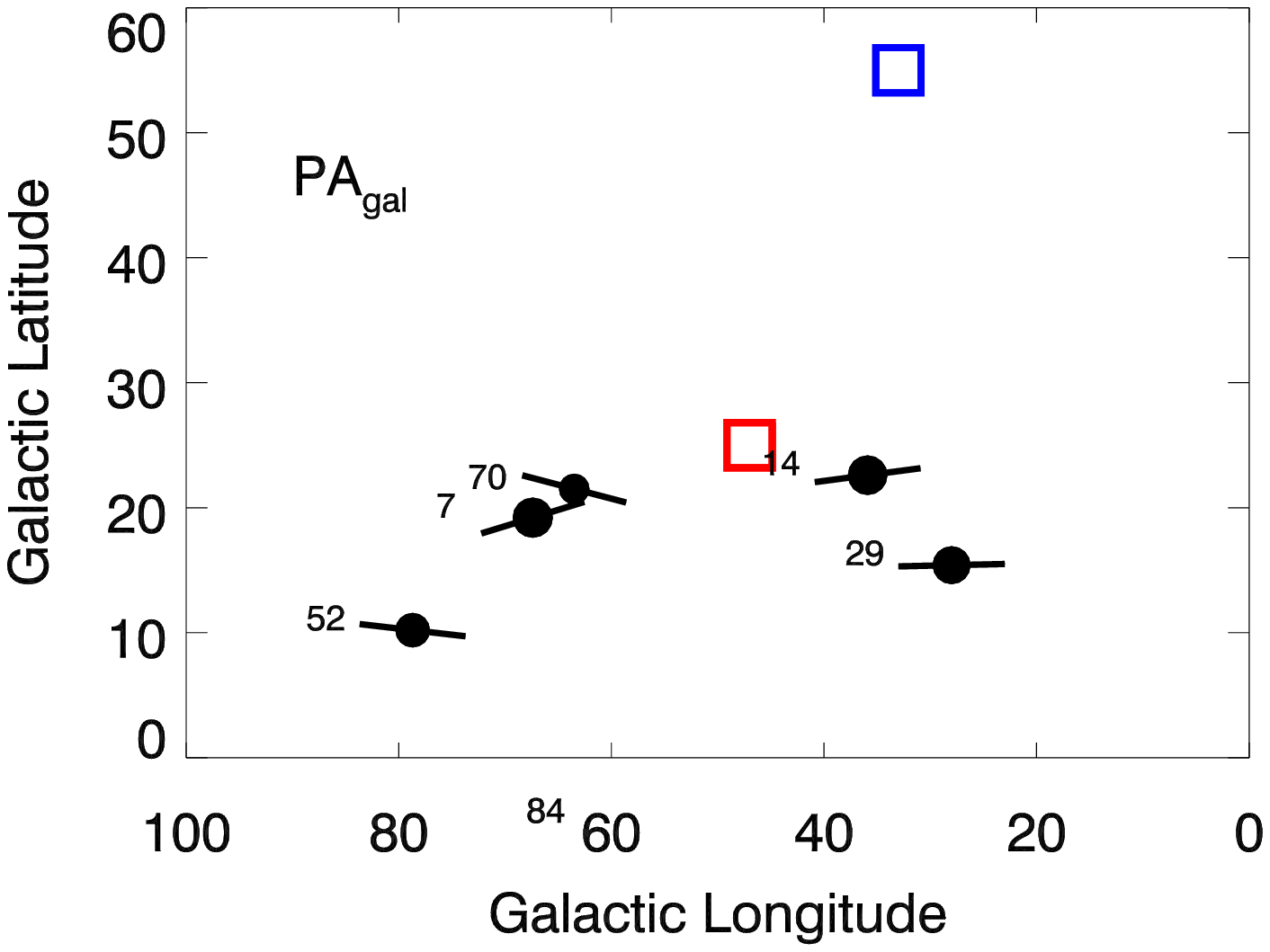}
\caption{Left:  Polarization position angles (\PAcel) in equatorial
coordinates are plotted against the star distance for the PlanetPol
data in the \sixteen\ region.  Filled circles indicate the seven stars
that form the upper envelope of the polarization envelope (Figure
\ref{fig:pp1}, right).  Except for the outlier data point HD 164058
(at 45 pc, \PAcel=145\deeg, \S \ref{sec:pp}), the polarization position
angles of the envelope stars vary smoothly for the stars within 55 pc.
A linear fit was made to the position angles of envelope stars within
55 pc, excluding the outlier HD 164058.  The result is \PAcel$= \mathrm{A} - \mathrm{B}
\cdot \mathrm{D_{pc}}$, where
A=$35.97 \pm 1.41$, B$=0.252 \pm 0.030$, $\chi^2 = 0.559$ and $D_{pc}$ 
is the distance in parsecs.  The
probability that the $\chi^2$ value is larger than 0.559 by chance is
0.76, indicating that a line provides a satisfactory fit to the
position angles of the nearest envelope stars.  A measure of magnetic
turbulence is provided by the 23\deeg\ standard deviation of the position
angles of all stars within 55 pc, excluding the outlier, with respect to
the linear fit to the nearest envelope stars.
Right:  Polarization position angles rotated into galactic coordinates are
plotted at the locations of the envelope stars within 70 pc
(except for the outlier). Data are labeled by
the star distance (Table \ref{tab:seven}).  The best-fitting ISMF from the weighted fit
(red box) and the center of the IBEX Ribbon arc (blue box) are
plotted for comparison.  Uncertainties on the position angles are
negligible.  Note that the position angle of the closest star HD 172167, 8 pc,
is directed towards the best-fitting ISMF.
}\label{fig:pp2}
\end{figure}

Optical polarization data that traces the \HI\ filamentary structure
of Loop I beyond $\sim 80$ parsecs reveal an ordered ISMF that follows
the axis of an \ion{H}{1} filament of Loop I \citep{Heiles:1998lb}.  Optical polarization vectors
(Figure \ref{fig:aitoff}), and the location of the \sixteen\ region
towards the North Polar Spur, suggest that the very local ISMF
direction may also be ordered.  Figure \ref{fig:pp2} displays the
polarization position angles, \PAcel, for PlanetPol stars with
\PdP$>3.0$ in this interval, as a function of distance.  With the
exception of the outlier star HD 164058, which has \PAcel=$145^\circ
\pm 0.5^\circ$ and a distance of 45 parsecs, the envelope stars within
55 parsecs display a smooth rotation of polarization position angle
with distance.  HD 164058 is a K5III star and belongs to a spectral
class where intrinsic polarization occurs because of chromospheric
magnetic activity and was omitted from the fit in eqn. \ref{eqn:pafit}
as an obvious outlier (see below).  A linear fit to position angles versus
distance gives
\begin{equation} \label{eqn:pafit}
{PA}_\mathrm{RA} \mathrm{(D)} = 35.97 (\pm 1.4) - 0.25 (\pm 0.03) \cdot \mathrm{D}
\end{equation}
and a $\chi^2$ of 0.559, where \PA$_\mathrm{RA}$ is in degrees and the
distance D is in parsecs.  The small value of $\chi^2$ per degree of
freedom (2) suggests that the uncertainties on \PAcel\ quoted by
\citet{BaileyLucas:2010planetpol} may be
generous. \footnote{\citet{BaileyLucas:2010planetpol} include several
different effects when calculating uncertainties for their
measurements, including counting statistics, possible intrinsic
telescope polarization, and in some cases atmospheric polarization
caused by Sahara dust.}  The probability that the computed $\chi^2$
value would have a value larger than 0.559 is 0.76, indicating that
parameters of the fit in eqn. \ref{eqn:pafit} are believable.  After
excluding HD 164058, the relation between the polarization direction
and star distances of the four envelope stars within 55 parsecs
appears to be linear.  The large difference of $7.8
\sigma_\mathrm{pp}$ between the position angle of the outlier star and
the best-fit line (eqn. \ref{eqn:pafit}), where $\sigma_\mathrm{pp}$
is the mean difference between the best-fit line and the position
angles of PlanetPol stars within 55 pc except for the outlier,
justifies omitting HD 164058 from the calcuation of
eqn. \ref{eqn:pafit}. The envelope stars are partly self-selected by
the fact they represent sightlines where intervening depolarization is
minimized, so the linear fit to the polarization position angles
suggests the existence of a nearby ordered ISMF that rotates $\sim
0.25^\circ$ per parsec.  This rotation is confined to the small region
of space (filled dots in Figure 7, left) where
\citet{BaileyLucas:2010planetpol} found the systematic increase in
polarization strengths with distance (Figure \ref{fig:pp1}, right).  The
envelope stars appear to sample the uniform part of an extended medium
that contains a magnetic field and reaches to within 8 pc of the Sun.

The standard deviation of position angles around the ordered component
is a measure of the magnetic turbulence in this region.  For all stars
in the \sixteen\ region and within 55 parsecs, excluding the outlier
star, the standard deviation of polarization position angles about the
best-fit line is 23\deeg, which we interpret as magnetic turbulence.

The smooth variation of the ISMF direction over distance suggests that
the four nearest envelope stars are embedded in the polarizing medium,
which extends between $\alpha$ Lyr (Vega, 8 parsecs) and $\delta$ Cyg
(52 parsecs). The sightlines to all of the envelope stars sample
either the S1 shell, the S2 shell, or both shells, according to the
\citet{Wolleben:2007} model and the shell configurations plotted in
\citet{Frisch:2010s1}.  The angular spread on the sky of the envelope
stars is 50\deeg, which suggests that the ISMF in the nominal S1 and S2
shells consists of a homogeneous "matrix" component of smooth material
that is sampled by the envelope stars, and patchy material and/or a
turbulent ISMF that produces depolarization of, and turbulence in, the
position angles of non-envelope stars.

If the polarizations of the envelope stars are tracing matrix ISM in
this region, this material should be detectable as interstellar
absorption features.  Absorption lines have been measured towards the
four envelope stars in the fit.  All four stars show absorption at or
close to the velocity of the Apex Cloud defined by
\citet{Frisch:2003apex},\footnote{This cloud vector is similar to the
``Panoramix'' cloud that was identified much earlier by
\citet{Lallementetal:1986}.} so it is tempting to assume that the
'Apex cloud' represents the smooth ISM filling the local region of the
S1 and S2 shells. Interstellar absorption lines have also been
measured towards a fifth polarized star, HD 187642, in region
\sixteen, where a component at the Apex cloud velocity is also found.
Towards the envelope star HD 159561, the Apex cloud component is
characterized by weak \CaII\ absorption and a typical ratio
\TiII/\CaII=0.41, suggesting it contains some neutral gas since \TiII\
and \HI\ have the same ionization potential
\citep{WeltyCrowther:2010Ti}.  A more recent and complete cloud
identification scheme was developed by \citet{RLIV:2008vel}, where
three of the Apex cloud components towards envelope stars are instead
assigned to the MIC cloud, while the fourth component is left as
unassigned.  Three of the four stars in \citet{RLIII} that are in the
\sixteen\ region have components at the Apex velocity, however the
cloud temperature ranges from $\sim 1,700 - 12,500$ K for these
components.  A second possible carrier of the ordered field is the 'G'
cloud since it has a wide angular extent in the upwind direction, is
within 1.3 parsecs of the Sun, and is observed towards the envelope
stars. The identification of the cloud carrying the ordered component
of the ISMF requires additional study.

\subsubsection{Polarization versus interstellar radiation field in
  $16^\mathrm{H} - 20^\mathrm{H}$ region } \label{sec:ppisrf}

Radiative torques are known to play
a role in the alignment of dust grains in dense clouds where the
radiation field has a preferred direction
\citep[e.g.][]{AnderssonPotter:2006,Draine:1997radiativetorques,AnderssonPotter:2011,LazarianHoang:2008},
but this does not appear to be the case locally.  The interstellar
radiation field extrapolated to the location of each star in the
\sixteen\ region appears to anticorrelate with polarization strengths.
We believe that this correlation is a chance coincidence of the
relative locations of the brightest radiation sources and polarized
stars.  Figure \ref{fig:pp3} shows polarizations for stars inside
(filled symbols) and outside (open symbols) of the \sixteen\ region,
plotted against the total far-UV 975\AA\ flux that has been
extrapolated to the position of each polarized star, while ignoring
any possible opacity effects.  The 975\AA\ radiation flux is given by
the 25 brightest stars at this wavelength \citep{OpalWeller:1984},
which are primarily located in the third and fourth galactic quadrants
because of the geometry and low opacity of the Local Bubble
\citep[see, e.g., the interstellar radiation field at the Sun from the
S2 UV survey,][]{Gondhalekaretal:1980}.  These polarization data show
a trend for linear polarization to increase as radiation flux
decreases.  The rough anticorrelation between radiation fluxes and
polarizations is shown by the fit log \Pol $=4.61 - 0.054 $*\fluxnine\
(dashed line in Figure \ref{fig:pp3}), where \Pol\ is in units of
$10^{-6}$ and \fluxnine\ is the radiation flux at 975 \AA\ in units of
$10^3$ photons \cmtwo\ \persec\ \AA$^{-1}$.  This apparent
anticorrelation appears to be the result of the relative locations of
the brightest hot UV-bright stars and the most distant polarized
stars, that define the low-flux region of the trend (see Appendix
\ref{app:isrf} for more details).  

\subsubsection{CMB dipole anisotropy and the local ISMF} \label{sec:cmb}

Large scale global deviations from a uniform blackbody spectrum of the
CMB radiation define a dipole anisotropy that is oriented towards
\glon$=263.85^\circ \pm 0.1^\circ$ and \glat$ = 48.25^\circ \pm
0.04^\circ$ \citep{BennettHalpernwmap:2003dipoledirection}.  The great
circle on the sky that is midway between the two poles of the CMB dipole displays
a striking symmetry around the direction of the heliosphere nose
direction defined by the flow of interstellar \HeI\ through the
heliosphere \citep{Frisch:2007cmb}.  The best direction of the
heliosphere nose is defined by the new IBEX consensus direction for the
interstellar \HeI\ flow, which includes correction
for propagation and ionization effects \citep[][see Appendix \ref{app:ibex}]{McComas:2012bow}.
This new consensus IBEX direction for the
heliosphere nose more closely aligns the nose direction with
the great circle that is midway between the hot and cold poles of the
CMB dipole moment.  The angle between the heliosphere nose and dipole
anisotropy direction is $91.5^\circ \pm 0.6^\circ$.  
The probability that the great circle that is
evenly spaced between the poles of the CMB dipole anisotropy, which is
defined to within $\sim \pm 0.10^\circ$, would
pass through the $\sim 3^\circ \times 3^\circ$ area on the sky
that includes the heliosphere nose direction, is less than 1/200.
Figure \ref{fig:cmb} shows the great circle in the sky that is
equidistant from the antipodes of the CMB dipole moment (gray line),
as well as the hot (red) and cold (blue) dipole directions.

The 90-degree great circle midway between the dipole directions also
passes through the best-fit ISMF direction, when uncertainties are
included (Table \ref{tab:ismf2}).  The ISMF direction from the
weighted fit us plotted as a purple dot in Figure \ref{fig:cmb},
together with the uncertainties on this direction (purple polygon) and
the 90-degree circle.  That the great-circle dividing the antipodes of
the CMB dipole moment would also pass through the ISMF pole, which is
defined by an area of $\sim 300$ square-degrees, including uncertainties,
is less than one in $ 10^4$.  A deeper knowledge
of the configuration and properties of the global heliosphere and
local ISMF will elucidate whether additional CMB foregrounds might be
present.

Although a possible relation between the CMB dipole and heliosphere
may seem far-fetched, a number of studies have shown that the
symmetries traced by the low-$\ell$ CMB moments are not consistent
with the predictions of gaussianity of the universe based on standard
cosmological models, and in particular the quadrupole and octopole
moments show evidence of the ecliptic geometry
\citep[e.g.][and citations in and to these articles]{Schwarzetal:2004,StarkmanCopietal:2009}.  Since the ISM
flows through the heliosphere from a direction 5$^\circ$ above the
ecliptic plane, these low-$\ell$ symmetries are equivalent to
symmetries around the heliosphere nose \citep[e.g. see Figures 6 and 7
in ][]{Frisch:2007cmb}.  Unrecognized CMB foregrounds related to the
heliosphere may be present, with the most likely candidate being
emission from interstellar dust inside the heliosphere since the dust
traces both the gas flow and ISMF direction
\citep{Slavinetal:2012,FrischSlavin:2012aogs,SlavinFrisch:2009sw12}.

\subsection{Speculative implications}\label{sec:dire}
\subsubsection{Local ISMF: Interarm or Arm? }\label{sec:fr}

Comparisons between the ISMF at the heliosphere and the ISMF observed
in other nearby regions provides insight into the uniform and random
components of the local ISMF, such as whether the field has the
large-scale coherence of interarm fields or of the small-scale random
fields seen in spiral arm regions.  Random components of the ISMF in
spiral arms are coherent over scale sizes of 5--50 parsecs; Faraday
rotation structure functions indicates a more turbulent ISMF with
outer scales of 10--20 parsecs
\citep{RandKulkarni:1989,Haverkorn:2006}.  In contrast, Faraday
rotation measures for interarm sightlines show a more uniform
component that is coherent over scales of 100-200 parsecs.  The ratio
of the uniform to random components of interstellar polarizations in
the plane of the sky is typically $\sim 0.8$, when global polarization
data are considered \citep{FosalbaLazarian:2002}.

If the local ISMF sampled by the IBEX Ribbon and optical polarization
data is part of an ISMF that is coherent over several hundred parsecs,
the coherence should appear in the ISMF direction obtained from
Faraday rotation measure (RM) data on nearby pulsars that yield an
electron-density weighted measurement of the parallel component of the
ISMF.  \citet{Salvati:2010} fit the RM and dispersion measures of four
nearby (distances 160--290 parsecs) pulsars that sample the low
density interior of the Local Bubble in the third galactic quadrant
(\glon$\sim 180^\circ - 270^\circ$).  The ISMF strength obtained from
these data is 3.3 \microG, and the field is directed towards
\glon,\glat$ \sim 5^\circ, 42^\circ$. The third galactic quadrant is
opposite to the \sixteen\ region in the sky. The difference in the
direction of the ISMF traced by the RM data and IBEX Ribbon, $\sim
23^\circ$, is easily accommodated by the field curvature indicated by
the envelope stars in the \sixteen\ region (including $\alpha$ Lyr 8
parsecs away), and may indicate that the ISMF around the Sun extends
into the third galactic quadrant.

The strength of the ISMF that shapes the heliosphere, $\sim 3$
\microG, is close to the field strength found from the pulsar data.
At the heliosphere, magnetic field strength is estimated from the
pressure equilibrium between the inner heliosheath ions traced by ENAs
and the ISM, and from the magnetic tension required to produce the
44\deeg\ offset between the observed ENA minimum and downwind gas flow
direction \citep{Schwadronetal:2011sep}, while in the LIC it is found
from equilibrium between gas and magnetic pressures
\citep{SlavinFrisch:2008}.  The polarity of the ISMF found from the pulsar RM
data is directed towards \elon,\elat=232\deeg,18\deeg, and therefore
is directed upwards through the ecliptic plane (see Table 1 for
galactic coordinates).  Although the field direction is consistent
with the local ISMF shaping the heliosphere, the polarity appears to
be the opposite of the polarity built into most MHD heliosphere models
(Appendix \ref{app:ibex}).

In the opposite direction of the sky from the pulsars, the \sixteen\
region tests the coherence of the ISMF along the sightlines that are tangent to
the Loop I shell.  The slow
rotation of the nearby ordered field component and its extension to within 8
parsecs pf the Sun, together with the similarity of ISMF directions
from the IBEX Ribbon and pulsar data, suggests that the Loop I ISMF
may extend into the third galactic quadrant.  Figure \ref{fig:loopI}
shows that the dust cavity caused by the Loop I expansion covers the
northern polar cap and extends into the third galactic quadrant at
high latitudes.  The coherence of the ISMF can be found by comparing
values from these three regions: (1) at the solar location, from the
center of the IBEX Ribbon arc (to within $\sim 15^\circ$, see Appendix \ref{app:ibex}); (2) the nearest 40 parsecs in the
galactic center region, including the tangential direction of Loop I
where the ordered component of the ISMF is found; (3) the low density
void of the Local Bubble in the third galactic quadrant.  The Sun is
located (by definition) in region 1, and between regions 2 and 3. The
difference in the ISMF directions determined from regions 1 and 2 is
$\sim 32^\circ$, and it is $\sim 22^\circ$ between regions 1 and
3. The field curvature must extend from region 2, through the solar
location in region 1, and into the third galactic quadrant
corresponding to region 3.  These data suggest that the local ISMF at
the heliosphere is part of a large scale field that extends several
hundred parsecs in space, such as is characteristics of interarm
regions.  The Loop I ISMF may be "opening out" into an interarm field
as it expands into the low density third galactic quadrant.


\subsubsection{Galactic cosmic rays and the local ISMF} \label{sec:gcr}

Galactic cosmic rays (CR) in the energy range of 50 GeV -- 40 TeV
trace the local ISMF because the gyroradius is on the order of several
hundred AU in a 1 \microG\ magnetic field.  Large scale anisotropies
and smaller scale excess regions, have been detected in the
distribution of CRs in this energy range
\citep{Nagashimaetal:1998,Halletal:1999gcr,Abdo:2008prlgcr,Abbasi:2011icecubegcr,SurdoArgogcrTeV:2011}.
It has been suggested that the Sun is located in a magnetic flux tube
in the ISM, and that such a flux tube would be a conduit for galactic
cosmic rays (GCRs) \citep{Frisch:1997fluxtube}.
\citet{CoxHelenius:2003} presented a model of the Local Bubble that
placed the Sun in a flux tube that has detached from the walls of the
Local Bubble.  \citet{Schwadronetal:2012gcr} have identified
anisotropies in the CR distribution towards the heliotail direction
caused by the propagation of CR's through the magnetic structure
surrounding the heliosphere.  Therefore it is an interesting exercise
to compare the GCR asymmetries with the local ISMF direction.

Using cosmic ray fluxes in the northern and southern hemispheres,
\citet{Nagashimaetal:1998} described two anisotropies in terms of a
broad sidereal galactic anisotropy, and a narrow anisotropy, the
``tail-in'' anisotropy, described by a maximum towards the anti-apex
direction of the solar motion.\footnote{The solar apex motion
corresponds to a velocity of $18 \pm 0.9$ \kms\ directed towards $\ell
\sim 47.9^\circ \pm 3.0^\circ, b=23.8^\circ \pm 2.0^\circ$, however
these values changes on a fairly regular basis with improved
understanding of the astrometric data for the comparison stars used to
establish the solar motion.}  The galactic anisotropy arrives from the
direction $\alpha,\delta = 0^\mathrm{H}, -20^\circ$ ($\ell, b \sim
61^\circ, -76^\circ$).  The tail-in excess corresponds to the
direction $\alpha, \delta = 6^\mathrm{H}, -24^\circ$ ($\ell, b \sim
230 ^\circ,-21^\circ$), with a half-width of $ \sim68^\circ$ and a
maximum effect near $\sim 10^3$ GeV.  The tail-in excess was
originally attributed to the heliotail because it is approximately
opposite to the direction of the solar motion through the LSR.
\citet{Halletal:1999gcr} fit the anisotropies as Gaussian features,
and found a direction of maximum anisotropy for the tail-in excess
towards $\alpha,\delta \sim 4.6^\mathrm{H}, -14^\circ$ ($\ell, b \sim
210^\circ, -36^\circ$), and a loss-cone centered at $\alpha, \delta
\sim \ 13^\mathrm{H}, 0^\circ$ ($\ell, b \sim 123.^\circ, 63.^\circ$).
To within the uncertainties, the direction of the tail-in excess found
by Hall et al. overlaps both the downwind \HeI\ flow vector that
defines the tail of neutral ISM gas behind the heliosphere, and the
axis of the ISMF direction determined from the weighted fit (towards
\glon,\glat $= 229^\circ \pm 30^\circ, -28^\circ \pm 30^\circ$).
Figure \ref{fig:cmb} shows the directions of the ISMF, tail-in excess
compared to the ISMF direction, and the true heliotail corresponding
to the ENA flux minimum centered $\sim 44^\circ$ west of the downwind
gas flow \citep{Schwadronetal:2011sep}.

\section{Summary and conclusions}\label{sec:conclusion}

We have determined the direction of the interstellar magnetic field
within 40 parsecs of the Sun and 90\deeg\ of the heliosphere nose
($\sim 15^\circ$ from the galactic center) using starlight polarized
by magnetically aligned interstellar dust grains.  The analysis is
based on the assumption that the polarization E-vector is parallel to
the direction of the ISMF and builds on the analysis in Paper I
\citep{Frisch:2010ismf1}.  The new technique for finding the ISMF
direction minimizes the mean of an ensemble of sines of polarization
position angles (\S \ref{sec:method}).  The use of an expanded data
set, including new polarization data that have been acquired at
several observatories for this study (\S \ref{sec:data}), permits the
use of weighted fits to the ISMF (\S \ref{sec:results}) as opposed to
the unweighted fit used in Paper I.  The best-fit ISMF from the
weighted fit is towards \glon,\glat$= 47^\circ \pm 20^\circ, 25^\circ
\pm 20^\circ$.  The direction obtained from the unweighted fit is
consistent with the results of the weighted fit (Table \ref{tab:ismf2}). 

The best-fit ISMF directions obtained from the
weighted fit and from the center of the IBEX Ribbon arc (Table 1) are separated
by $32^{+27}_{-30}$ degrees (\S \ref{sec:ibex}).  Although heliosphere
models indicate that there could be a small offset between the direction of 
the ISMF shaping the heliosphere and the Ribbon arc center ($0-15^\circ$),
the agreement between these two field directions is remarkable considering
that the polarization-based measurement is determined from light of stars up to
40 pc away.  The similarity of
these directions encourages the conclusion that the nearby ISMF is
coherent over decades of parsecs, while the difference indicates that
the local ISMF either has a curvature or is turbulent.

Interstellar optical polarizations reach maximum strengths in
directions perpendicular to the ISMF, and the IBEX Ribbon appears in
directions perpendicular to the ISMF draping over the heliosphere.
Hence if the directions of the best-fit ISMF and magnetic field
traced by the IBEX Ribbon are related, the polarization
strengths and ENA fluxes will be proportional.  In one region of the
sky that touches the Ribbon, where the ISMF is observed within 8 pc of
the Sun, ENA fluxes and polarization strengths are
found to increase together (\S \ref{sec:ibex}).  This effect is seen
only in the most sensitive polarization data.

In one region of the sky in the right-ascension interval \sixteen,
\citet{BaileyLucas:2010planetpol} used their high-sensitivity
PlanetPol data to conclude that polarization systematically increases
with distance (\S \ref{sec:ppdispol}).  Using stars that form the
upper envelope of the polarization versus distance relation in that
region, we find that $\mathrm{log}P = 1.200 + 0.013 *D$.  For standard relations
between polarization and color excess, and color excess versus \NH,
the distant-dependent increase in column density in this spatial
interval becomes log $N$(H)=18.01+0.013*D; the relation
overpredicts the actual \NHI\ measured for several stars by $\sim
35\%$.

This same region, which overlaps a low-latitude portion of Loop I,
appears to contain an ordered ISMF that extends to within 8 parsecs of
the Sun (\S \ref{sec:pppa}).  Stars forming the upper envelope of the
polarization-distance relation define a unique group where foreground
depolarization is minimized. For the nearby envelope stars, the
polarization angle depends on the star distance with the relation
\PAcel$ = 35.97 (\pm 1.4) - 0.25 (\pm 0.03) \cdot \mathrm{D} $ (\S
\ref{sec:pp}).  The ISMF rotates by $\sim 0.25^\circ$ pc$^{-1}$,
yielding a 10\deeg\ ISMF rotation over a 40 parsecs thick magnetic
layer.  The nearest polarized star showing the ordered ISMF is HD
172167 ($\alpha$ Lyr, 8 parsecs), where the polarization direction is
consistent with the direction of best-fit ISMF field. The dispersion
of position angles about this nearby ordered ISMF component suggests
that the turbulent ISMF component is $\sim \pm 23^\circ$.  These
results are conditioned on the exclusion of a 7.8$\sigma$ outlier star
from the analysis, and on the small number of stars (4) available for
defining the position angle rotation. Confirmation of this result
would require observations at ppm accuracy towards very faint stars,
V$>6$ mag, in the spatial interval \glon$\sim 25^\circ - 70^\circ$,
\glat$\sim 0^\circ - 325^\circ$.

The best-fit magnetic field direction, the flow of local ISM past
the Sun, and variable abundances of refractory elements in the local
ISM all suggest that the local ISM is associated with an expanding
fragment of the S1 shell of the Loop I superbubble (\S
\ref{sec:loopI}).  The bulk LSR flow of the local ISM past the Sun has
an upwind direction within 15\deeg\ of the center of the S1 shell.
The best-fit ISMF makes an angle of $\approx 76^\circ$ with the flow
direction, suggesting the field is parallel to the rim and
perpendicular to the flow velocity.  Such a configuration would be
expected for an ISMF swept up and compressed in an expanding
superbubble shell.  Variable abundances of refractory elements in
local ISM indicate recent shock destruction of some of the local dust
grains.
A consistent interpretation of the kinematics of the local ISM and the
direction of the best-fit ISMF would be that the ordered component
represents the ISMF that was been swept up in the expanding S1
superbubble shell. The
flow of ISM past the Sun is decelerating, as shown by the deviations
of velocity components from a rigid-body flow
\citep{Frisch:2011araa}. Cloud collisions in the decelerating flow may
generate the observed magnetic turbulence of $\pm 23^\circ$, with the
ordered component representing the matrix ISM between the clouds.

Polarization strengths for the envelope stars in region \sixteen\
anticorrelate with the interstellar radiation field at 975\AA\ (\S
\ref{sec:ppisrf}).  This anticorrelation is probably due to the fact
that those stars are in the first galactic quadrant and, by chance,
the most distant of those stars are also the most distant from the
bright stellar far ultraviolet and extreme ultraviolet radiation
sources that are mainly located in the third galactic quadrant.

The ISMF direction from pulsars in the third galactic quadrant is
within 22\deeg\ of the ISMF direction from the IBEX Ribbon.  A common
field strength of $\sim 3$ \microG\ is found from the pulsar data, a
range of heliosphere diagnostics including plasma pressure and the
deflection of the heliotail, and models of the ionization of the ISM
surrounding the heliosphere.  This suggests that the ISMF near the Sun
is coherent over large spatial scales, such as expected for interarm
regions.  One source of uncertainty is the polarity of the field,
since the polarity derived from pulsar data is opposite in direction
from the polarity that is assumed by several of the MHD heliosphere
models.

The new ISMF direction, together with the new consensus direction for the ISM
flow through the heliosphere derived from IBEX observations of
interstellar \HeI, strengthens the spatial
coincidence between the geometry of the CMB dipole moment, the
heliosphere nose, and the local ISMF (\S \ref{sec:cmb}).  The great
circle that divides the two poles of the CMB dipole moment passes
within $1.5^\circ \pm 0.6^\circ$ of the heliosphere nose defined by
the IBEX measurements of the flow of interstellar \HeI\ in the
heliosphere, and passes through the best-fit ISMF determined by
weighted fits to polarization position angles, to within the
uncertainties.  

The axis of the best-fit ISMF direction from polarization data
extends through the direction of the 'tail-in' excess of GeV--TeV
cosmic rays that has been attributed to the heliotail.  The direction
of the tail-in excess coincides poorly with the flow of interstellar
\HeI\ through the heliosphere.

High-sensitivity interstellar polarization data provide an opportunity
to explore the ISMF and magnetic turbulence in the very diffuse ISM
close to the Sun, and our very local galactic environment.  The
implications of understanding this field may affect our understanding
of the most distant reaches of the Universe.  The domain of the
heliosphere and the local ISM are filters through which all
observations about our more distant universe must pass.

\acknowledgments This work was supported by the IBEX mission as part
of NASA's Explorer Program, and by NASA grants NNX09AH50G and
NNX08AJ33G to the University of Chicago.  We are grateful for
observations made with the Nordic Optical Telescope, operated on the
island of La Palma jointly by Denmark, Finland, Iceland, Norway, and
Sweden, in the Spanish Observatorio del Roque de los Muchachos of the
Instituto de Astrofisica de Canarias.  We are also grateful for
observations made with telescopes at the Observatorio do Pico dos Dias
of the Laboratorio Nacional de Astrofisica of Brazil.


\begin{thebibliography}{112}
\expandafter\ifx\csname natexlab\endcsname\relax\def\natexlab#1{#1}\fi

\bibitem[{{Abbasi} {et~al.}(2011){Abbasi}, {Abdou}, {Abu-Zayyad}, {Adams},
  {Aguilar}, {Ahlers}, {Altmann}, {Andeen}, {Auffenberg}, {Bai}, \&
  et~al.}]{Abbasi:2011icecubegcr}
{Abbasi}, R., {Abdou}, Y., {Abu-Zayyad}, T., {Adams}, J., {Aguilar}, J.~A.,
  {Ahlers}, M., {Altmann}, D., {Andeen}, K., {Auffenberg}, J., {Bai}, X., \&
  et~al. 2011, \apj, 740, 16

\bibitem[{{Abdo} {et~al.}(2008){Abdo}, {Allen}, {Aune}, {Berley}, {Blaufuss},
  {Casanova}, {Chen}, {Dingus}, {Ellsworth}, {Fleysher}, {Fleysher},
  {Gonzalez}, {Goodman}, {Hoffman}, {H{\"u}ntemeyer}, {Kolterman}, {Lansdell},
  {Linnemann}, {McEnery}, {Mincer}, {Nemethy}, {Noyes}, {Pretz}, {Ryan},
  {Parkinson}, {Shoup}, {Sinnis}, {Smith}, {Sullivan}, {Vasileiou}, {Walker},
  {Williams}, \& {Yodh}}]{Abdo:2008prlgcr}
{Abdo}, A.~A., {Allen}, B., {Aune}, T., {Berley}, D., {Blaufuss}, E.,
  {Casanova}, S., {Chen}, C., {Dingus}, B.~L., {Ellsworth}, R.~W., {Fleysher},
  L., {Fleysher}, R., {Gonzalez}, M.~M., {Goodman}, J.~A., {Hoffman}, C.~M.,
  {H{\"u}ntemeyer}, P.~H., {Kolterman}, B.~E., {Lansdell}, C.~P., {Linnemann},
  J.~T., {McEnery}, J.~E., {Mincer}, A.~I., {Nemethy}, P., {Noyes}, D.,
  {Pretz}, J., {Ryan}, J.~M., {Parkinson}, P.~M.~S., {Shoup}, A., {Sinnis}, G.,
  {Smith}, A.~J., {Sullivan}, G.~W., {Vasileiou}, V., {Walker}, G.~P.,
  {Williams}, D.~A., \& {Yodh}, G.~B. 2008, Physical Review Letters, 101,
  221101

\bibitem[{{Adams} \& {Frisch}(1977)}]{AdamsFrisch:1977}
{Adams}, T.~F. \& {Frisch}, P.~C. 1977, \apj, 212, 300

\bibitem[{{Andersson}(2012)}]{Andersson:2012rev}
{Andersson}, B.-G. 2012, ArXiv e-prints/astro-ph:1208.4393

\bibitem[{{Andersson} {et~al.}(2011){Andersson}, {Pintado}, {Potter}, {Strai{\v
  z}ys}, \& {Charcos-Llorens}}]{AnderssonPotter:2011}
{Andersson}, B.-G., {Pintado}, O., {Potter}, S.~B., {Strai{\v z}ys}, V., \&
  {Charcos-Llorens}, M. 2011, \aap, 534, A19

\bibitem[{{Andersson} \& {Potter}(2006)}]{AnderssonPotter:2006}
{Andersson}, B.-G. \& {Potter}, S.~B. 2006, \apjl, 640, L51

\bibitem[{{Bailey} {et~al.}(2010){Bailey}, {Lucas}, \&
  {Hough}}]{BaileyLucas:2010planetpol}
{Bailey}, J., {Lucas}, P.~W., \& {Hough}, J.~H. 2010, \mnras, 405, 2570

\bibitem[{{Bennett} {et~al.}(2003){Bennett}, {Halpern}, {Hinshaw}, {Jarosik},
  {Kogut}, {Limon}, {Meyer}, {Page}, {Spergel}, {Tucker}, {Wollack}, {Wright},
  {Barnes}, {Greason}, {Hill}, {Komatsu}, {Nolta}, {Odegard}, {Peiris},
  {Verde}, \& {Weiland}}]{BennettHalpernwmap:2003dipoledirection}
{Bennett}, C.~L., {Halpern}, M., {Hinshaw}, G., {Jarosik}, N., {Kogut}, A.,
  {Limon}, M., {Meyer}, S.~S., {Page}, L., {Spergel}, D.~N., {Tucker}, G.~S.,
  {Wollack}, E., {Wright}, E.~L., {Barnes}, C., {Greason}, M.~R., {Hill},
  R.~S., {Komatsu}, E., {Nolta}, M.~R., {Odegard}, N., {Peiris}, H.~V.,
  {Verde}, L., \& {Weiland}, J.~L. 2003, \apjs, 148, 1

\bibitem[{{Berdyugin} {et~al.}(2011){Berdyugin}, {Piirola}, \&
  {Teerikorpi}}]{Berdyuginetal:2011}
{Berdyugin}, A., {Piirola}, V., \& {Teerikorpi}, P. 2011, in Astronomical
  Society of the Pacific Conference Series, Vol. 449, Astronomical Society of
  the Pacific Conference Series, ed. P.~{Bastien}, 157

\bibitem[{{Berkhuijsen}(1973)}]{Berkhuijsen:1973}
{Berkhuijsen}, E.~M. 1973, \aap, 24, 143

\bibitem[{{Bochsler} {et~al.}(2012){Bochsler}, {Petersen}, {M{\"o}bius},
  {Schwadron}, {Wurz}, {Scheer}, {Fuselier}, {McComas}, {Bzowski}, \&
  {Frisch}}]{Bochsler:2012isn}
{Bochsler}, P., {Petersen}, L., {M{\"o}bius}, E., {Schwadron}, N.~A., {Wurz},
  P., {Scheer}, J.~A., {Fuselier}, S.~A., {McComas}, D.~J., {Bzowski}, M., \&
  {Frisch}, P.~C. 2012, \apjs, 198, 13

\bibitem[{{Bohlin} {et~al.}(1978){Bohlin}, {Savage}, \&
  {Drake}}]{Bohlin_etal_1978}
{Bohlin}, R.~C., {Savage}, B.~D., \& {Drake}, J.~F. 1978, \apj, 224, 132

\bibitem[{{Butters} {et~al.}(2009){Butters}, {Katajainen}, {Norton}, {Lehto},
  \& {Piirola}}]{ButtersPiirola:2009}
{Butters}, O.~W., {Katajainen}, S., {Norton}, A.~J., {Lehto}, H.~J., \&
  {Piirola}, V. 2009, \aap, 496, 891

\bibitem[{{Chalov} {et~al.}(2010){Chalov}, {Alexashov}, {McComas}, {Izmodenov},
  {Malama}, \& {Schwadron}}]{Chalov:2010ribbon}
{Chalov}, S.~V., {Alexashov}, D.~B., {McComas}, D., {Izmodenov}, V.~V.,
  {Malama}, Y.~G., \& {Schwadron}, N. 2010, \apjl, 716, L99

\bibitem[{{Clarke} \& {Fullerton}(1996)}]{ClarkeFullerton:1996sun}
{Clarke}, D. \& {Fullerton}, S.~R. 1996, \aap, 310, 331

\bibitem[{{Cox}(2000)}]{Cox_2000}
{Cox}, A.~N. 2000, {A}llen's {A}strophysical {Q}uantities (AIP Press), 29--30

\bibitem[{{Cox} \& {Helenius}(2003)}]{CoxHelenius:2003}
{Cox}, D.~P. \& {Helenius}, L. 2003, \apj, 583, 205

\bibitem[{{Davis}(1955)}]{Davis:1955}
{Davis}, L. 1955, Physical Review, 100, 1440

\bibitem[{{Davis} \& {Greenstein}(1951)}]{DavisGreenstein:1951}
{Davis}, L.~J. \& {Greenstein}, J.~L. 1951, \apj, 114, 206

\bibitem[{{de Geus}(1992)}]{deGeus:1992}
{de Geus}, E.~J. 1992, \aap, 262, 258

\bibitem[{{Draine} \& {Weingartner}(1997)}]{Draine:1997radiativetorques}
{Draine}, B.~T. \& {Weingartner}, J.~C. 1997, \apj, 480, 633

\bibitem[{{Florinski} {et~al.}(2010){Florinski}, {Zank}, {Heerikhuisen}, {Hu},
  \& {Khazanov}}]{Florinski:2010ribbon}
{Florinski}, V., {Zank}, G.~P., {Heerikhuisen}, J., {Hu}, Q., \& {Khazanov}, I.
  2010, \apj, 719, 1097

\bibitem[{{Fosalba} {et~al.}(2002){Fosalba}, {Lazarian}, {Prunet}, \&
  {Tauber}}]{FosalbaLazarian:2002}
{Fosalba}, P., {Lazarian}, A., {Prunet}, S., \& {Tauber}, J.~A. 2002, \apj,
  564, 762

\bibitem[{{Frisch}(1990)}]{Frisch:1990}
{Frisch}, P.~C. 1990, in Physics of the Outer Heliosphere, ed. S.~Grzedzielski
  \& D.~E. Page, 19--22

\bibitem[{{Frisch}(1995)}]{Frisch:1995rev}
{Frisch}, P.~C. 1995, \ssr, 72, 499

\bibitem[{{Frisch}(1996)}]{Frisch:1996}
---. 1996, \ssr, 78, 213

\bibitem[{Frisch(1997)}]{Frisch:1997fluxtube}
Frisch, P.~C. 1997

\bibitem[{{Frisch}(2003)}]{Frisch:2003apex}
{Frisch}, P.~C. 2003, \apj, 593, 868

\bibitem[{{Frisch}(2007)}]{Frisch:2007cmb}
---. 2007, ArXiv e-prints:arXiv:0707.2970v2

\bibitem[{{Frisch}(2010)}]{Frisch:2010s1}
---. 2010, \apj, 714, 1679

\bibitem[{{Frisch} {et~al.}(2010{\natexlab{a}}){Frisch}, {Andersson},
  {Berdyugin}, {Funsten}, {Magalhaes}, {McComas}, {Piirola}, {Schwadron},
  {Slavin}, \& {Wiktorowicz}}]{Frisch:2010ismf1}
{Frisch}, P.~C., {Andersson}, B., {Berdyugin}, A., {Funsten}, H.~O.,
  {Magalhaes}, M., {McComas}, D.~J., {Piirola}, V., {Schwadron}, N.~A.,
  {Slavin}, J.~D., \& {Wiktorowicz}, S.~J. 2010{\natexlab{a}}, \apj, 724, 1473

\bibitem[{{Frisch} {et~al.}(1999){Frisch}, {Dorschner}, {Geiss}, {Greenberg},
  {Gr\"un}, {Landgraf}, {Hoppe}, {Jones}, {Kr{\"{a}}tschmer}, {Linde},
  {Morfill}, {Reach}, {Slavin}, {Svestka}, {Witt}, \& {Zank}}]{Frischetal:1999}
{Frisch}, P.~C., {Dorschner}, J.~M., {Geiss}, J., {Greenberg}, J.~M., {Gr\"un},
  E., {Landgraf}, M., {Hoppe}, P., {Jones}, A.~P., {Kr{\"{a}}tschmer}, W.,
  {Linde}, T.~J., {Morfill}, G.~E., {Reach}, W., {Slavin}, J.~D., {Svestka},
  J., {Witt}, A.~N., \& {Zank}, G.~P. 1999, \apj, 525, 492

\bibitem[{{Frisch} {et~al.}(2002){Frisch}, {Grodnicki}, \& {Welty}}]{FGW:2002}
{Frisch}, P.~C., {Grodnicki}, L., \& {Welty}, D.~E. 2002, \apj, 574, 834

\bibitem[{{Frisch} {et~al.}(2010{\natexlab{b}}){Frisch}, {Heerikhuisen},
  {Pogorelov}, {DeMajistre}, {Crew}, {Funsten}, {Janzen}, {McComas}, {Moebius},
  {Mueller}, {Reisenfeld}, {Schwadron}, {Slavin}, \& {Zank}}]{Frisch:2010next}
{Frisch}, P.~C., {Heerikhuisen}, J., {Pogorelov}, N.~V., {DeMajistre}, B.,
  {Crew}, G.~B., {Funsten}, H.~O., {Janzen}, P., {McComas}, D.~J., {Moebius},
  E., {Mueller}, H., {Reisenfeld}, D.~B., {Schwadron}, N.~A., {Slavin}, J.~D.,
  \& {Zank}, G.~P. 2010{\natexlab{b}}, \apj, 719, 1984

\bibitem[{{Frisch} {et~al.}(2011){Frisch}, Redfield, \&
  Slavin}]{Frisch:2011araa}
{Frisch}, P.~C., Redfield, S., \& Slavin, J. 2011, \araa, 49

\bibitem[{{Frisch} \& {Slavin}(2012)}]{FrischSlavin:2012aogs}
{Frisch}, P.~C. \& {Slavin}, J.~D. 2012, ArXiv e-prints/astro-ph:1205.4017

\bibitem[{{Funsten} {et~al.}(2009){Funsten}, {Allegrini}, {Crew}, {DeMajistre},
  {Frisch}, {Fuselier}, {Gruntman}, {Janzen}, {McComas}, {M{\"o}bius},
  {Randol}, {Reisenfeld}, {Roelof}, \& {Schwadron}}]{Funsten:2009sci}
{Funsten}, H.~O., {Allegrini}, F., {Crew}, G.~B., {DeMajistre}, R., {Frisch},
  P.~C., {Fuselier}, S.~A., {Gruntman}, M., {Janzen}, P., {McComas}, D.~J.,
  {M{\"o}bius}, E., {Randol}, B., {Reisenfeld}, D.~B., {Roelof}, E.~C., \&
  {Schwadron}, N.~A. 2009, Science, 326, 964

\bibitem[{{Gamayunov} {et~al.}(2010){Gamayunov}, {Zhang}, \&
  {Rassoul}}]{GamayunovZhang:2010ribbon}
{Gamayunov}, K., {Zhang}, M., \& {Rassoul}, H. 2010, \apj, 725, 2251

\bibitem[{{Gondhalekar} {et~al.}(1980){Gondhalekar}, {Phillips}, \&
  {Wilson}}]{Gondhalekaretal:1980}
{Gondhalekar}, P.~M., {Phillips}, A.~P., \& {Wilson}, R. 1980, \aap, 85, 272

\bibitem[{{Gray} {et~al.}(2006){Gray}, {Corbally}, {Garrison}, {McFadden},
  {Bubar}, {McGahee}, {O'Donoghue}, \& {Knox}}]{Gray:2006}
{Gray}, R.~O., {Corbally}, C.~J., {Garrison}, R.~F., {McFadden}, M.~T.,
  {Bubar}, E.~J., {McGahee}, C.~E., {O'Donoghue}, A.~A., \& {Knox}, E.~R. 2006,
  \aj, 132, 161

\bibitem[{{Grygorczuk} {et~al.}(2011){Grygorczuk}, {Ratkiewicz}, {Strumik}, \&
  {Grzedzielski}}]{Ratkiewicz:2011ribbon}
{Grygorczuk}, J., {Ratkiewicz}, R., {Strumik}, M., \& {Grzedzielski}, S. 2011,
  \apjl, 727, L48+

\bibitem[{{Grzedzielski} {et~al.}(2010){Grzedzielski}, {Bzowski}, {Czechowski},
  {Funsten}, {McComas}, \& {Schwadron}}]{GrzedzielskiBzowski:2010ribbon}
{Grzedzielski}, S., {Bzowski}, M., {Czechowski}, A., {Funsten}, H.~O.,
  {McComas}, D.~J., \& {Schwadron}, N.~A. 2010, \apjl, 715, L84

\bibitem[{{Hall} {et~al.}(1999){Hall}, {Munakata}, {Yasue}, {Mori}, {Kato},
  {Koyama}, {Akahane}, {Fujii}, {Fujimoto}, {Humble}, {Fenton}, {Fenton}, \&
  {Duldig}}]{Halletal:1999gcr}
{Hall}, D.~L., {Munakata}, K., {Yasue}, S., {Mori}, S., {Kato}, C., {Koyama},
  M., {Akahane}, S., {Fujii}, Z., {Fujimoto}, K., {Humble}, J.~E., {Fenton},
  A.~G., {Fenton}, K.~B., \& {Duldig}, M.~L. 1999, \jgr, 104, 6737

\bibitem[{{Hanbury Brown} {et~al.}(1960){Hanbury Brown}, {Davies}, \&
  {Hazard}}]{HanburyBrown:1960nps}
{Hanbury Brown}, R., {Davies}, R.~D., \& {Hazard}, C. 1960, The Observatory,
  80, 191

\bibitem[{{Haverkorn} {et~al.}(2006){Haverkorn}, {Gaensler}, {Brown},
  {Bizunok}, {McClure-Griffiths}, {Dickey}, \& {Green}}]{Haverkorn:2006}
{Haverkorn}, M., {Gaensler}, B.~M., {Brown}, J.~C., {Bizunok}, N.~S.,
  {McClure-Griffiths}, N.~M., {Dickey}, J.~M., \& {Green}, A.~J. 2006, \apjl,
  637, L33

\bibitem[{{Heerikhuisen} \& {Pogorelov}(2011)}]{HeerikhuisenPogorelov:2011}
{Heerikhuisen}, J. \& {Pogorelov}, N.~V. 2011, \apj, 738, 29

\bibitem[{{Heerikhuisen} {et~al.}(2010){Heerikhuisen}, {Pogorelov}, {Zank},
  {Crew}, {Frisch}, {Funsten}, {Janzen}, {McComas}, {Reisenfeld}, \&
  {Schwadron}}]{Heerikhuisen:2010ribbon}
{Heerikhuisen}, J., {Pogorelov}, N.~V., {Zank}, G.~P., {Crew}, G.~B., {Frisch},
  P.~C., {Funsten}, H.~O., {Janzen}, P.~H., {McComas}, D.~J., {Reisenfeld},
  D.~B., \& {Schwadron}, N.~A. 2010, \apjl, 708, L126

\bibitem[{{Heiles}(1976)}]{Heiles:1976araa}
{Heiles}, C. 1976, \araa, 14, 1

\bibitem[{{Heiles}(1998{\natexlab{a}})}]{Heiles:1998lb}
{Heiles}, C. 1998{\natexlab{a}}, in Lecture Notes in Physics, Berlin Springer
  Verlag, Vol. 506, IAU Colloq. 166: The Local Bubble and Beyond, ed.
  D.~{Breitschwerdt}, M.~J. {Freyberg}, \& J.~{Truemper}, 229--238

\bibitem[{{Heiles}(1998{\natexlab{b}})}]{Heiles:1998whence}
---. 1998{\natexlab{b}}, \apj, 498, 689

\bibitem[{{Heiles}(2000)}]{Heiles:2000pol}
---. 2000, \aj, 119, 923

\bibitem[{{Heiles}(2001)}]{Heiles:2001}
---. 2001, \apjl, 551, L105

\bibitem[{{Heiles}(2009)}]{Heiles:2009}
{Heiles}, C. 2009, in American Institute of Physics Conference Series, Vol.
  1156, American Institute of Physics Conference Series, ed. {R.~K.~Smith,
  S.~L.~Snowden, \& K.~D.~Kuntz}, 199--207

\bibitem[{{Jones} {et~al.}(1992){Jones}, {Klebe}, \&
  {Dickey}}]{JonesKlebe:1992}
{Jones}, T.~J., {Klebe}, D., \& {Dickey}, J.~M. 1992, \apj, 389, 602

\bibitem[{{Kimura} {et~al.}(2003){Kimura}, {Mann}, \&
  {Jessberger}}]{KimuraMann:2003clic24p5}
{Kimura}, H., {Mann}, I., \& {Jessberger}, E.~K. 2003, \apj, 582, 846

\bibitem[{{Kurth} \& {Gurnett}(2003)}]{KurthGurnett:2003}
{Kurth}, W.~S. \& {Gurnett}, D.~A. 2003, \jgr, 108, 2

\bibitem[{{Lallement} {et~al.}(2010){Lallement}, {Qu{\'e}merais}, {Koutroumpa},
  {Bertaux}, {Ferron}, {Schmidt}, \& {Lamy}}]{Lallement:2010soho}
{Lallement}, R., {Qu{\'e}merais}, E., {Koutroumpa}, D., {Bertaux}, J.,
  {Ferron}, S., {Schmidt}, W., \& {Lamy}, P. 2010, Twelfth International Solar
  Wind Conference, 1216, 555

\bibitem[{{Lallement} {et~al.}(1986){Lallement}, {Vidal-Madjar}, \&
  {Ferlet}}]{Lallementetal:1986}
{Lallement}, R., {Vidal-Madjar}, A., \& {Ferlet}, R. 1986, \aap, 168, 225

\bibitem[{{Lazarian}(2007)}]{Lazarian:2007rev}
{Lazarian}, A. 2007, \jqsrt, 106, 225

\bibitem[{{Lazarian} \& {Hoang}(2008)}]{LazarianHoang:2008}
{Lazarian}, A. \& {Hoang}, T. 2008, \apjl, 676, L25

\bibitem[{{Leroy}(1993)}]{Leroy:1993lism}
{Leroy}, J.~L. 1993, \aaps, 101, 551

\bibitem[{{Leroy}(1999)}]{Leroy:1999}
---. 1999, \aap, 346, 955

\bibitem[{{Magalhaes} {et~al.}(1996){Magalhaes}, {Rodrigues}, {Margoniner},
  {Pereyra}, \& {Heathcote}}]{Magalhaes:1996}
{Magalhaes}, A.~M., {Rodrigues}, C.~V., {Margoniner}, V.~E., {Pereyra}, A., \&
  {Heathcote}, S. 1996, in Astronomical Society of the Pacific Conference
  Series, Vol.~97, Polarimetry of the Interstellar Medium, ed. {W.~G.~Roberge
  \& D.~C.~B.~Whittet}, 118--+

\bibitem[{{Mao} {et~al.}(2010){Mao}, {Gaensler}, {Haverkorn}, {Zweibel},
  {Madsen}, {McClure-Griffiths}, {Shukurov}, \&
  {Kronberg}}]{MaoZweibel:2010ismf}
{Mao}, S.~A., {Gaensler}, B.~M., {Haverkorn}, M., {Zweibel}, E.~G., {Madsen},
  G.~J., {McClure-Griffiths}, N.~M., {Shukurov}, A., \& {Kronberg}, P.~P. 2010,
  \apj, 714, 1170

\bibitem[{{Martin}(1971)}]{Martin:1971}
{Martin}, P.~G. 1971, \mnras, 153, 279

\bibitem[{{Mathewson} \& {Ford}(1970)}]{MathewsonFord:1970}
{Mathewson}, D.~S. \& {Ford}, V.~L. 1970, \memras, 74, 139

\bibitem[{{Mathis}(1986)}]{Mathis:1986}
{Mathis}, J.~S. 1986, \apj, 308, 281

\bibitem[{{McComas} {et~al.}(2012){McComas}, {Alexashov}, {Bzowski}, {Fahr},
  {Heerikhuisen}, {Izmodenov}, {Lee}, {M{\"o}bius}, {Pogorelov}, {Schwadron},
  \& {Zank}}]{McComas:2012bow}
{McComas}, D.~J., {Alexashov}, D., {Bzowski}, M., {Fahr}, H., {Heerikhuisen},
  J., {Izmodenov}, V., {Lee}, M.~A., {M{\"o}bius}, E., {Pogorelov}, N.~V.,
  {Schwadron}, N.~A., \& {Zank}, G.~P. 2012, Science, 000, 0

\bibitem[{{McComas} {et~al.}(2009){McComas}, {Allegrini}, {Bochsler},
  {Bzowski}, {Christian}, {Crew}, {DeMajistre}, {Fahr}, {Fichtner}, {Frisch},
  {Funsten}, {Fuselier}, {Gloeckler}, {Gruntman}, {Heerikhuisen}, {Izmodenov},
  {Janzen}, {Knappenberger}, {Krimigis}, {Kucharek}, {Lee}, {Livadiotis},
  {Livi}, {MacDowall}, {Mitchell}, {M{\"o}bius}, {Moore}, {Pogorelov},
  {Reisenfeld}, {Roelof}, {Saul}, {Schwadron}, {Valek}, {Vanderspek}, {Wurz},
  \& {Zank}}]{McComas:2009sci}
{McComas}, D.~J., {Allegrini}, F., {Bochsler}, P., {Bzowski}, M., {Christian},
  E.~R., {Crew}, G.~B., {DeMajistre}, R., {Fahr}, H., {Fichtner}, H., {Frisch},
  P.~C., {Funsten}, H.~O., {Fuselier}, S.~A., {Gloeckler}, G., {Gruntman}, M.,
  {Heerikhuisen}, J., {Izmodenov}, V., {Janzen}, P., {Knappenberger}, P.,
  {Krimigis}, S., {Kucharek}, H., {Lee}, M., {Livadiotis}, G., {Livi}, S.,
  {MacDowall}, R.~J., {Mitchell}, D., {M{\"o}bius}, E., {Moore}, T.,
  {Pogorelov}, N.~V., {Reisenfeld}, D., {Roelof}, E., {Saul}, L., {Schwadron},
  N.~A., {Valek}, P.~W., {Vanderspek}, R., {Wurz}, P., \& {Zank}, G.~P. 2009,
  Science, 326, 959

\bibitem[{{McComas} {et~al.}(2010){McComas}, {Bzowski}, {Frisch}, {Crew},
  {Dayeh}, {DeMajistre}, {Funsten}, {Fuselier}, {Gruntman}, {Janzen}, {Kubiac},
  {Livadiotis}, {M{\"o}bius}, {Reisenfeld}, \&
  {Schwadron}}]{McComasetal:2010var}
{McComas}, D.~J., {Bzowski}, M., {Frisch}, P.~C., {Crew}, G.~B., {Dayeh},
  M.~A., {DeMajistre}, R., {Funsten}, H.~O., {Fuselier}, S.~A., {Gruntman}, M.,
  {Janzen}, P., {Kubiac}, M.~A., {Livadiotis}, G., {M{\"o}bius}, E.,
  {Reisenfeld}, D., \& {Schwadron}, N.~A. 2010, \jgr, 115, A09113

\bibitem[{{McComas} {et~al.}(2011){McComas}, {Funsten}, {Fuselier}, {Lewis},
  {M{\"o}bius}, \& {Schwadron}}]{McComas:2011grlrev}
{McComas}, D.~J., {Funsten}, H.~O., {Fuselier}, S.~A., {Lewis}, W.~S.,
  {M{\"o}bius}, E., \& {Schwadron}, N.~A. 2011, \grl, 381, 18101

\bibitem[{{Nagashima} {et~al.}(1998){Nagashima}, {Fujimoto}, \&
  {Jacklyn}}]{Nagashimaetal:1998}
{Nagashima}, K., {Fujimoto}, K., \& {Jacklyn}, R.~M. 1998, \jgr, 103, 17429

\bibitem[{{Naghizadeh-Khouei} \&
  {Clarke}(1993{\natexlab{a}})}]{Naghizadeh-KhoueiClarke:1993}
{Naghizadeh-Khouei}, J. \& {Clarke}, D. 1993{\natexlab{a}}, \aap, 274, 968

\bibitem[{{Naghizadeh-Khouei} \&
  {Clarke}(1993{\natexlab{b}})}]{NaghizadehClarke:1993stats}
---. 1993{\natexlab{b}}, \aap, 274, 968

\bibitem[{{Opal} \& {Weller}(1984)}]{OpalWeller:1984}
{Opal}, C.~B. \& {Weller}, C.~S. 1984, \apj, 282, 445

\bibitem[{{Opher} {et~al.}(2009){Opher}, {Bibi}, {Toth}, {Richardson},
  {Izmodenov}, \& {Gombosi}}]{OpherBibi:2009nature}
{Opher}, M., {Bibi}, A., {Toth}, G., {Richardson}, J.~D., {Izmodenov}, V.~V.,
  \& {Gombosi}, T.~I. 2009, Nature, 000, 000

\bibitem[{{Ostriker} {et~al.}(2001){Ostriker}, {Stone}, \&
  {Gammie}}]{OstrikerGammie:2001ismf}
{Ostriker}, E.~C., {Stone}, J.~M., \& {Gammie}, C.~F. 2001, \apj, 546, 980

\bibitem[{{Perryman}(1997)}]{Perrymanetal:1997}
{Perryman}, M.~A.~C. 1997, \aap, 323, L49

\bibitem[{{Piirola}(1973)}]{Piirola:1973}
{Piirola}, V. 1973, \aap, 27, 383

\bibitem[{{Piirola}(1977)}]{Piirola:1977}
---. 1977, \aaps, 30, 213

\bibitem[{{Pogorelov} {et~al.}(2009{\natexlab{a}}){Pogorelov}, {Heerikhuisen},
  {Mitchell}, {Cairns}, \& {Zank}}]{Pogorelovetal:2009asymmetries}
{Pogorelov}, N.~V., {Heerikhuisen}, J., {Mitchell}, J.~J., {Cairns}, I.~H., \&
  {Zank}, G.~P. 2009{\natexlab{a}}, \apjl, 695, L31

\bibitem[{{Pogorelov} {et~al.}(2009{\natexlab{b}}){Pogorelov}, {Heerikhuisen},
  {Mitchell}, {Cairns}, \& {Zank}}]{Pogorelov:2009Lasymmetry}
---. 2009{\natexlab{b}}, \apjl, 695, L31

\bibitem[{{Prested} {et~al.}(2010){Prested}, {Opher}, \&
  {Schwadron}}]{Prestedetal:2010}
{Prested}, C., {Opher}, M., \& {Schwadron}, N. 2010, \apj, 716, 550

\bibitem[{{Rand} \& {Kulkarni}(1989)}]{RandKulkarni:1989}
{Rand}, R.~J. \& {Kulkarni}, S.~R. 1989, \apj, 343, 760

\bibitem[{{Ratkiewicz} {et~al.}(2008){Ratkiewicz}, {Ben-Jaffel}, \&
  {Grygorczuk}}]{Ratkiewicz_etal_2008}
{Ratkiewicz}, R., {Ben-Jaffel}, L., \& {Grygorczuk}, J. 2008, in Astronomical
  Society of the Pacific Conference Series, Vol. 385, Numerical Modeling of
  Space Plasma Flows, ed. {N.~V.~Pogorelov, E.~Audit, \& G.~P.~Zank}, 189--+

\bibitem[{{Ratkiewicz} {et~al.}(2012){Ratkiewicz}, {Strumik}, \&
  {Grygorczuk}}]{Ratkiewicz:2012ribbon}
{Ratkiewicz}, R., {Strumik}, M., \& {Grygorczuk}, J. 2012, \apj, 756, 3

\bibitem[{{Redfield} \& {Linsky}(2004{\natexlab{a}})}]{RLII}
{Redfield}, S. \& {Linsky}, J.~L. 2004{\natexlab{a}}, \apj, 602, 776

\bibitem[{{Redfield} \& {Linsky}(2004{\natexlab{b}})}]{RLIII}
---. 2004{\natexlab{b}}, \apj, 613, 1004

\bibitem[{{Redfield} \& {Linsky}(2008)}]{RLIV:2008vel}
---. 2008, \apj, 673, 283

\bibitem[{{Richardson} \& {Stone}(2009)}]{Richardson+Stone_2009}
{Richardson}, J.~D. \& {Stone}, E.~C. 2009, \ssr, 143, 7

\bibitem[{{Roberge}(2004)}]{Roberge:2004}
{Roberge}, W.~G. 2004, in Astronomical Society of the Pacific Conference
  Series, Vol. 309, Astrophysics of Dust, ed. {A.~N.~Witt, G.~C.~Clayton, \&
  B.~T.~Draine}, 467--+

\bibitem[{{Salvati}(2010)}]{Salvati:2010}
{Salvati}, M. 2010, \aap, 513, A28+

\bibitem[{{Santos} {et~al.}(2011){Santos}, {Corradi}, \&
  {Reis}}]{Santosetal:2010}
{Santos}, F.~P., {Corradi}, W., \& {Reis}, W. 2011, \apj, 728, 104

\bibitem[{{Schwadron} {et~al.}(2012){Schwadron}, {Adams}, {Dingus}, {Funsten},
  {Desiati}, {McComas}, \& {Frisch}}]{Schwadronetal:2012gcr}
{Schwadron}, N.~A., {Adams}, F.~C., {Dingus}, B., {Funsten}, H.~O., {Desiati},
  P., {McComas}, D.~J., \& {Frisch}, P.~C. 2012, \apj, submitted, 00, 00

\bibitem[{{Schwadron} {et~al.}(2011){Schwadron}, {Allegrini}, {Bzowski},
  {Christian}, {Crew}, {Dayeh}, {DeMajistre}, {Frisch}, {Funsten}, {Fuselier},
  {Goodrich}, {Gruntman}, {Janzen}, {Kucharek}, {Livadiotis}, {McComas},
  {Moebius}, {Prested}, {Reisenfeld}, {Reno}, {Roelof}, {Siegel}, \&
  {Vanderspek}}]{Schwadronetal:2011sep}
{Schwadron}, N.~A., {Allegrini}, F., {Bzowski}, M., {Christian}, E.~R., {Crew},
  G.~B., {Dayeh}, M., {DeMajistre}, R., {Frisch}, P., {Funsten}, H.~O.,
  {Fuselier}, S.~A., {Goodrich}, K., {Gruntman}, M., {Janzen}, P., {Kucharek},
  H., {Livadiotis}, G., {McComas}, D.~J., {Moebius}, E., {Prested}, C.,
  {Reisenfeld}, D., {Reno}, M., {Roelof}, E., {Siegel}, J., \& {Vanderspek}, R.
  2011, \apj, 731, 56

\bibitem[{{Schwadron} {et~al.}(2009){Schwadron}, {Bzowski}, {Crew}, {Gruntman},
  {Fahr}, {Fichtner}, {Frisch}, {Funsten}, {Fuselier}, {Heerikhuisen},
  {Izmodenov}, {Kucharek}, {Lee}, {Livadiotis}, {McComas}, {Moebius}, {Moore},
  {Mukherjee}, {Pogorelov}, {Prested}, {Reisenfeld}, {Roelof}, \&
  {Zank}}]{Schwadron:2009sci}
{Schwadron}, N.~A., {Bzowski}, M., {Crew}, G.~B., {Gruntman}, M., {Fahr}, H.,
  {Fichtner}, H., {Frisch}, P.~C., {Funsten}, H.~O., {Fuselier}, S.,
  {Heerikhuisen}, J., {Izmodenov}, V., {Kucharek}, H., {Lee}, M., {Livadiotis},
  G., {McComas}, D.~J., {Moebius}, E., {Moore}, T., {Mukherjee}, J.,
  {Pogorelov}, N.~V., {Prested}, C., {Reisenfeld}, D., {Roelof}, E., \& {Zank},
  G.~P. 2009, Science, 326, 966

\bibitem[{{Schwarz} {et~al.}(2004){Schwarz}, {Starkman}, {Huterer}, \&
  {Copi}}]{Schwarzetal:2004}
{Schwarz}, D.~J., {Starkman}, G.~D., {Huterer}, D., \& {Copi}, C.~J. 2004,
  Phys. Rev. Let., 93, 221301

\bibitem[{{Serkowski} {et~al.}(1975){Serkowski}, {Mathewson}, \&
  {Ford}}]{Serkowskietal:1975}
{Serkowski}, K., {Mathewson}, D.~S., \& {Ford}, V.~L. 1975, \apj, 196, 261

\bibitem[{{Slavin} \& {Frisch}(2008)}]{SlavinFrisch:2008}
{Slavin}, J.~D. \& {Frisch}, P.~C. 2008, \aap, 491, 53

\bibitem[{{Slavin} {et~al.}(2009){Slavin}, {Frisch}, {Heerikhuisen},
  {Pogorelov}, {Mueller}, {Reach}, {Zank}, {Dasgupta}, \&
  {Avinash}}]{SlavinFrisch:2009sw12}
{Slavin}, J.~D., {Frisch}, P.~C., {Heerikhuisen}, J., {Pogorelov}, N.~V.,
  {Mueller}, H., {Reach}, W.~T., {Zank}, G.~P., {Dasgupta}, B., \& {Avinash},
  K. 2009, Space Science Reviews, (\rm Proceedings of Solar Wind 12). in press
  (http://adsabs.harvard.edu/abs/2009arXiv0911.1492S)

\bibitem[{{Slavin} {et~al.}(2012){Slavin}, {Frisch}, {Mueller}, {Heerikhuisen},
  {Pogorelov}, {Reach}, \& {Zank}}]{Slavinetal:2012}
{Slavin}, J.~D., {Frisch}, P.~C., {Mueller}, H.-R., {Heerikhuisen}, J.,
  {Pogorelov}, N.~V., {Reach}, W.~T., \& {Zank}, G.~P. 2012, \apj, submitted

\bibitem[{{Spoelstra}(1972)}]{Spoelstra:1972lb}
{Spoelstra}, T.~A.~T. 1972, \aap, 21, 61

\bibitem[{{Starkman} {et~al.}(2009){Starkman}, {Copi}, {Huterer}, \&
  {Schwarz}}]{StarkmanCopietal:2009}
{Starkman}, G., {Copi}, C.~J., {Huterer}, D., \& {Schwarz}, D.~J. 2009, in
  Cosmic Structure and Evolution

\bibitem[{{Stone} {et~al.}(2008){Stone}, {Cummings}, {McDonald}, {Heikkila},
  {Lal}, \& {Webber}}]{Stoneetal:2008nature}
{Stone}, E.~C., {Cummings}, A.~C., {McDonald}, F.~B., {Heikkila}, B.~C., {Lal},
  N., \& {Webber}, W.~R. 2008, \nat, 454, 71

\bibitem[{{Surdo} \& {Argo-Ybj Collaboration}(2011)}]{SurdoArgogcrTeV:2011}
{Surdo}, A. \& {Argo-Ybj Collaboration}. 2011, Astrophysics and Space Sciences
  Transactions, 7, 131

\bibitem[{{Taylor} {et~al.}(2009){Taylor}, {Stil}, \&
  {Sunstrum}}]{TaylorStilSunstrum:2010}
{Taylor}, A.~R., {Stil}, J.~M., \& {Sunstrum}, C. 2009, \apj, 702, 1230

\bibitem[{{Tilley} {et~al.}(2006){Tilley}, {Balsara}, \&
  {Howk}}]{TilleyBalsaraHowk:2006}
{Tilley}, D.~A., {Balsara}, D.~S., \& {Howk}, J.~C. 2006, \mnras, 371, 1106

\bibitem[{Tinbergen(1982)}]{Tinbergen:1982}
Tinbergen, J. 1982, \aap, 105, 53

\bibitem[{{Welty} \& {Crowther}(2010)}]{WeltyCrowther:2010Ti}
{Welty}, D.~E. \& {Crowther}, P.~A. 2010, \mnras, 404, 1321

\bibitem[{{Witte}(2004)}]{Witte:2004}
{Witte}, M. 2004, \aap, 426, 835

\bibitem[{{Wolleben}(2007)}]{Wolleben:2007}
{Wolleben}, M. 2007, \apj, 664, 349

\bibitem[{{Wood} {et~al.}(2005){Wood}, {Redfield}, {Linsky}, {M{\"u}ller}, \&
  {Zank}}]{Woodetal:2005lya}
{Wood}, B.~E., {Redfield}, S., {Linsky}, J.~L., {M{\"u}ller}, H.-R., \& {Zank},
  G.~P. 2005, \apjs, 159, 118

\end{thebibliography}


\appendix

\section{Appendix:  Heliosphere models and the IBEX Ribbon \label{app:ibex}}

Heliosphere models predict the density and ionization of the LIC and
the ISMF that shapes the heliosphere
\citep[e.g.][]{OpherBibi:2009nature,Pogorelovetal:2009asymmetries,Ratkiewicz_etal_2008,Prestedetal:2010},
and originally showed that the Ribbon forms where the sightline is
perpendicular to the direction of the interstellar field draping over
the heliosphere
\citep{Schwadron:2009sci,Heerikhuisen:2010ribbon,Chalov:2010ribbon,Ratkiewicz:2011ribbon,HeerikhuisenPogorelov:2011,Ratkiewicz:2012ribbon}.
The exact mechanism generating the Ribbon is a conundrum, partly
because the properties of turbulence upstream of the heliopause are
unknown \citep{GamayunovZhang:2010ribbon,Florinski:2010ribbon}, so
that several possible scenarios for the formation of the Ribbon have
been suggested
\citep[e.g.][]{McComasetal:2010var,Schwadronetal:2011sep,GrzedzielskiBzowski:2010ribbon}.
One possible Ribbon formation mechanism requires outflowing ENAs to
escape from the heliosphere, and charge exchange with interstellar
protons in the outer heliosheath.  These ENAs subsequently create
secondary inflowing ENAs from another quick charge exchange with
interstellar neutrals before the ring-beam distribution of the ions
(with a pitch angle of $\sim 90^\circ$) is disrupted by turbulence,
and therefore giving the \BdotR $=0$ alignment.  Testing for such a
model, \citet{HeerikhuisenPogorelov:2011} varied the strength and
direction of the ISMF, and the interstellar neutral H and proton
densities, to obtain a best-fit to the geometry of the Ribbon.  These
models suggest the Ribbon is formed roughly $\sim 100$ AU beyond the
heliopause, by a local ISMF that is directed away from the galactic
coordinates \glon,\glat=$ 33^\circ \pm 4^\circ, 53^\circ \pm 2^\circ$,
for a field strength of 2--3 \microG.  For comparison, an earlier MHD
model of the heliosphere asymmetries implied by the deflection of the
solar wind in the inner heliosheath gave an ISMF direction of
\glon$=10^\circ - 22^\circ$, \glat$=28^\circ - 38^\circ$, and field
strength $\sim 3.7 - 5.5$ \microG\ \citep{OpherBibi:2009nature}.  For
the \citet{HeerikhuisenPogorelov:2011} models, the angle between the
center of the Ribbon arc and the ISMF direction is $\sim 0^\circ -
15^\circ$.

The global heliosphere provides an \emph{in situ} diagnostic of the
magnetic field and plasma shaping the heliosphere.  The $\sim
48^\circ$ offset between ISMF direction given by the Ribbon arc center and the velocity of the
inflowing neutral interstellar gas creates asymmetries that can be
predicted using MHD models
\citep[e.g.][]{Pogorelov:2009Lasymmetry,OpherBibi:2009nature,Ratkiewicz_etal_2008,Prestedetal:2010}.
The data tracing the heliosphere asymmetries include the different
distances of the solar wind termination shock found by Voyager 1 in
the northern ecliptic hemisphere and that found by Voyager 2 in the
south \citep[94 AU vs.\ 84
AU,][]{Stoneetal:2008nature,Richardson+Stone_2009}, the $\sim 8^\circ$
offset between the upwind directions of the neutral interstellar \HI\
and \HeI\ flowing into the heliosphere \citep[using data from
][]{Lallement:2010soho,McComas:2012bow}, and the preferential
alignment exhibited by the 3 kHz plasma emissions from the ISM beyond
the heliopause that were detected by both Voyagers
\citep[e.g.][]{KurthGurnett:2003}.  The ISMF directions from these
models are generally consistent with the field direction from the IBEX
ribbon.

\citet{McComas:2012bow} have determined a new consensus direction for
the velocity vector of interstellar \HeI\ flowing through the
heliosphere.  The \HeI\ gas velocity, $23.2 \pm 0.3$ \kms\ (Table
\ref{tab:ismf2}), agrees with the (less precise) velocity of
interstellar dust traveling through the heliosphere, after correction
for propagation effects, of $24.5^{+1.1}_{-1.2}$ \kms\ from galactic
coordinates \glon,\glat$=8^\circ \pm 15^\circ, 14^\circ \pm 4^\circ$
\citep{Frischetal:1999,KimuraMann:2003clic24p5}.  It also agrees with
the velocity of interstellar \HI\ interacting with the heliosphere
found from the first spectrum of \Lya\ emission from interstellar \HI\
inside of the heliosphere, $22.5 \pm 2.8 $ \kms, after correction for
\HI\ propagation and to the consensus IBEX upwind direction
\citep{AdamsFrisch:1977}.

 The outer boundary conditions of the heliosphere models are set by
the physical properties of the LIC, which must be reconstructed from
ISM observations using photoionization models because of the low
opacity of the cloud.  Ribbon models are highly sensitive to the
interstellar boundary conditions of the heliosphere; 15\% variations
in the interstellar parameters lead to pronounced differences in the
location and width of the Ribbon \citep{Frisch:2010next}.
Photoionization models indicate that the LIC is a low density, ${n}
\sim 0.27$ \cc, partially ionized cloud, $\mathrm{H^+/H} \sim 22$\%
and $\mathrm{He^+/He} \sim 39$\% \citep[Model 26
in][]{SlavinFrisch:2008}.  Observations of ISM inside of the
heliosphere also probe the LIC properties. When the IBEX measurements
of interstellar \OI, and \NeI\ are corrected for propagation effects
through and entering the heliosphere, and ionization in the LIC using
photoionization models, the interstellar abundance ratio O/Ne$=0.27\pm
0.10$ that is recovered agrees with other local ISM measurements and
suggests that over 50\% of the O is depleted onto interstellar dust
grains \citep{Bochsler:2012isn}.  IBEX and Ulysses measurements of
interstellar \HeI\ indicate the LIC is warm, $6300 \pm 390$ K
\citep{Witte:2004,McComas:2012bow}.

New IBEX-LO observations of the flow of interstellar \HeI\ through the
inner heliosphere \citep{McComas:2012bow} indicate that ISM flow is
$\sim 12$\% slower than found from earlier Ulysses data
\citep{Witte:2004}, which lowers the interstellar ram pressure on the
heliosphere by $\sim 22$\% and may affect the models of the Ribbon
location.  The value of the \HeI\ flow vector is given in Table 1,
together with the $1\sigma$ uncertainties for longitude, latitude, and
velocity.  The new IBEX upwind direction has shifted to 3.6\deeg\ east
of the Ulysses direction.  The uncertainties are somewhat larger when
the bounding range constrains are included; these constraints set the
limits as \glon$ = 5.25^{+1.48}_{-1.37}$ degrees,
\glat$=12.03^{+2.67}_{-3.19}$ degrees, the velocity to $
-23.20^{+2.5}_{-1.9}$ \kms, and the temperature to 5000--8300 K.

\section{Appendix:  List of stars from the literature included in study \label{app:A}}

The stars within 40 parsecs from the PlanetPol catalog included
in this study are: HD 97603 (HR
4357), HD 140573 (HR 5854), HD 139006 (HR 5793), HD 116656 (HR 5054),
HD 156164 (HR 6410), HD 95418 (HR 4295), HD 112185 (HR 4905), HD
161096 (HR 6603), HD 177724 (HR 7235), HD 127762 (HR 5435), HD 153210
(HR 6299), HD 120315 (HR 5191), HD 113226 (HR 4932), HD 112413 (HR
4915), HD 163588 (HR 6688), HD 95689 (HR 4301) and HD 131873 (HR
5563).

The nearby stars from \citet{Santosetal:2010} catalog that have been
used in this study include: HD 223889 (HIP
117828), HD 120467 (HIP 67487), HD 173818 (HIP 92200), HIP 65520, HD
127339 (HIP 70956), HIP 82283, HIP 50808, HIP 106803, HD 162283 (HIP
87322), HIP 114859 HIP 83405 HIP 96710 HIP 90035 HIP 87745 HD 176986
(HIP 93540), HD 175726 (HIP 92984), HIP 61872, HIP 95417, HD 155802
(HIP 84303), HD 166184 (HIP 88961), HD 161098 (HIP 86765), HD 182085
(HIP 95299), HD 164651 (HIP 88324), HD 119638 (HIP 67069), HD 125184
(HIP 69881), HD 95521 (HIP 53837), HIP 66918, HD 177409 (HIP 94154),
HD 144660 (HIP 78983), HD 151528 (HIP 82260), HD 95338 (HIP 53719), HD
127352 (HIP 70973), HD 183063 (HIP 95722), HD 105328 (HIP 59143), HD
204385 (HIP 106213), HD 145809 (HIP 79524), HD 172675 (HIP 91645), HD
149013 (HIP 81010), HD 185615 (HIP 96854) and HD 138885 (HIP 76292).

The seven PlanetPol stars that define the upper end of the region \sixteen\
polarization envelope (\S \ref{sec:pp}) are listed in Table \ref{tab:seven}.

\section{Polarization vrs Distance relation } \label{app:pol}

Polarization of optical starlight in the interstellar medium is due to
magnetically aligned dichroic dust grains, so that there is a
correlation between the upper envelope of polarization and
interstellar extinction for a given set of observations
\citep{Serkowskietal:1975,FosalbaLazarian:2002,Andersson:2012rev}.
The usual term ``upper envelope'' describes the relation between the
maximum observed polarizations for a set of stars as a function of
extinction, i.e. for a given extinction no polarization is found above
the upper envelope value by definition.  Figure \ref{fig:pp1} shows
that an upper envelope is also present in the polarization-distance
plot.  The scatter of polarization strengths for a given extinction is
due to the patchy nature of the ISM and magnetic turbulence.
Likewise, the presence of an upper envelope to the
polarization-distance plot in Figure \ref{fig:pp1} likely results from
either the patchy nature of nearby ISM \citep{Frisch:2011araa}, or
magnetic turbulence, or both.

The polarization data represented in Figure
\ref{fig:dvrspolcent} were collected over a timespan of a half of a
century, using a diverse set of instruments with unknown systematics.
The reasonableness of the statement that polarization increases with
distance has been tested for the stars shown in the figure using the
nonparametric Kendall's tau measure of bivariate correlation.  For the
set of stars where Pol/dPol$\ge3$, the rank correlation coefficiet is
$\tau = 0.86$, where a value of one indicates that there is a perfect
association of increasing polarization with increasing distance.
Selecting only the most accurate data points, Pol/dPol$\ge5$, yields
$\tau=0.89$.  However for the nearest stars used in this analysis, $<
40$ pc, this association is not demonstrated ($\tau=0.14$ \PdP$\ge3$),
an effect that we attribute both to influence of systematic
uncertainties on weak polarization measurements and to the patchy
nature of nearby ISM. In \S \ref{sec:ppdispol} we discuss a special
nearby region where increasing interstellar polarization is shown to
be associated with increasing distance for stars within 40 pc.

\section{Asymmetry of very local interstellar radiation field } \label{app:isrf}

\begin{figure}
\plotone{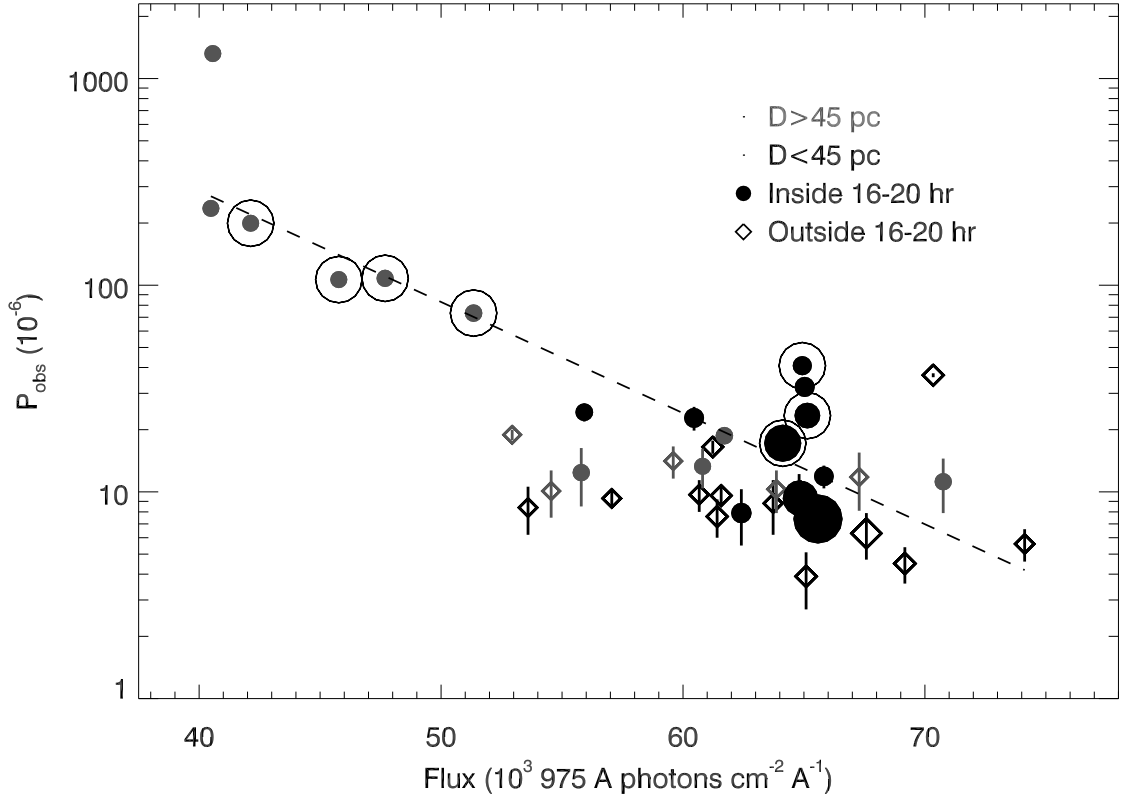}
\caption{Stellar polarizations with $3 \sigma$ detections are plotted
against the FUV radiation field at each star for the 975 \AA\ fluxes
from the 25 brightest stars measured by \citet{OpalWeller:1984}.  Dots
show stars in the \sixteen\ region, and diamonds show stars outside of
this region.  Black/gray symbols indicate stars within/beyond 45 pc.
Symbol sizes are inversely proportional to the distance of
the star.  The seven envelope stars in Figure \ref{fig:pp1} are
circled.  The dashed/dotted lines show fits between the
polarizations and 975 A fluxes for stars inside/outside of region
\sixteen.  Polarization increases strongly at low FUV fluxes, but the
effect is probably due to the relative locations of the most polarized
distant stars in the first galactic quadrant, and the bright FUV
sources that are mainly located in the third galactic quadrant
(see Appendix \ref{app:isrf}).  }\label{fig:pp3}
\end{figure}

\begin{figure}
\plottwo{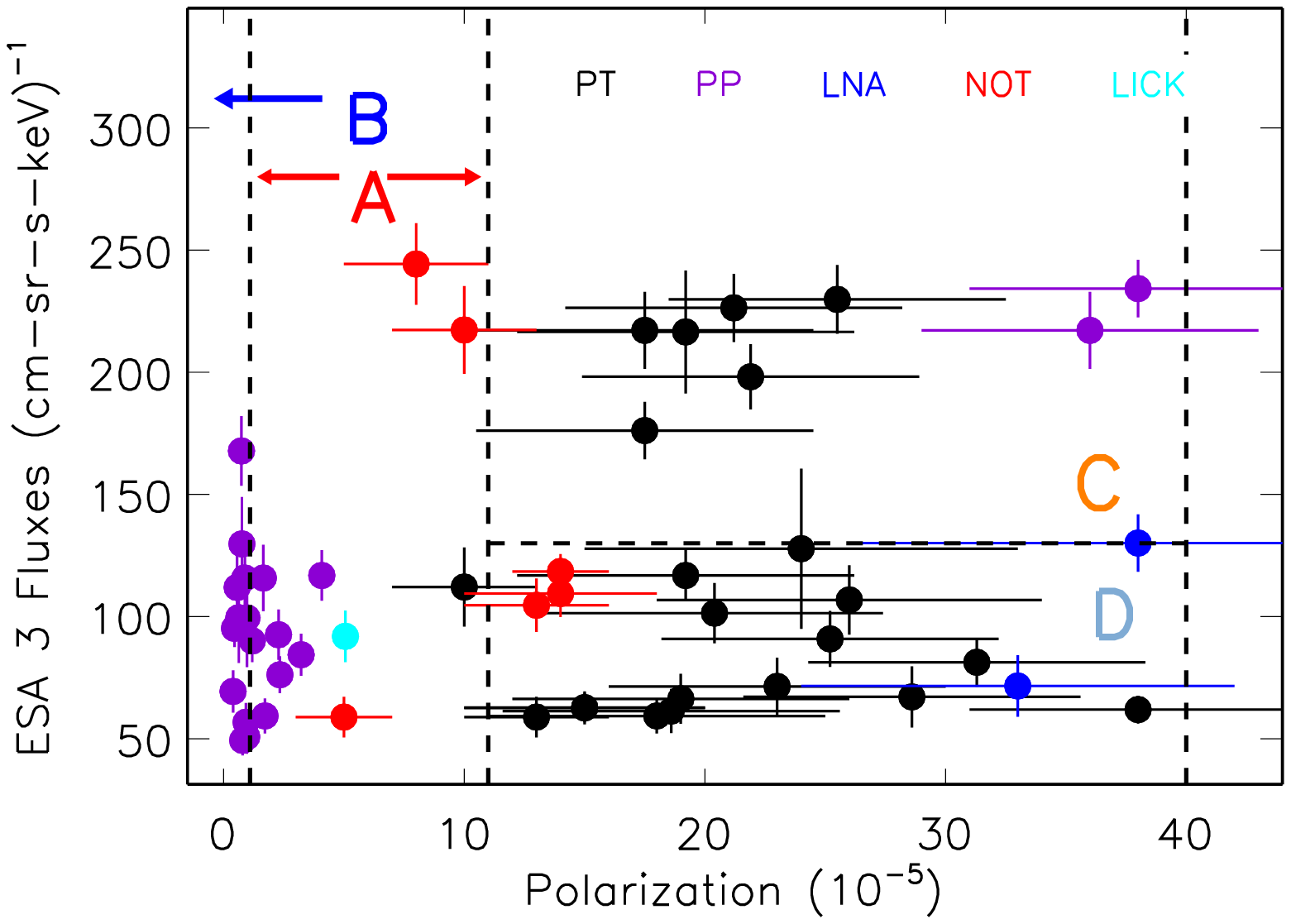}{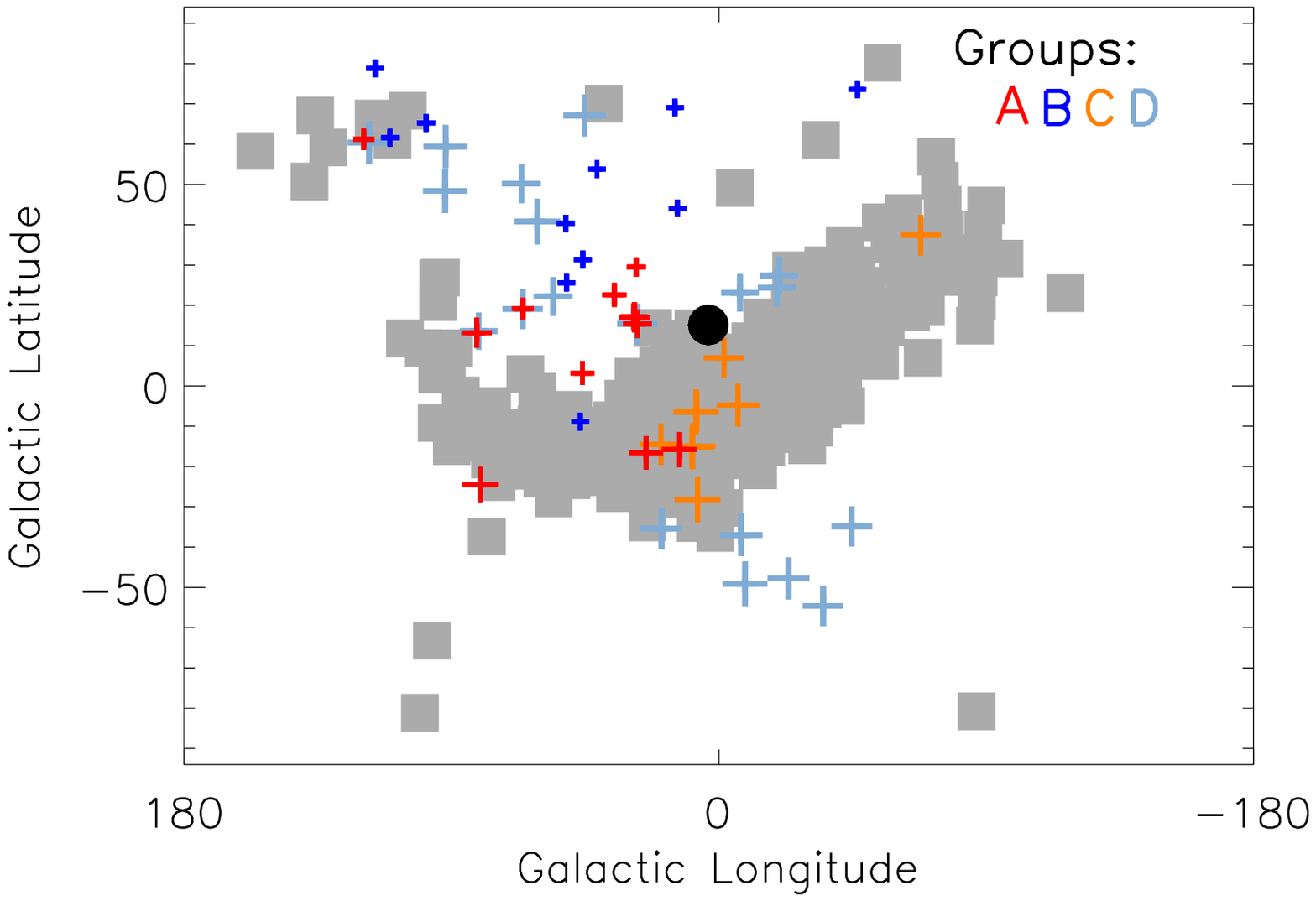}
\caption{Polarization strengths tend to increase with the ENA fluxes
for the polarizations with the highest statistical significance.
Left: The abscissa shows the polarization for stars within 40 pc and
$90^\circ$ of the heliosphere nose, for $P/\sigma>2.5$ where
$P,~\sigma$ are the polarization and $1\sigma$ uncertainty.  The
ordinate shows the mean IBEX-HI 1.1 keV ENA fluxes towards the same
stars, where the IBEX fluxes are calculated as the average value for
pixels that have S/N$>3.5$ and are within 6 degrees of the star.  The
polarization sources are identified by color, with PT (black) representing
20th century data, PP (purple) representing
PlanetPol data, and LNA, NOT, and LICK
data representing data acquired for this study (blue, red, cyan, see \S \ref{sec:data}).
The data shown in the left figure are divided into four groups, denoted by
A, B, C, and D, given by (A) $1 \times 10^{-5} < P < 11 \times 10^{-5} $, 
all ENA flux levels;(B) $P< 1 \times 10^{-5}$, all ENA flux
levels; (C) $ 11 \times 10^{-5} < P < 40 \times 10^{-5} $ and ENA
fluxes $>130$; (D) $ 11 \times 10^{-5} < P < 40 \times 10^{-5} $ and
ENA fluxes $<130$. The ENA fluxes increase with
polarization strength for stars in group A. PlanetPol stars in group A are located in
the \sixteen\ region.  Right: Stars in the four groups are overplotted
on the IBEX Ribbon (1 keV), and color-coded according to the group in the
left figure: red=A, B=blue, C=orange, and D=light-blue.
Symbol size is proportional
to polarization strength.  The polarizations in groups A
and B are recent data with lower uncertainties; group A in particular
shows polarization increasing with ENA fluxes.  The light-blue data
points (group D) tend to be older data with large uncertainties.
}\label{fig:pvrsena}
\end{figure}

\begin{figure}
\plottwo{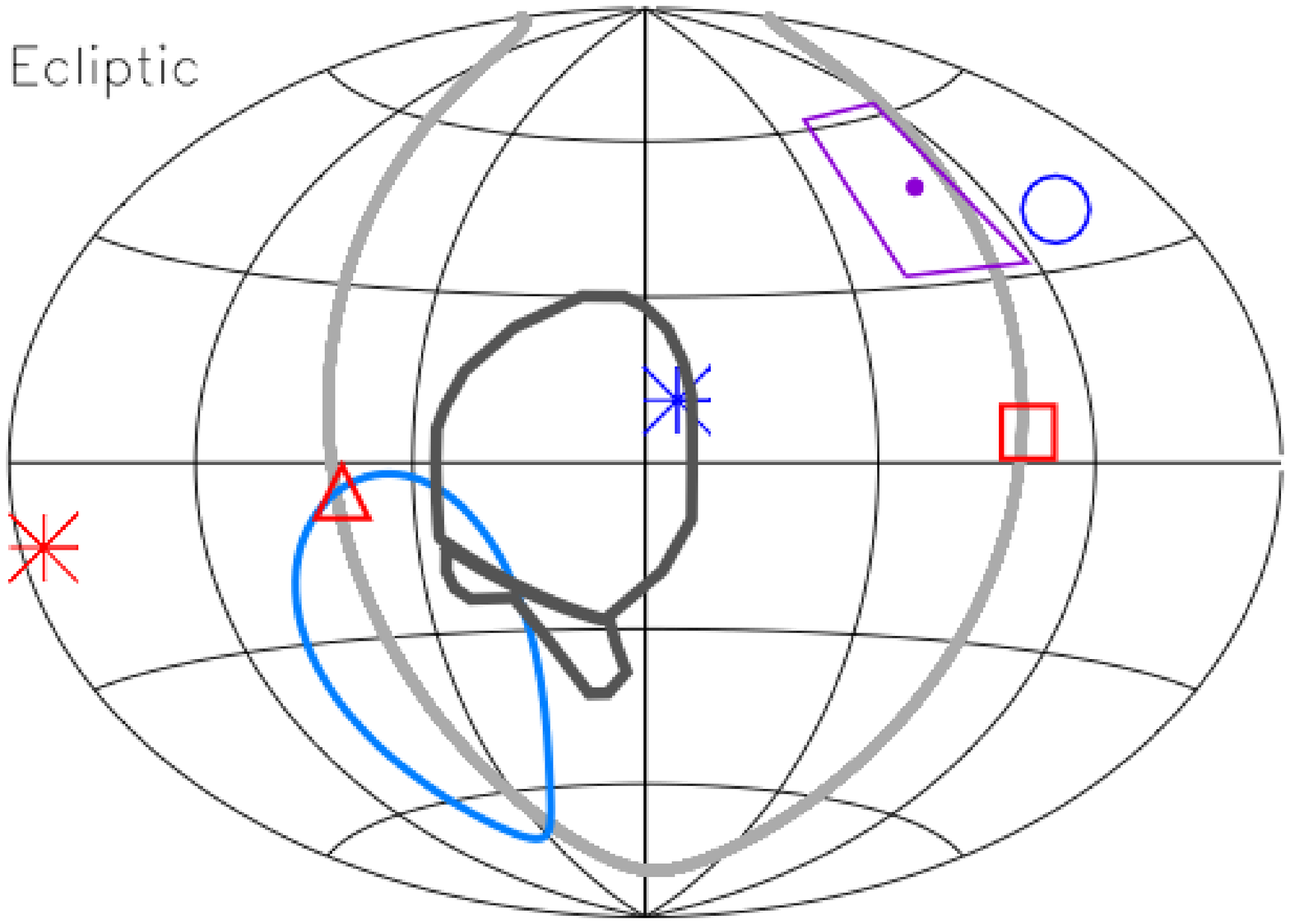}{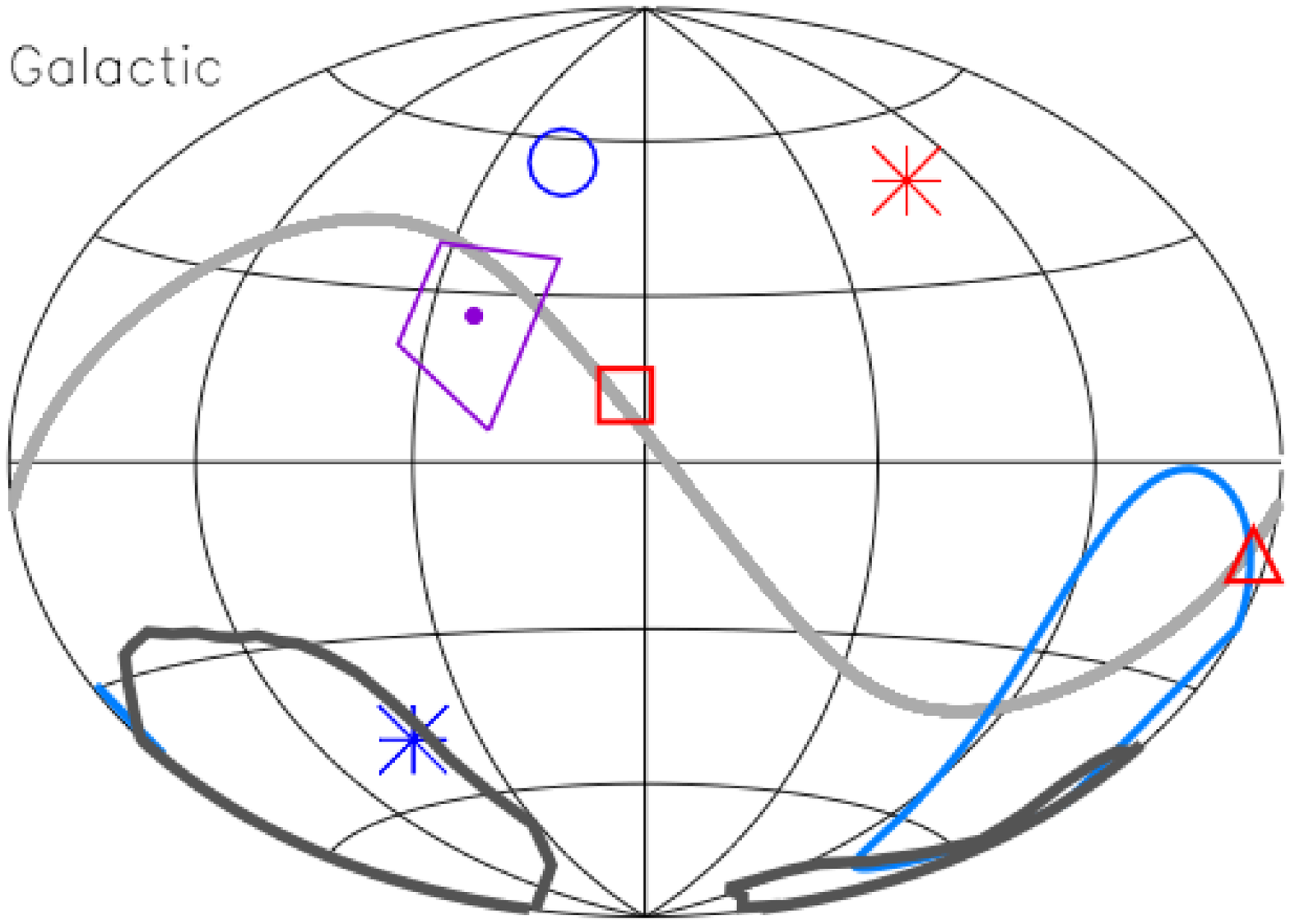}
\caption{The great circle (gray line) in the sky that evenly separates
the hot (red asterisk) and cold (blue asterisk) poles of the cosmic
microwave background dipole anisotropy is plotted in ecliptic (left)
and galactic (right coordinates).  The best-fitting
ISMF is shown in purple on each plot, with the dot/polygon giving the
direction/uncertainties of the weighted fit in Table \ref{tab:ismf2}.
The great circle equidistant between the CMB dipoles intersects both the
best-fitting local ISMF direction 
and the heliosphere nose (red square, see Table \ref{tab:ismf2}), to
within the uncertainties.  The blue open circle gives the ISMF
direction derived from the center of the IBEX Ribbon arc.  The large
blue circular region (light gray) shows the location of the tail-in anisotropy for
$\sim 500$ GCRs GeV modeled by \citet{Halletal:1999gcr}, with the
extent defined by the 68\deeg\ width from
\citet{Nagashimaetal:1998}. The regions outlined by thick black lines
show the observed heliotail that is deflected $\sim 44 ^\circ$ to the
west of the downwind gas direction (red triangle), and is prominent in
maps of the globally distributed ENAs \citep{Schwadronetal:2011sep}.
Each plot is centered on the longitude of 0\deeg\ in the respective
coordinate system, with longitude increasing towards the left.
}\label{fig:cmb}
\end{figure}

The asymmetry of the local interstellar radiation field is related to
the overall opacity of the Local Bubble.  The brightest of the far-UV
sources are located in the third and fourth galactic quadrants
\citep[\glon$=180^\circ - 360^\circ$,][]{Gondhalekaretal:1980} because
of the distribution of massive stars with respect to the Local Bubble
void. Those same stars are bright at the 13.6 eV ionization edge of
\HI\ so that in the absence of interstellar opacity the
\glon$=180^\circ - 360^\circ$ interval will be more highly ionized
\citep{Frisch:2010s1}.  This distribution of radiation sources
explains the relation between polarization and the radiation field
at each star that is displayed in Figure \ref{fig:pp3}.

\begin{figure}
\plottwo{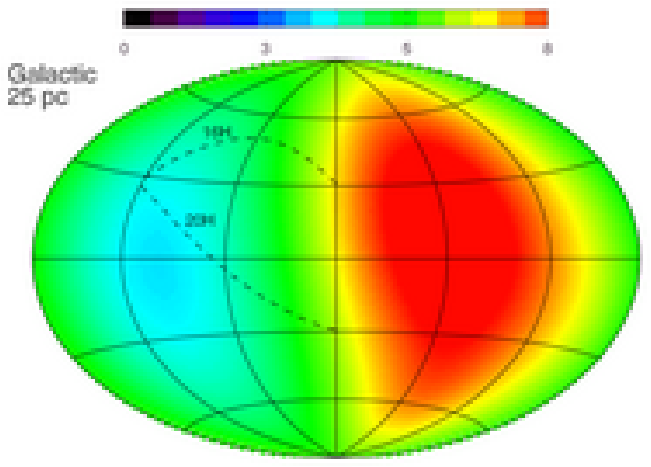}{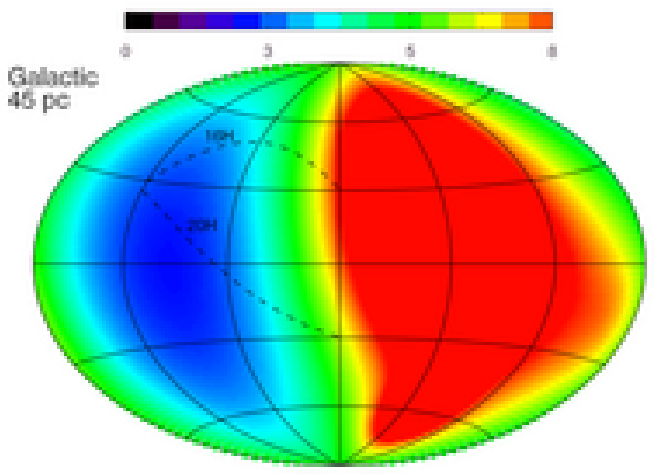}
\caption{Diffuse interstellar fluxes at 975 \AA\ are plotted for
spheres of radius 25 pc (left) and 45 pc (right) around the Sun.  The
figure is centered on the galactic center with galactic longitude
increasing towards the left.  These fluxes are calculated from the
combined fluxes of the 25 brightest stars in the sky at 975 \AA\ based
on the measurements of \citet{OpalWeller:1984}.  Most of that flux
originates in the third and fourth galactic quadrants
\citep{Gondhalekaretal:1980}, which leads to the strong radiation
fluxes in quadrants III and IV (right figure, \glon$>180^\circ$) when
compared to weak fluxes in quadrants I and II (\glon$<180^\circ$).
This asymmetry in the diffuse interstellar radiation field is due to
the distribution of massive stars around the Local Bubble void.  The
more distant stars in the \sixteen\ interval are exposed to a lower
ambient interstellar radiation field than in regions closer to the
Sun.  Color bars are the same for each figure, and are plotted with
arbitrary units that are scaled between the minimum and maximum fluxes
of $4.0 \times 10^4$ and $8.0 \times 10^4$ photons \cmtwo\ \persec\
\AA$^{-1}$, respectively. }\label{fig:pp4}
\end{figure}

Figure \ref{fig:pp4} shows the different levels of far UV radiation at
975 \AA\ between galactic quadrants I and II (\glon$=0^\circ -
180^\circ$) and III and IV (\glon$=180^\circ - 360^\circ$).  The sum
of the stellar fluxes are shown at spherical surfaces around the Sun
with radii of 25 parsecs and 45 parsecs, and displayed using an aitoff
projection.  The mapped fluxes represent the sum of stellar fluxes at
each position, based on the twenty-five brightest OB stars at 975 \AA\
\citep{OpalWeller:1984}.  Flux units are photons \cmtwo\ \persec\
\AA$^{-1}$. The fluxes are plotted with the same color-coding in both
figures.  The minimum and maximum flux values on the color bar
correspond to $4.0 \times 10^4$ and $8.0 \times 10^4$ \cmtwo\ \persec\
\AA$^{-1}$, respectively.  Low radiation fluxes in the region of large
PlanetPol polarizations are the result of the locations of hot stars.

\begin{deluxetable}{lccc}

\tablecaption{Various Directions \label{tab:ismf2} }
\tablewidth{0pt} 
\tabletypesize{\small}
\footnotesize{\tiny}
\tablehead{ 
\colhead{Source} & \colhead{Longitude} & \colhead{Latitude} & \colhead{Notes} \\ 
\colhead{coordinates} & \colhead{deg.} & \colhead{deg.} & \colhead{(and references)}  
}

\startdata 

\multicolumn{3}{l}{\bf Direction of best-fitting Magnetic Field } & \\
\multicolumn{2}{l}{\emph{Polarization, Paper I, unweighted:}}  &&\\
Ecliptic & $ 263$ & $37 $ & uncertainties $\pm 35$ \\
Galactic  & $37$ & $23 $ & uncertainties $\pm 35$ \\ 

\multicolumn{2}{l}{\emph{Polarization, this Paper, unweighted:}}  &&\\
Ecliptic & $263 ^{+10}_{-5} $ & $37 \pm 15 $  &  \\
Galactic & $37\pm 15 $ & $22 \pm 15 $  & \\ 

\multicolumn{2}{l}{\emph{Polarization, this Paper, weighted: }}  &&\\
Ecliptic & $ 263 ^{+15}_{-20}$ & $47 \pm 15$  &  \\
Galactic & $47\pm 20$ & $25\pm 20 $  & \\ 

\multicolumn{2}{l}{\bf ISMF from Center of Ribbon arc:} & & [2]  \\
Ecliptic & $ 221 \pm 4$ & $39  \pm 4$  & (see text)\\ 
Galactic & $33 \pm 4$ & $55 \pm 4 $  &\\ 

\multicolumn{3}{l}{\bf Upwind direction of interstellar \HeI\ flow through heliosphere } & 
		V=$23.2 \pm 0.3$ \kms, T$=6300 \pm 390$ K [1] \\
Ecliptic &  $259.00 \pm 0.47$  & $ 4.98 \pm 0.21$  &  \\
Galactic  &  $5.25 \pm 0.24$ & $12.03 \pm 0.51$ & \\

\multicolumn{3}{l}{\bf Quadrant III pulsars} & \\
Ecliptic & $232 $ & $18$ &  [6]  \\
Galactic & $5 $ & $42 $  & [6]  \\

\multicolumn{3}{l}{\bf Heliotail in globally distributed  IBEX ENA flux} & \\
Ecliptic & $30 \pm 30$ & $0 \pm 30$ &  [5]  \\
Galactic & $146 \pm 30 $ & $-49 \pm  30$  & [5]  \\

\multicolumn{3}{l}{\bf Direction of tail-in cosmic ray asymmetries } & \\
\multicolumn{2}{l}{\emph{Sub-TeV anisotropies:}} & & \\

Ecliptic & $90$ & $-47$ & cone half-width=$68$ [3]  \\
Galactic & $230$ & $-21$ & cone half-width=$68$ [3]  \\

Ecliptic & $66$ & $-36$ & center of Gaussian fit [4]  \\
Galactic & $211$ & $-35$ & center of  Gaussian fit [4] \\




\enddata 
\footnotesize{\tiny}
\tablerefs{
[1] \citet{McComas:2012bow}
[2] \citet{Funsten:2009sci}
[3] \citet{Nagashimaetal:1998}
[4] \citet{Halletal:1999gcr}
[5] \citet{Schwadronetal:2011sep}
[6] \citet{Salvati:2010}
}

\end{deluxetable}
		
\begin{deluxetable}{c  ccc ccc  c}
\tabletypesize{\footnotesize}
\tablewidth{0pt}
\tablecaption{New LNA and NOT Polarization Data \label{tab:lnanot}}
\tablehead{\colhead{HD} & \colhead{$\ell$}  & \colhead{$b$} & \colhead{Dist.} & \colhead{Pol} & \colhead{PA$_\mathrm{cel}$} & \colhead{PA$_\mathrm{gal}$ }& \colhead{Source}  \\
\colhead{} &  \colhead{deg}  & \colhead{deg} & \colhead{pc} & \colhead{$10^{-5}$} & \colhead{deg} & \colhead{deg} &\colhead{}   \\ 
}
\startdata
\tableline

     1581 & 309 & -53 &   8 & $    15 \pm     16$ & $    73 \pm     52$  & $    60 \pm     52$  & LNA   \\
     2025 &  32 & -83 &  18 & $   142 \pm     88$ & $   125 \pm     17$  & $    34 \pm     17$  & LNA   \\
     3443 &  69 & -84 &  15 & $    19 \pm     13$ & $   138 \pm     52$  & $     2 \pm     52$  & LNA   \\
     4628 & 120 & -57 &   7 & $     7 \pm     11$ & $    86 \pm     52$  & $    88 \pm     52$  & LNA   \\
     5133 & 297 & -87 &  14 & $    45 \pm     73$ & $    75 \pm     52$  & $    72 \pm     52$  & LNA   \\
     7570 & 290 & -71 &  15 & $    41 \pm     14$ & $    70 \pm      9$  & $    86 \pm      9$  & LNA   \\
    10360 & 289 & -59 &   8 & $    22 \pm     11$ & $   158 \pm     14$  & $   179 \pm     14$  & LNA   \\
    14412 & 209 & -70 &  12 & $     8 \pm     17$ & $    28 \pm     52$  & $   131 \pm     52$  & LNA   \\
    16160 & 162 & -48 &   7 & $    20 \pm     15$ & $   131 \pm     52$  & $    95 \pm     52$  & LNA   \\
    23356 & 210 & -50 &  14 & $   449 \pm     53$ & $     7 \pm      3$  & $   117 \pm      3$  & LNA   \\
   125072 & 313 &   2 &  11 & $    29 \pm    135$ & $   156 \pm     52$  & $   174 \pm     52$  & LNA   \\
   129502 & 346 &  47 &  18 & $     3 \pm      3$ & $   110 \pm     24$  & $   148 \pm     24$  & NOT   \\
   130819 & 340 &  38 &  23 & $     6 \pm      3$ & $    81 \pm     14$  & $   114 \pm     14$  & NOT   \\
   131923 & 323 &  10 &  24 & $    47 \pm     24$ & $    41 \pm     14$  & $    68 \pm     14$  & LNA   \\
   131976 & 338 &  33 &   7 & $    63 \pm     65$ & $    30 \pm     52$  & $    63 \pm     52$  & LNA   \\
   131977 & 338 &  33 &   5 & $    55 \pm     24$ & $    66 \pm     12$  & $    99 \pm     12$  & LNA   \\
   134987 & 339 &  27 &  25 & $    14 \pm      2$ & $    65 \pm      4$  & $   100 \pm      4$  & NOT   \\
   136894 & 340 &  24 &  28 & $    14 \pm      4$ & $    70 \pm      8$  & $   106 \pm      8$  & NOT   \\
   141272 & 369 &  40 &  21 & $    63 \pm     32$ & $    56 \pm     13$  & $   110 \pm     13$  & KVA   \\
   144253 & 352 &  23 &  18 & $    13 \pm      3$ & $    69 \pm      7$  & $   115 \pm      7$  & NOT   \\
   144585 & 358 &  27 &  28 & $     2 \pm      2$ & $    56 \pm     23$  & $   104 \pm     23$  & NOT   \\
   152311 &   0 &  14 &  28 & $     3 \pm      3$ & $    39 \pm     20$  & $    91 \pm     20$  & NOT   \\
   153631 &   7 &  17 &  26 & $     5 \pm      3$ & $     6 \pm     13$  & $    61 \pm     13$  & NOT   \\
   154088 & 355 &   7 &  18 & $     6 \pm      3$ & $    60 \pm     11$  & $   113 \pm     11$  & NOT   \\
   156897 &   3 &   8 &  17 & $     5 \pm      3$ & $   127 \pm     13$  & $     3 \pm     13$  & NOT   \\
   160346 &  27 &  18 &  10 & $    30 \pm     28$ & $   173 \pm     52$  & $    55 \pm     52$  & LNA   \\
   161096 &  29 &  17 &  25 & $     5 \pm      0$ & $   166 \pm      0$  & $    49 \pm      0$  & Wik   \\
   165222 &  24 &   9 &   7 & $    17 \pm     18$ & $    23 \pm     52$  & $    84 \pm     52$  & LNA   \\
   167665 &   4 &  -5 &  29 & $     4 \pm      3$ & $   113 \pm     16$  & $   174 \pm     16$  & NOT   \\
   169916 &   7 &  -6 &  23 & $     7 \pm      3$ & $   109 \pm     12$  & $   171 \pm     12$  & NOT   \\
   169916 &   7 &  -6 &  24 & $    36 \pm      7$ & $   109 \pm      7$  & $   172 \pm      7$  & KVA   \\
   170657 &  13 &  -4 &  13 & $     4 \pm      3$ & $    66 \pm     18$  & $   128 \pm     18$  & NOT   \\
   172051 &  12 &  -6 &  13 & $     2 \pm      3$ & $   110 \pm     26$  & $   173 \pm     26$  & NOT   \\
   176029 &  38 &   1 &  11 & $    25 \pm     23$ & $   141 \pm     52$  & $    23 \pm     52$  & LNA   \\
   177716 &   9 & -15 &  37 & $    38 \pm      7$ & $   113 \pm      7$  & $   179 \pm      7$  & KVA   \\
   178428 &  49 &   4 &  21 & $     2 \pm      3$ & $    63 \pm     23$  & $   125 \pm     23$  & NOT   \\
   179949 &  13 & -15 &  27 & $    10 \pm      3$ & $    77 \pm      7$  & $   144 \pm      7$  & NOT   \\
   180409 &  25 & -10 &  28 & $     4 \pm      2$ & $    95 \pm     16$  & $   159 \pm     16$  & NOT   \\
   180409 &  27 &  -9 &  28 & $    23 \pm     14$ & $    50 \pm     16$  & $   113 \pm     16$  & LNA   \\
   180617 &  40 &  -2 &   5 & $    37 \pm     82$ & $    25 \pm     52$  & $    87 \pm     52$  & LNA   \\
   184489 &  41 &  -7 &  14 & $     7 \pm     16$ & $    73 \pm     52$  & $   135 \pm     52$  & LNA   \\
   184509 &  18 & -18 &  31 & $     7 \pm      3$ & $   108 \pm     11$  & $   175 \pm     11$  & NOT   \\
   184985 &  25 & -16 &  30 & $     8 \pm      3$ & $    88 \pm      9$  & $   153 \pm      9$  & NOT   \\
   186427 &  83 &  13 &  21 & $     5 \pm      2$ & $   111 \pm      9$  & $   174 \pm      9$  & NOT   \\
   188088 &  18 & -22 &  14 & $    10 \pm     12$ & $   138 \pm     52$  & $    26 \pm     52$  & LNA   \\
   189245 &   8 & -27 &  20 & $    38 \pm     12$ & $   144 \pm      9$  & $    37 \pm      9$  & LNA   \\
   189340 &  33 & -18 &  24 & $    28 \pm     21$ & $   166 \pm     52$  & $    50 \pm     52$  & LNA   \\
   189931 &   4 & -29 &  27 & $    17 \pm     16$ & $   108 \pm     52$  & $     3 \pm     52$  & LNA   \\
   191408 &   5 & -30 &   6 & $    15 \pm     35$ & $    65 \pm     52$  & $   141 \pm     52$  & LNA   \\
   191849 & 355 & -32 &   6 & $    16 \pm      5$ & $   167 \pm      8$  & $    69 \pm      8$  & LNA   \\
   191862 &  30 & -23 &  28 & $     1 \pm      3$ & $   112 \pm     34$  & $   178 \pm     34$  & NOT   \\
   192310 &  15 & -28 &   8 & $    36 \pm     21$ & $   156 \pm     16$  & $    48 \pm     16$  & LNA   \\
   194640 &  13 & -32 &  19 & $    47 \pm   1760$ & $    17 \pm     52$  & $    91 \pm     52$  & LNA   \\
   194943 &  26 & -29 &  30 & $     1 \pm      3$ & $   170 \pm     35$  & $    58 \pm     35$  & NOT   \\
   196761 &  22 & -32 &  14 & $    25 \pm     18$ & $   128 \pm     52$  & $    19 \pm     52$  & LNA   \\
   202560 &   6 & -43 &   3 & $    10 \pm     32$ & $    89 \pm     52$  & $   175 \pm     52$  & LNA   \\
   207129 & 351 & -48 &  15 & $    32 \pm      9$ & $    93 \pm      8$  & $    14 \pm      8$  & LNA   \\
   211415 & 340 & -51 &  13 & $    13 \pm      5$ & $    41 \pm     10$  & $   156 \pm     10$  & LNA   \\
   213042 &  18 & -58 &  15 & $    18 \pm     18$ & $   108 \pm     52$  & $    14 \pm     52$  & LNA   \\
   222237 & 311 & -44 &  11 & $    86 \pm     97$ & $    52 \pm     52$  & $    28 \pm     52$  & LNA   \\
\tableline
\enddata
\end{deluxetable}  
		
\begin{deluxetable}{lccccc}
\tablecaption{PlanetPol\tablenotemark{*} stars in polarization envelope \label{tab:seven}}
\tablewidth{0pt} 
\tabletypesize{\small}
\footnotesize{\tiny}
\tablehead{ 
\colhead{HD} & \colhead{hame} & \colhead{Coordinates} & \colhead{Distance} & \colhead{Polarization} & \colhead{\PAcel} \\ 
\colhead{} & \colhead{} & \colhead{deg} & \colhead{pc} & \colhead{ppm} & \colhead{deg}  
}
\startdata 
172167   &  $\alpha$ Lyr  &    67.4,  19.2   &  7.8  &   $17.2 \pm 1.1 $ &  $34.5 \pm  1.4$ \nl
159561   &  $\alpha$ Oph  &  35.9,  22.6  &  14.3  &   $23.4 \pm  2.0 $ &  $30.8 \pm  2.4 $ \nl
161868  &  $\gamma$ Oph  &   28.0,  15.4  &    29.1  &  $40.8 \pm 3.1 $ &   $28.5 \pm 2.1 $ \nl
164058  &  $\gamma$ Dra  &     79.1,  29.2  &  45.2  &  $ 73.3 \pm 1.2  $   &  $145.0 \pm  0.5 $ \nl
186882   &   \nodata  &   78.7,  10.2   &   52.4   &   $108.0 \pm  2.0 $  &   $22.8 \pm 0.5$ \nl 
168775   &  \nodata  &     63.5 , 21.5  &     70.4  &  $106.5 \pm  2.7 $  &  $ 6.1 \pm  0.7 $  \nl
189319  & \nodata & 58.0 , --5.2 & 84.0 &  $199.5 \pm  1.4 $ &  $56.1  \pm  0.2$ \nl
\enddata 
\footnotesize{\tiny}
\tablenotetext{*}{\citet{BaileyLucas:2010planetpol}}
\end{deluxetable}
		

\clearpage

\clearpage

\clearpage

\end{document}